\documentclass[aps,amssymb,showkeys]{revtex4}

\usepackage{array}
\usepackage{tabularx}
\usepackage{multirow}
\usepackage{amsmath}
\usepackage{amsthm}
\usepackage{graphicx} 
\usepackage{amssymb}
\usepackage[pdftex, bookmarks, colorlinks=true, plainpages = false, citecolor = red, urlcolor = blue, filecolor = blue]{hyperref}
\usepackage[usenames,dvipsnames]{color}
\usepackage{mathrsfs}
\usepackage[shortlabels]{enumitem}
\setenumerate{listparindent=\parindent}
\usepackage{subeqnarray}
\usepackage{stmaryrd}

\newcommand{\red}{\textcolor{red}}

\newcommand{\be}{\begin{equation}}
\newcommand{\ee}{\end{equation}}
\newcommand{\bea}{\begin{eqnarray}}
\newcommand{\eea}{\end{eqnarray}}
\newcommand{\bean}{\begin{eqnarray*}}
\newcommand{\eean}{\end{eqnarray*}}

\theoremstyle{plain}

\theoremstyle{definition}

\allowdisplaybreaks

\begin{document}

\title{Multiple-SLE$_\kappa$ connectivity weights for rectangles, hexagons, and octagons}

\date{\today}

\author{Steven M. Flores}
\email{steven.miguel.flores@gmail.com} 
\affiliation{Department of Mathematics \& Statistics, University of Helsinki, P.O. Box 68, 00014, Finland,\\
and\\
Department of Mathematics, University of Michigan, Ann Arbor, Michigan, 48109-2136, USA}

\author{Jacob J.\ H.\ Simmons}
\email{jacob.simmons@mma.edu}
\affiliation{Maine Maritime Academy, Pleasant Street, Castine, ME, 04420, USA}

\author{Peter Kleban}
\email{kleban@maine.edu} 
\affiliation{FIRST and Department of Physics \& Astronomy, University of Maine, Orono, Maine, 04469-5708, USA}

\begin{abstract}  
In previous work, two of the authors determined, completely and rigorously, a solution space $\mathcal{S}_N$ for a homogeneous system of $2N+3$ linear partial differential equations (PDEs) in $2N$ variables that arises in conformal field theory (CFT) and multiple Schramm-L\"owner evolution (SLE$_\kappa$). The system comprises $2N$ null-state equations and three conformal Ward identities that govern CFT correlation functions of $2N$ one-leg boundary operators or SLE$_\kappa$ partition functions.  M.\ Bauer et al.\ conjectured a formula, expressed in terms of ``pure SLE$_\kappa$ partition functions," for the probability that the growing curves of a multiple-SLE$_\kappa$ process join in a particular connectivity.  In a previous article, we rigorously define certain elements of $\mathcal{S}_N$, which we call ``connectivity weights," argue that they are in fact pure SLE$_\kappa$ partition functions, and show how to find explicit  formulas for them in terms of Coulomb gas contour integrals.

Our formal definition of the connectivity weights immediately leads to a method for finding explicit expressions for them.  However, this method gives very complicated formulas where simpler versions may be available, and it is not applicable for certain values of $\kappa\in(0,8)$ corresponding to well-known critical lattice models in statistical mechanics.  In this article, we determine expressions for all connectivity weights in $\mathcal{S}_N$ for $N\in\{1,2,3,4\}$ (those with $N\in\{3,4\}$ are new) and for so-called ``rainbow connectivity weights" in $\mathcal{S}_N$ for all $N\in\mathbb{Z}^++1$. We verify these formulas by explicitly showing that they satisfy the formal definition of a connectivity weight.  In appendix \ref{appendixB}, we investigate logarithmic singularities of some of these expressions, appearing for certain values of $\kappa$ predicted by logarithmic CFT.

\end{abstract}

\maketitle

\section{Introduction}\label{intro}

This work is dedicated to Robert Ziff, on the occasion of his 70th birthday. Friend, collaborator, valued mentor, renowned expert on percolation, and a tireless unearther of citations and relevant new publications, he has for decades led and inspired others with his ideas, insights, and counsel.

This article proposes explicit formulas for special functions that arise in multiple SLE$_\kappa$ \cite{bbk,dub2,graham,kl,sakai} (and that also have a CFT interpretation \cite{florkleb,florkleb2,florkleb3,florkleb4}).  Called ``connectivity weights" \cite{florkleb4}, they are the main ingredients of a formula conjectured in \cite{bber,bbk,florkleb4} to give a \emph{crossing probability}, or the probability that the growing curves of a multiple-SLE$_\kappa$ process join together pairwise in some specified connectivity.  In this introduction, we briefly discuss crossing probabilities as natural observables of certain critical models in statistical physics, review the definition of a connectivity weight given in \cite{florkleb4}, recall some results from \cite{florkleb,florkleb2,florkleb3} that support this definition, and describe the organization of this article.

\begin{figure}[t!]
\centering
\includegraphics[scale=0.43]{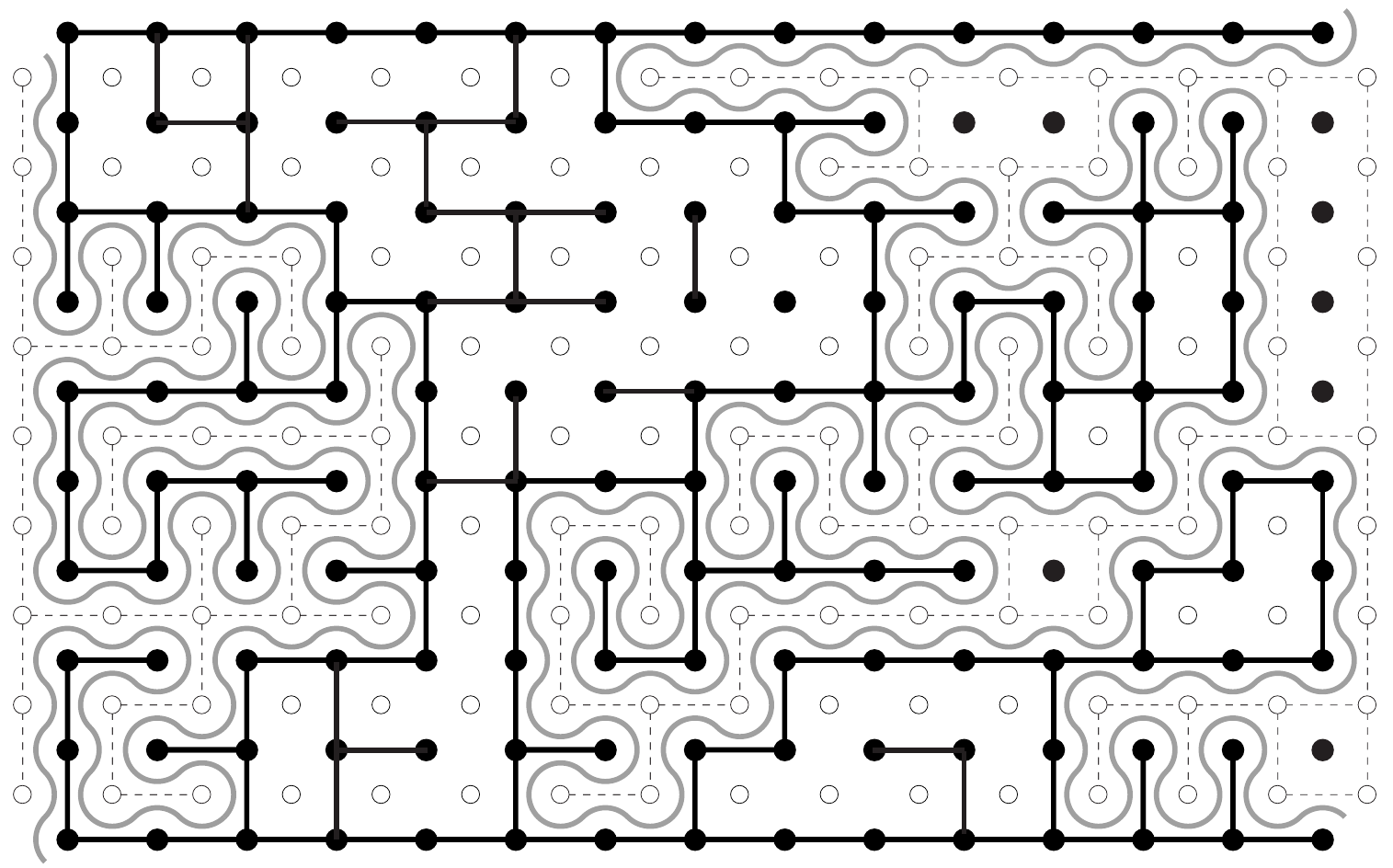}
\caption{A bond percolation configuration on a square lattice (black) in a rectangle, with a percolation cluster touching the top and bottom sides.  Activated bonds are black, and activated dual bonds are dashed.  Gray curves trace the cluster boundary.}
\label{CrossingFig}
\end{figure}

Crossing probabilities are natural observables of various random walks and critical models of statistical physics that possess a conformally invariant continuum limit.  Indeed, multiple SLE$_\kappa$, a generalization of ordinary SLE$_\kappa$, is expected to give the continuum limit of various mutually-avoiding random walks, including the loop-erased random walk $(\kappa=2)$ \cite{lsw} and the self-avoiding random walk $(\kappa=8/3)$ \cite{lsw2}, in a simply-connected domain.  Crossing probabilities are also natural observables of various statistical mechanics models.  Indeed, the archetypical example is the probability of a cluster-crossing event in critical percolation ($\kappa=6$) \cite{lsw3}.  Figure \ref{CrossingFig} illustrates this phenomenon for bond percolation on a discrete square lattice in a rectangle.   In the continuum limit of this crossing event (where we send the lattice spacing to zero but simultaneously increase the system size so it always fills the rectangle), the boundaries of the percolation cluster touching both the rectangle's top and bottom sides are conjectured to approach random curves that fluctuate to the law of multiple SLE$_6$ (figure \ref{CrossingFig}). (For site percolation on the triangular lattice, the marginal law for one of these curves is known to converge to SLE$_6$ \cite{smir2}, thanks in part to the ``locality" property of cluster interfaces \cite{glaw,card}.)  As such, the percolation vertical crossing event is, in the continuum limit, identically the event that the four multiple-SLE$_6$ curves exploring the inside of the rectangle, each with an endpoint at its own corner, eventually join pairwise to form two random curves.  One curve joins the two left corners of the rectangle, and the other curve joins the two right corners.  The probability of this event is given by \emph{Cardy's formula},
\be\label{Cardy}\mathbb{P}\{\text{top-bottom crossing}\}=\frac{\Gamma(2/3)}{\Gamma(4/3)\Gamma(1/3)}\lambda^{1/3}\,_2F_1\bigg(\frac{1}{3},\frac{2}{3};\frac{4}{3}\,\bigg|\,\lambda\bigg), \quad R=K(1-\lambda)/K(\lambda).\ee
Here, $\lambda\in(0,1)$ corresponds one-to-one with the aspect ratio $R\in(0,\infty)$ (that is, the ratio of the length of the bottom side to length of the left side) of the rectangle via the second equation in (\ref{Cardy}), with $K$ the complete elliptic integral of the first kind \cite{morsefesh,absteg}.  This formula (\ref{Cardy}) was first predicted by J.\ Cardy \cite{c3} using CFT methods.  It was subsequently proven by S.\ Smirnov \cite{smir2}, and later again by M.\ Khristoforov and S.\ Smirnov \cite{smir5}, for critical site percolation on the triangular lattice.  The bond percolation crossing event, and its relation to multiple SLE$_\kappa$, generalizes to similar crossing events in the critical Potts model, the closely-related random cluster model \cite{smir4,smir,gamsacardy}, and level lines of various height models \cite{schrsheff}.  We study some of these generalizations for polygon crossing events in the companion article \cite{fkz}.

Connectivity weights satisfy the system of differential equations that govern a correlation function comprising $2N$ one-leg boundary operators \cite{florkleb} in a conformal field theory (CFT) \cite{bpz,fms,henkel} with central charge $c\leq1$.  With $\boldsymbol{x}:=(x_1,x_2,\ldots,x_{2N})\in\Omega_0:=\{\boldsymbol{x}\in\mathbb{R}^{2N}\,|\,x_1<x_2<\ldots< x_{2N-1}< x_{2N}\}$ and $c=(6-\kappa)(3\kappa-8)/2\kappa$ \cite{bauber}, where $\kappa>0$ is the multiple-SLE$_\kappa$ speed, this system is
\begin{gather}\label{nullstate}\left[\frac{\kappa}{4}\partial_j^2+\sum_{k\neq j}^{2N}\left(\frac{\partial_k}{x_k-x_j}-\frac{(6-\kappa)/2\kappa}{(x_k-x_j)^2}\right)\right]F(\boldsymbol{x})=0,\quad j\in\{1,2,\ldots,2N\},\\
\label{wardid}\sum_{k=1}^{2N}\partial_kF(\boldsymbol{x})=0,\quad \sum_{k=1}^{2N}\left[x_k\partial_k+\frac{(6-\kappa)}{2\kappa}\right]F(\boldsymbol{x})=0,\quad \sum_{k=1}^{2N}\left[x_k^2\partial_k+\frac{(6-\kappa)x_k}{\kappa}\right]F(\boldsymbol{x})=0.\end{gather}
Following CFT nomenclature, we call the first $2N$ equations (\ref{nullstate}) \emph{null-state equations}, and we call the last three equations (\ref{wardid}) \emph{conformal Ward identities}.  In \cite{florkleb,florkleb2,florkleb3,florkleb4}, we study the vector space $\mathcal{S}_N$ (over the real numbers) comprising all (classical) real-valued solutions $F:\Omega_0\rightarrow\mathbb{R}$ of this system (\ref{nullstate}, \ref{wardid}) with the following property: there are positive constants $C$ and $p$ (possibly depending on $F$) such that
\be\label{powerlaw} |F(\boldsymbol{x})|\leq C\prod_{i<j}^{2N}|x_j-x_i|^{\mu_{ij}(p)},\quad\text{with}\quad\mu_{ij}(p):=\begin{cases}-p, & |x_j-x_i|<1 \\ +p, & |x_j-x_i|\geq1\end{cases}\quad\text{for all $\boldsymbol{x}\in\Omega_0.$}\ee
In \cite{florkleb,florkleb2,florkleb3,florkleb4}, two authors of this article rigorously prove the following facts concerning the solution space $\mathcal{S}_N$ for all $\kappa\in(0,8)$:
\begin{enumerate}
\item\label{item1} $\dim\mathcal{S}_N=C_N$, with $C_N$ the $N$th Catalan number: 
\be\label{catalan}C_N=\frac{(2N)!}{N!(N+1)!}.\ee
\item\label{item2} $\mathcal{S}_N$ is spanned by real-valued \emph{Coulomb gas solutions} \cite{florkleb3}.  These are linear combinations of (or limits as $\varkappa\rightarrow\kappa$ of linear combinations of) any functions that have the explicit formula \cite{df1,df2}
\be\label{CGsolns}F(\kappa\,|\,\boldsymbol{x})=\Bigg(\prod_{\substack{j<k \\ j,k\neq c}}^{2N}(x_k-x_j)^{2/\kappa}\Bigg)\Bigg(\prod_{\substack{k=1 \\ k\neq c}}^{2N}|x_c-x_k|^{1-6/\kappa}\Bigg)\mathcal{J}\Big(\kappa\,\Big|\,\Gamma_1,\Gamma_2,\ldots,\Gamma_{N-1}\,\Big|\,\boldsymbol{x}\Big),\ee
where $c\in\{1,2,\ldots,2N\}$ (we call $x_c$ \emph{the point bearing the conjugate charge}), $\mathcal{J}$ is the \emph{Coulomb gas integral} or \emph{Dotsenko-Fateev integral} and is given by
\begin{multline}\label{eulerintegral}
\mathcal{J}\Big(\kappa\,\Big|\,\Gamma_1,\Gamma_2,\ldots,\Gamma_{N-1}\,\Big|\,\boldsymbol{x}\Big)=\oint_{\Gamma_{N-1}}{\rm d}u_{N-1}\oint_{\Gamma_{N-2}}{\rm d}u_{N-2}\dotsm\\
\dotsm\,\,\oint_{\Gamma_2}{\rm d}u_2\,\,\oint_{\Gamma_1}{\rm d}u_1\,\Bigg(\prod_{\substack{l=1 \\ l\neq c}}^{2N}\prod_{m=1}^{N-1}(x_l-u_m)^{-4/\kappa}\Bigg)\Bigg(\prod_{m=1}^{N-1}(x_c-u_m)^{12/\kappa-2}\Bigg)\Bigg(\prod_{p<q}^{N-1}(u_p-u_q)^{8/\kappa}\Bigg),\end{multline}
and $\Gamma_1,$ $\Gamma_2,\ldots,\Gamma_{N-1}$ are nonintersecting closed contours in the complex plane.
\item\label{item3} The dual space $\mathcal{S}_N^*$ has a basis $\mathscr{B}_N^*:=\{[\mathscr{L}_1],[\mathscr{L}_2],\ldots,[\mathscr{L}_{C_N}]\}$ comprising equivalence classes of linear functionals $\mathscr{L}_\varsigma:\mathcal{S}_N\rightarrow\mathbb{R}$ called \emph{allowable sequences of limits} \cite{florkleb}.  We explain what these are below.
\end{enumerate}
One may construct the Coulomb gas solutions (\ref{CGsolns}, \ref{eulerintegral}) via the CFT Coulomb gas formalism, introduced by V.S.\ Dotsenko and V.A.\ Fateev \cite{df1,df2}. In \cite{dub}, J.\ Dub\'edat proves that these putative solutions indeed solve the system (\ref{nullstate}--\ref{wardid}).

\begin{figure}[b]
\includegraphics[scale=0.3]{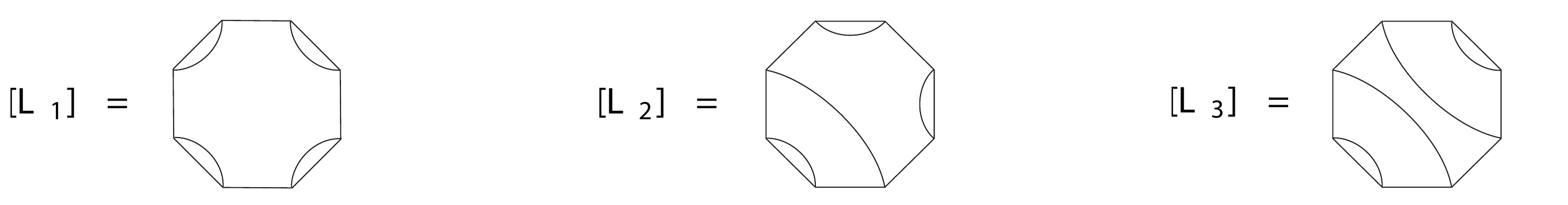}
\caption{Polygon diagrams for three different equivalence classes of allowable sequences of $N=4$ limits.  We find the other $C_4-3=11$ diagrams by rotating each of these three.}
\label{Csls}
\end{figure}

In \cite{florkleb,florkleb2,florkleb3,florkleb4}, we use certain elements of the dual space $\mathcal{S}_N^*$ to prove, rigorously, items \ref{item1}--\ref{item3} above.  To construct these linear functionals $\mathscr{L}:\mathcal{S}_N\rightarrow\mathbb{R}$, we prove in \cite{florkleb} that for all $F\in\mathcal{S}_N$ and all $i\in\{1,2,\ldots,2N-1\}$, the limits
\begin{align}\label{lim}\bar{\ell}_1F(x_1,x_2,\ldots,x_i,x_{i+2},\ldots,x_{2N})&:=\lim_{x_{i+1}\rightarrow x_i}(x_{i+1}-x_i)^{6/\kappa-1}F(x_1,x_2,\ldots,x_{2N}),\\
\label{lim2}\underline{\ell}_1F(x_2,x_3,\ldots,x_{2N-1})&:=\lim_{R\rightarrow\infty}(2R)^{6/\kappa-1}F(-R,x_2,x_3\ldots,x_{2N-1},R)\end{align}
exist, (\ref{lim}) is independent of $x_i$, and, after trivially sending $x_i\rightarrow x_{i-1}$ in (\ref{lim}), both are in $\mathcal{S}_{N-1}$.  Then, we let $\mathscr{L}$ be a composition of $N$ such limits.  These functionals gather into equivalence classes $[\mathscr{L}]$ whose elements differ only by the order in which we take their limits.  For convenience, we represent every equivalence class $[\mathscr{L}]$ by a unique \emph{polygon diagram}, in which $N$ nonintersecting arcs inside a $2N$-sided polygon $\mathcal{P}$ join pairwise the vertices of $\mathcal{P}$ (enumerated in counterclockwise order). The endpoints of the $j$th arc are the $i_{2j-1}$th and $i_{2j}$th vertices, where $x_{i_{2j-1}}<x_{i_{2j}}$ are the two points brought together by the $j$th limit of $\mathscr{L}$.  There are $C_N$ such diagrams, and they correspond one-to-one with the available equivalence classes (figure \ref{Csls}).  We enumerate the equivalence classes $[\mathscr{L}_1]$, $[\mathscr{L}_2],\ldots,[\mathscr{L}_{C_N}]$.

\begin{figure}[t]
\includegraphics[scale=0.3]{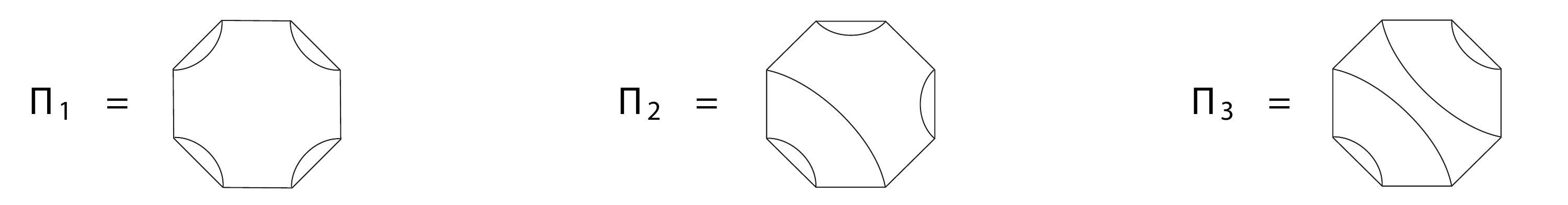}
\caption{Polygon diagrams for three different connectivity weights in $\mathscr{B}_4\subset\mathcal{S}_4$.  We find the other $C_4-3=11$ diagrams by rotating each of these three.  Compare with the polygon diagrams in figure \ref{Csls}.}
\label{PiDiagrams}
\end{figure}

We conclude our analysis in \cite{florkleb} with a rigorous proof that the linear mapping $v:\mathcal{S}_N\rightarrow\mathbb{R}^{C_N}$ with $v(F)_\varsigma:=[\mathscr{L}_\varsigma]F$ is well-defined and injective, so $\dim\mathcal{S}_N \leq C_N$.  Then in \cite{florkleb3}, we use this map again to show that a certain subset of $\mathcal{S}_N$ comprising $C_N$ distinct Coulomb gas solutions is linearly independent, thus establishing items \ref{item1} and \ref{item2} above.  These results imply that $v$ is an isomorphism, and item \ref{item3} above almost immediately follows from this fact.

Item \ref{item3} naturally leads us to consider the basis $\mathscr{B}_N$ for $\mathcal{S}_N$ dual to $\mathscr{B}_N^*\subset\mathcal{S}_N^*$.  We call the elements of $\mathscr{B}_N$ \emph{connectivity weights}, and we denote them as $\Pi_1,$ $\Pi_2,\ldots,\Pi_{C_N}$, where $\Pi_\vartheta$ is defined through  the duality relation
\begin{gather}\label{duality}\text{$[\mathscr{L}_{\varsigma}]\Pi_\vartheta=\delta_{\varsigma,\vartheta}$ for all $\varsigma,\vartheta\in\{1,2,\ldots,C_N\}$}, \\ 
\label{BNbasis}\mathscr{B}_N^*=\{[\mathscr{L}_1],[\mathscr{L}_2],\ldots,[\mathscr{L}_{C_N}]\},\quad\mathscr{B}_N=\{\Pi_1,\Pi_2,\ldots,\Pi_{C_N}\}.\end{gather}
We also define the \emph{polygon diagram} of $\Pi_\varsigma$ to be that of its corresponding equivalence class $[\mathscr{L}_\varsigma]$ (figure \ref{PiDiagrams}).  

The basis $\mathscr{B}_N$ (\ref{BNbasis}) is natural.  Indeed, because of the duality relation (\ref{duality}), any element $F\in\mathcal{S}_N$ has the decomposition over $\mathscr{B}_N$
\be\label{decompose}F=a_1\Pi_1+a_2\Pi_2+\dotsm+a_{C_N}\Pi_{C_N},\quad a_{\varsigma}=[\mathscr{L}_\varsigma]F.\ee

\begin{figure}[b]
\centering
\includegraphics[scale=0.3]{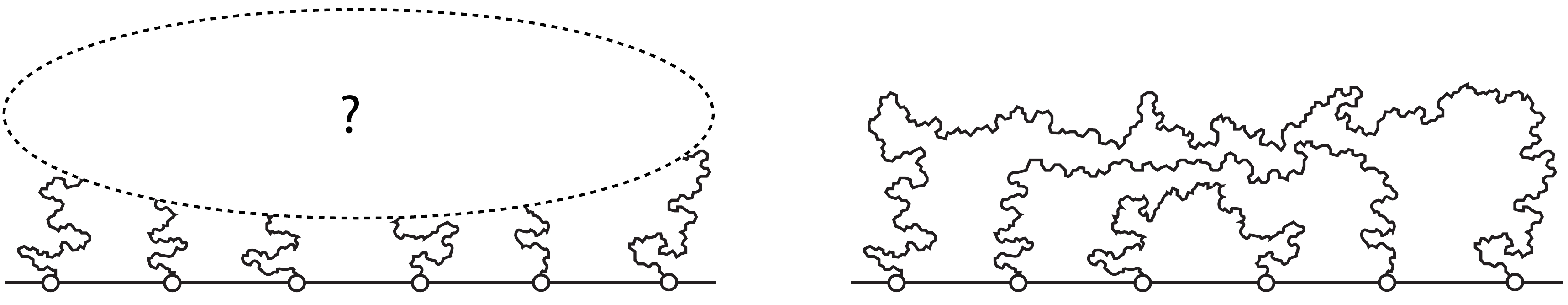}
\caption{Refs.\ \cite{florkleb4,bber} conjecture a formula for the probability that the growing curves of a multiple-SLE$_\kappa$ process (left) eventually join pairwise in the $\varsigma$th connectivity (right).}
\label{Connect}
\end{figure}

As we mentioned, the connectivity weights $\Pi_1$, $\Pi_2,\ldots,\Pi_{C_N}$ naturally arise in a certain stochastic process called \emph{multiple Schramm L\"owner evolution (SLE$_\kappa$)} \cite{bbk,dub2,graham,kl,sakai}.  In this process, $2N$ distinct self/mutually-avoiding fractal curves grow from the points $x_1<x_2<\ldots<x_{2N}$ in the real axis and explore the upper half-plane.  This process stops once the tips of these $2N$ curves join pairwise to form $N$ fractal curves called \emph{boundary arcs}.  The non-intersecting boundary arcs join the points $x_1<x_2<\ldots<x_{2N}$ together pairwise in any one of $C_N$ distinct connectivities (figure \ref{Connect}), and we enumerate these connectivities in such a way that every element of $[\mathscr{L}_\varsigma]$ brings together the endpoints of each arc in the $\varsigma$th connectivity.  The boundary arcs are conjectured, and in some cases proven, to be conformally invariant scaling limits of cluster perimeters in various critical lattice models or of certain random walks \cite{lsw,lsw2,smir3,smir,schrsheff,smir2}. Thus, multiple SLE$_\kappa$ is useful for studying these systems.

The multiple-SLE$_\kappa$ process is completely determined up to an arbitrary nonzero solution of the system (\ref{nullstate}--\ref{wardid}), called an \emph{SLE$_\kappa$ partition function} \cite{bbk,florkleb4}.  Without loss of generality, we suppose that an SLE$_\kappa$ partition function is always positive-valued.  Refs.\ \cite{florkleb4,bber} conjecture and give supporting arguments that 
\be\label{xing} P_\varsigma=\frac{a_\varsigma\Pi_\varsigma}{F}=\frac{a_\varsigma\Pi_\varsigma}{a_1\Pi_1+a_2\Pi_2+\dotsm+a_{C_N}\Pi_{C_N}},\quad a_{\varsigma}:=[\mathscr{L}_\varsigma]F\geq0
\ee  
gives the probability of the growing curves in the multiple-SLE$_\kappa$ process, with partition function $F\in\mathcal{S}_N$, almost surely joining pairwise in the $\varsigma$th connectivity (figure \ref{Connect}).

The proposed crossing-probability formula (\ref{xing}) predicts some fundamental properties of the connectivity weights that are not apparent from their formal definition (\ref{duality}).  For each $\varsigma\in\{1,2,\ldots,C_N\}$, if we choose $F$ so $a_\varsigma>0$ and  assume, as is natural, that $P_\varsigma$ is positive-valued, then  (\ref{xing}) immediately implies that $\Pi_\varsigma$ is positive-valued and is therefore itself an SLE$_\kappa$ partition function.  In fact, (\ref{xing}) gives $P_\varsigma=\delta_{\varsigma,\vartheta}$ for the multiple-SLE$_\kappa$ process with partition function $F=\Pi_\vartheta$, meaning that the growing curves of this process join pairwise in the $\vartheta$th connectivity almost surely.  Thanks to this special property, connectivity weights are also called \emph{pure SLE$_\kappa$ partition functions}, or more succinctly, \emph{pure partition functions}, in the literature \cite{bbk,bber,kype}.

Explicit formulas for the connectivity weights $\Pi_1,$ $\Pi_2,\ldots,\Pi_{C_N}$ are of interest because they are natural to use as a basis for $\mathcal{S}_N$ and are important to multiple SLE$_\kappa$.  One way to determine such formulas is through finding another basis $\mathcal{B}_N=\{\mathcal{F}_1,\mathcal{F}_2,\ldots,\mathcal{F}_{C_N}\}$ comprising the explicit solutions (\ref{CGsolns}, \ref{eulerintegral}), computing the coefficients in the decomposition (\ref{decompose}) of each basis element over $\mathscr{B}_N$, and inverting the collection of $C_N$ resulting equations to find $\Pi_1,$ $\Pi_2,\ldots,\Pi_{C_N}$.  In fact, we use this approach in \cite{florkleb4}.  However, determining a convenient basis for $\mathcal{S}_N$ and calculating the coefficients in (\ref{decompose}) may be difficult.  Moreover, the connectivity weight formulas that result may be unnecessarily complicated.  In addition, the so-called basis may fail to span $\mathcal{S}_N$ for certain $\kappa\in(0,8)$, causing this approach to fail for those values.

Independently of this work, K.\ Kyt\"ol\"a and E.\ Peltola developed a completely different approach for finding explicit solutions of the system (\ref{nullstate}, \ref{wardid}) that uses quantum group methods \cite{kype2}.  Called the \emph{spin-chain Coulomb gas correspondence}, their formalism gives another means for determining connectivity weight formulas (which they call ``pure partition functions") for all $N\in\mathbb{Z}^+$ and irrational $\kappa>0$ \cite{kype}.  Moreover, it has the advantage of explicitly determining the asymptotic properties of these formulas as two or more points $x_j$ approach each other simultaneously.  This information is expected to be useful for determining some anticipated, important properties of these functions, such as positivity \cite{florkleb4,kype2}.

But perhaps the most straightforward (if not the most elegant) approach for finding formulas of connectivity weights is to simply guess and verify them directly through the duality relation (\ref{duality}).  This approach is practical for small $N\in\mathbb{Z}^++1$ but very unwieldy for large $N$.  Fortunately, in most practical applications such as those involving lattice models or random walks inside $2N$-sided polygons, $N$ is small.  In section \ref{analysis}, we apply this approach to find explicit formulas for all connectivity weights in $\mathcal{S}_N$ with $N\in\{1,2,3,4\}$ and for special connectivity weights called ``rainbow connectivity weights" in $\mathcal{S}_N$ for all $N\in\mathbb{Z}^++1$.  The cases  $N\in\{1,2,3,4\}$ respectively pertain to multiple SLE$_\kappa$ in (or really, conformally mapped from the upper half-plane onto) the two-sided polygon, the rectangle, the hexagon, and the octagon.  For $N\in\{1,2\}$, these connectivity weight formulas are already known \cite{bbk,dub}, but for $N\in\{3,4\}$, these formulas are new.  In section \ref{summary}, we summarize our results.  In appendix \ref{appendix}, two authors of this article explicitly show that all singularities in $\kappa\in(0,8)$ of factors in the connectivity weight formulas are removable singularities, and the connectivity weights are therefore analytic functions of $\kappa\in(0,8)$ (and, actually, of $\kappa\in(0,8)\times i\mathbb{R}$ too).  Finally, in appendix \ref{appendixB}, we study logarithmic singularities of some of the connectivity weight formulas as one or more points $x_j$ approach a common point $x_i$.  These logarithmic singularities may arise only if $\kappa$ is an \emph{exceptional speed} \cite{florkleb3,florkleb4} and $N$ is sufficiently large.  Such $\kappa$ correspond with CFT minimal models \cite{florkleb4}.  In appendix \red{A} of \cite{florkleb4}, we investigate the existence of logarithmic singularities in certain elements of $\mathcal{S}_N$ for $8/\kappa\in2\mathbb{Z}^++1$, and in appendix \ref{appendixB} of this article, we uncover a similar logarithmic behavior of the hexagon connectivity weight (\ref{Pi1N3}) for $12/\kappa\in\mathbb{Z}^++1$ coprime with three.  For both cases, we briefly discuss how logarithmic CFT predicts the appearance of these logarithmic singularities.

In \cite{fkz}, we conformally map the multiple-SLE$_\kappa$ process with $N=3$ onto a hexagon, and we use (\ref{xing}) and the results of section \ref{HexXingSect} to find explicit formulas for crossing probabilities as functions of the hexagon's shape.  To verify this formula for $\kappa=\{16/3,24/5,4\}$  we measure via computer simulation $Q\in\{2,3,4\}$ random cluster model crossing probabilities in a hexagon with a free/fixed side-alternating boundary condition \cite{florkleb}.  (Multiple-SLE$_\kappa$ at these values of $\kappa$ corresponds with these respective models.  See \cite{florkleb} for more details.)   The $Q=1$ random cluster model corresponds with critical percolation $(\kappa=6)$, and in \cite{fzs}, we provide similar verification for this case.

Actually, the case $\kappa=6$ has a unique, interesting feature.  In percolation, the free/fixed side-alternating boundary condition \cite{florkleb} of the $2N$-sided polygon does not influence the probabilities of the configurations of the interior sites or bonds.  As such, the partition function for the system conditioned on this boundary-condition event approaches the free partition function, summing over all bond configurations, as we approach the continuum limit.  According to \cite{bbk,fkz}, it is natural to interpret $F$ as the ratio of these two partition functions in this limit, which is then one.  Upon inserting $F=1$ into (\ref{xing}) with $\kappa=6$, we find that $P_\varsigma=\Pi_\varsigma$.  That is, if $\kappa=6$, then the $\varsigma$th connectivity weight gives the probability that percolation clusters join the fixed sides of the polygon in the $\varsigma$th connectivity.

\section{Analysis}\label{analysis}

In this section, we present formulas for all connectivity weights in $\mathcal{S}_N$ with $N\in\{1,2,3,4\}$ and for all so-called ``rainbow connectivity weights" in $\mathcal{S}_N$ with $N\in\mathbb{Z}^++1$.  The $N\in\{3,4\}$ and ``rainbow" results are, to our knowledge, new.  Our derivations proceed in two steps.  First, we assume that each sought formula has the form (\ref{CGsolns}) and prudently choose integration contours for the Coulomb gas integral (\ref{eulerintegral}) of that formula.  Second, we verify that our ansatz is correct.  Because our  candidate formula gives an element of $\mathcal{S}_N$, we must only verify that it satisfies the duality condition (\ref{duality}) of a connectivity weight.

\begin{figure}[b]
\centering
\includegraphics[scale=0.28]{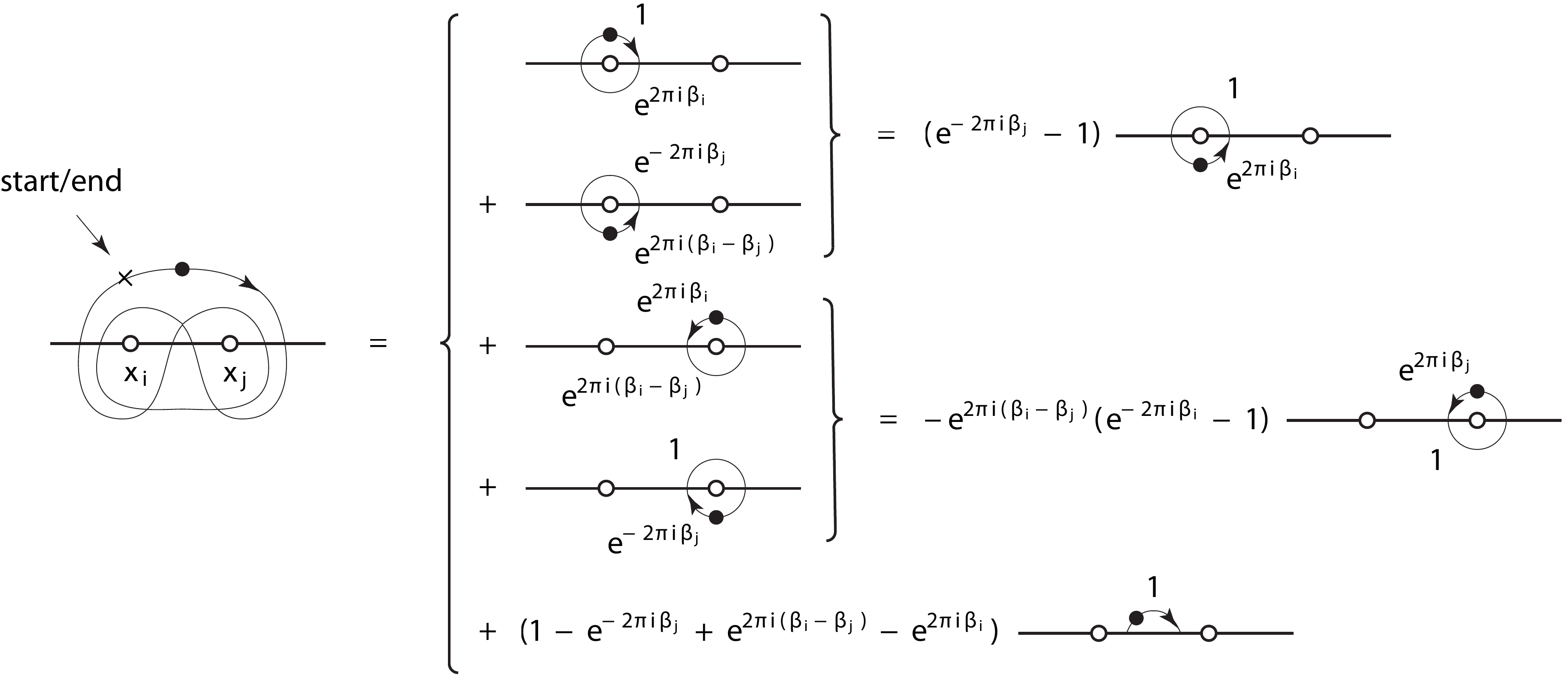}
\caption{The elementary Pochhammer contour $\mathscr{P}(x_i,x_j)$ (left) with ``endpoints" $x_i$ and $x_j$, and its decomposition (\ref{PochDecomp}) (right).  The phase factor of the integrand at the start point and end point (at the tip of the arrow) of each contour is shown.}
\label{BreakDown}
\end{figure}

\begin{figure}[t]
\centering
\includegraphics[scale=0.28]{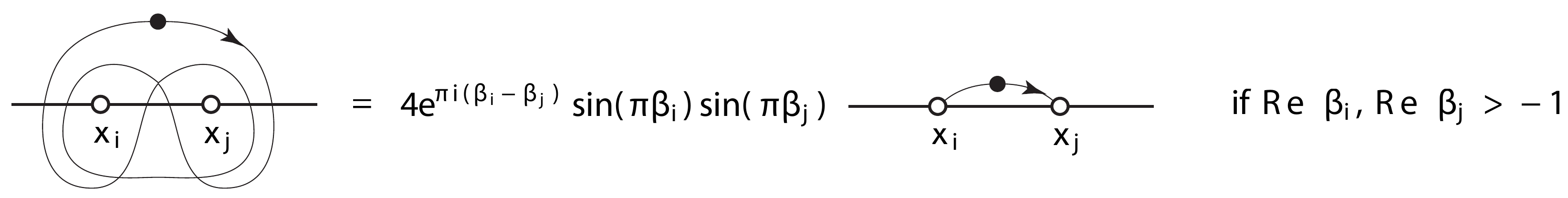}
\caption{If $e^{2\pi i\beta_i}$ and $e^{2\pi i\beta_j}$ are the monodromy factors associated with the endpoints $x_i$ and $x_j$ respectively, and $\text{Re}\,\beta_i,\text{Re}\,\beta_j>-1$, then we may replace the elementary Pochhammer contour $\mathscr{P}(x_i,x_j)$ with the simple contour on the right via (\ref{Pochtostraight}).}
\label{PochhammerContour}
\end{figure}

Certain constraints limit the various possible integration contours available for use in (\ref{eulerintegral}).  To begin, each contour must close in order for (\ref{CGsolns}) to solve the system (\ref{nullstate}, \ref{wardid}).  Furthermore, by Cauchy's theorem, each contour must circle around some of the branch points $x_1$, $x_2,\ldots,x_{2N}$, thereby passing onto different Riemann sheets of the integrand, in order for the integration to give something nontrivial.  And finally, each contour must have a winding number of zero around these branch points in order to end on the same Riemann sheet on which it started.  The simplest contour that satisfies these criteria is the \emph{Pochhammer contour}, shown in the left illustration of figures \ref{BreakDown} and \ref{PochhammerContour}.  In \cite{florkleb3,florkleb4}, we only consider Pochhammer contours that entwine together two branch points, but in this article, we consider Pochhammer contours that entwine together more than two branch points or that surround other integration contours or both.

Both theorem \red{5}  and section \red{IV B} of \cite{florkleb4} motivate our integration-contour selections for the Coulomb gas integral (\ref{eulerintegral}).  In particular, theorem \red{5} says that for any $\Pi_\varsigma\in\mathscr{B}_N$ and $i\in\{1,2,\ldots,2N-1\}$, 
\begin{enumerate}[label=\Roman*.,ref=\Roman*]
\item\label{twolegitem} $(x_i,x_{i+1})$ is a \emph{two-leg interval} of $\Pi_\varsigma$ (meaning that the limit $\bar{\ell}_1\Pi_\varsigma$ (\ref{lim})  vanishes) if no arc joins the $i$th vertex with the $(i+1)$th vertex in the polygon diagram for $\Pi_\varsigma$.
\item $(x_i,x_{i+1})$ is not a two-leg interval of $\Pi_\varsigma$ if an arc joins the $i$th vertex with the $(i+1)$th vertex in the polygon diagram for $\Pi_\varsigma$.
\end{enumerate}
On the other hand, the discussion in section \red{IV B} of \cite{florkleb4} specifies two scenarios for the interaction of the integration contours with the interval $(x_i,x_{i+1})$ in which this interval is a two-leg interval of $\Pi_\varsigma$.
\begin{enumerate}[label=(\alph*),ref=\alph*]
\item\label{itb} If no integration contour in a formula for $\Pi_\varsigma$ of the form (\ref{CGsolns}, \ref{eulerintegral}) either crosses $(x_i,x_{i+1})$ or touches its endpoints and if $c\not\in\{i,i+1\}$, then $(x_i,x_{i+1})$ is a two-leg interval of $\Pi_\varsigma$.
\item\label{ita} If a Pochhammer contour in a formula for $\Pi_\varsigma$ of the form (\ref{CGsolns}, \ref{eulerintegral}) entwines together both endpoints of $(x_i,x_{i+1})$ (as in figure \ref{BreakDown} with $j=i+1$) and if $c\in\{i,i+1\}$, then $(x_i,x_{i+1})$ is a two-leg interval of $\Pi_\varsigma$.
\end{enumerate}
Selecting integration contours for connectivity weights based on items \ref{itb} and \ref{ita} above is equivalent to constructing conformal blocks in CFT via Coulomb gas methods \cite{florkleb4,fms,henkel,df1,df2}.

In order for us to obtain numerical values of crossing probabilities (so we may, for example, compare these predicted values to computer simulation measurements), we must numerically evaluate the connectivity weights that appear in (\ref{xing}).  Formulas for connectivity weights typically involve integration around a Pochhammer contour, and although this integration is not entirely straightforward, we may perform it by decomposing the contour into a collection of loops and line segments.  This decomposition is simplest if a Pochhammer contour is \emph{elementary}, meaning that it entwines together only two ``endpoints" $x_i$ and $x_j$.  We denote such a contour by $\mathscr{P}(x_i,x_j)$.  As figure \ref{BreakDown} shows, for any function $f(u)$ that is analytic in the interior of a region containing $\mathscr{P}(x_i,x_j)$, any $\beta_i,\beta_j\in\mathbb{C}$, and any positive $\epsilon\ll |x_j-x_i|$, we have (as we account for the phase factors, we take $-\pi<\arg(z)\leq\pi$ for all complex $z$)
\begin{align}\sideset{}{_{\mathscr{P}(x_i,x_j)}}\oint(u-x_i)^{\beta_i}(x_j-u)^{\beta_j}f(u)\,{\rm d}u
&=(e^{-2\pi i\beta_j}-1)\oint_{x_i}(u-x_i)^{\beta_i}(x_j-u)^{\beta_j}f(u)\,{\rm d}u\nonumber\\
\label{PochDecomp}&-e^{2\pi i(\beta_i-\beta_j)}(e^{-2\pi i\beta_i}-1)\oint_{x_j}(u-x_i)^{\beta_i}(x_j-u)^{\beta_j}f(u)\,{\rm d}u\\
&+4e^{\pi i(\beta_i-\beta_j)}\sin\pi \beta_i\sin\pi \beta_j\sideset{}{_{x_i+\epsilon}^{x_j-\epsilon}}\int (u-x_i)^{\beta_i}(x_j-u)^{\beta_j}f(u)\,{\rm d}u,\nonumber\end{align}
where the subscript $x_i$ (resp.\ $x_j$) on the first (resp.\ second) integral on the right side of (\ref{PochDecomp}) indicates that $u$ traces counterclockwise a circle centered on $x_i$ (resp.\ $x_j$) with radius $\epsilon$, starting just above $x_i+\epsilon$ (resp.\ below $x_j-\epsilon$) where the integrand's phase is zero.  If $\text{Re}\,\beta_i,\text{Re}\,\beta_j>-1$, then sending $\epsilon\rightarrow0$ in (\ref{PochDecomp}) gives the useful identity
\begin{multline}\label{Pochtostraight}\sideset{}{_{\mathscr{P}(x_i,x_j)}}\oint(u-x_i)^{\beta_i}(x_j-u)^{\beta_j}f(u)\,{\rm d}u=\\
 4e^{\pi i(\beta_i-\beta_j)}\sin\pi \beta_i\sin\pi \beta_j\sideset{}{_{x_i}^{x_j}}\int (u-x_i)^{\beta_i}(x_j-u)^{\beta_j}f(u)\,{\rm d}u,\quad\text{Re}\,\beta_i,\text{Re}\,\beta_j>-1\end{multline}
(figure \ref{PochhammerContour}).  In other words, if $\text{Re}\,\beta_i,\text{Re}\,\beta_j>-1$, then we may discard the integrations in (\ref{PochDecomp}) around the loops.  Because the right side of (\ref{Pochtostraight}) is easier to evaluate numerically than the left side, we use the former whenever it converges.  We note that $\beta_i,\beta_j\in\mathbb{Z}^+\cup\{0\}$ are zeros of (\ref{PochDecomp}, \ref{Pochtostraight}).

\begin{figure}[b]
\centering
\includegraphics[scale=0.27]{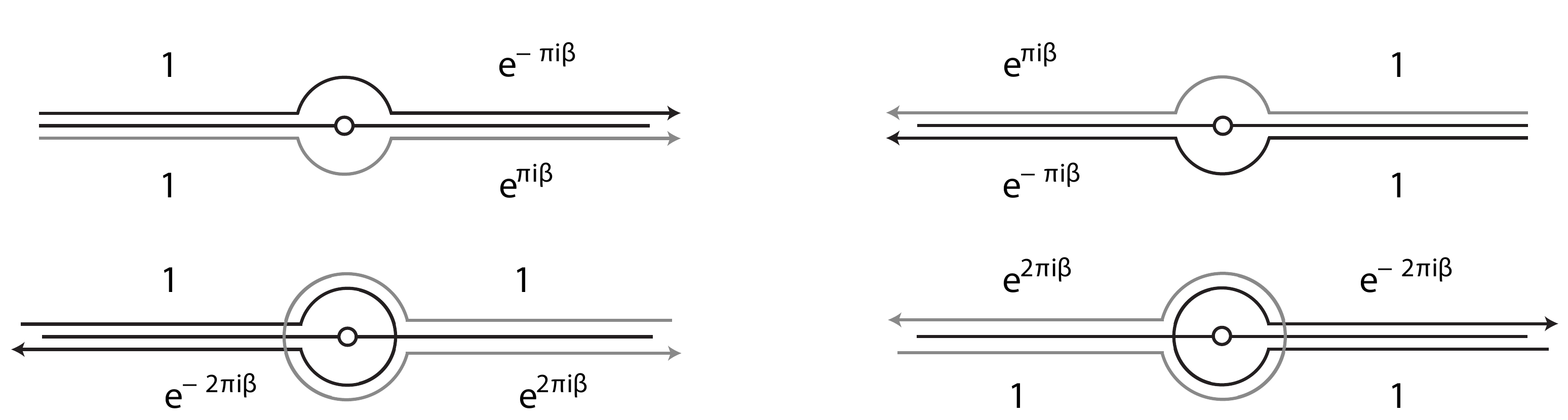}
\caption{Monodromy factors accrued by the function $(x-u)^\beta$ as $u$ passes over or under (upper illustration) or encircles around (lower illustration) the branch point $x$ (open circle on the real axis).}
\label{Phases}
\end{figure}

If a Pochhammer contour is not elementary but surrounds several branch points and/or integration contours, then we may decompose it into a collection of loops, each of which surrounds a branch point of the integrand, and line segments that join these loops, although the decomposition is more complicated than (\ref{PochDecomp}).  Furthermore, if the integrations along these segments converge, then we may shrink the radius of the loops to zero so the line segments terminate at the branch points.  (This is convenient for numerical evaluation because the resulting decomposition only contains integrations along line segments.)  As we account for the phase factors that arise in these decompositions, we take $-\pi<\arg(z)\leq\pi$ for all complex $z$.  This sets the conventions of figure \ref{Phases} for use throughout this article.

Ignoring the subtle issue of convergence, we express all formulas for connectivity weights $\Pi_\varsigma$ as sums of improper integrations along line segments that terminate at branch points, discarding the contributions that arise from integrating around loops.  If an improper integral does diverge, then we use identity (\ref{Pochtostraight}) to make the replacement 
\be\label{replacePoch}\sideset{}{_{x_i}^{x_j}}\int (u-x_i)^{\beta_i}(x_j-u)^{\beta_j}f(u)\,{\rm d}u\quad\longmapsto\quad\frac{1}{4e^{\pi i(\beta_i-\beta_j)}\sin\pi \beta_i\sin\pi \beta_j}\sideset{}{_{\mathscr{P}(x_i,x_j)}}\oint(u-x_i)^{\beta_i}(x_j-u)^{\beta_j}f(u)\,{\rm d}u.\ee
Although this replacement reintroduces a Pochhammer contour, this new Pochhammer contour is elementary, so we may use (\ref{PochDecomp}) to numerically integrate around it.  To summarize, the decomposition has the form
\begin{align}\label{form0}\Pi_\varsigma&\propto\oint_{\Gamma_1}\oint_{\Gamma_2}\dotsm\,\,\oint_{\Gamma_{N-1}}\big[\ldots\text{integrand of (\ref{eulerintegral})}\ldots\big]\,{\rm d}u_{N-1}\,\dotsm\,{\rm d}u_2\,{\rm d}u_1,\quad\Gamma_m=\text{Pochhammer contour},\\
\label{form1}&=\sum\,\,\big[\ldots\text{integration along line segments $[x_i,x_j]$}\ldots\big]\\
\label{form2}&=\sum\,\,\big[\ldots\text{integration around elementary $\mathscr{P}(x_i,x_j)$}\ldots\big],\end{align}
and we give the formula for each $\Pi_\varsigma$ in the form (\ref{form1}), with the form (\ref{form2}) implicitly used, when needed, by making the replacement (\ref{Pochtostraight}) for all contours.  Every definite integral that appears in the sum (\ref{form1}) has $\beta_i,\beta_j\in\{-4/\kappa,8/\kappa,12/\kappa-2\}$, with $\beta_i=-4/\kappa$ or $\beta_j=-4/\kappa$, so all (resp.\ none) of them diverge if $\kappa\in(0,4]$ (resp.\ $\kappa\in(4,8)$).  Because the right sides of (\ref{form0}--\ref{form2}) are equal for $\kappa\in(4,8)$ and (\ref{form0}) is analytic in $\kappa\in(0,8)\times i\mathbb{R}$, it follows that (\ref{form2}) is also analytic in this region, in particular at the poles $\beta_i,\beta_j\in\mathbb{Z}^-$ of the right side of (\ref{replacePoch}), so (\ref{form2}) gives the analytic continuation of (\ref{form1}) to the region $\kappa\in(0,8)\times i\mathbb{R}$.  This includes the line segment $\kappa\in(0,8)$ of exclusive interest here.

\begin{figure}[t]
\centering
\includegraphics[scale=0.27]{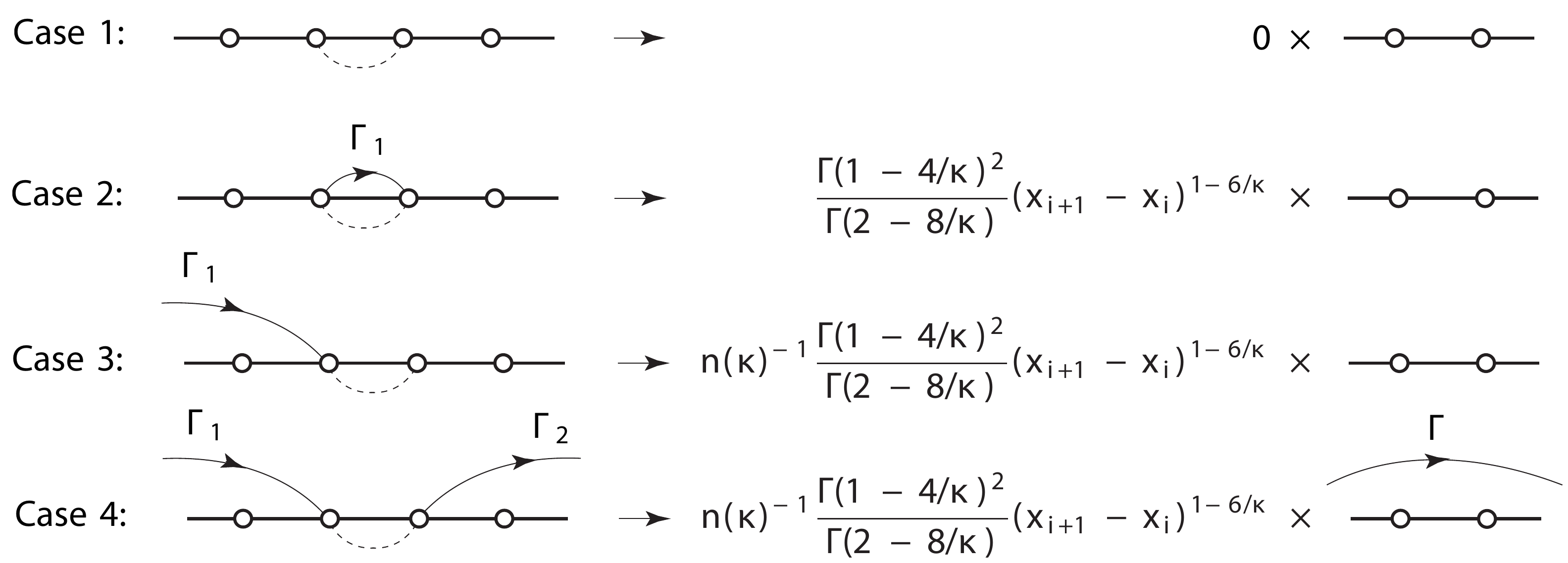}
\caption{Cases \ref{sc1}--\ref{sc4} (left) and their asymptotic behavior as $x_{i+1}\rightarrow x_i$ (right).  The dashed curve connects $x_i$ with $x_{i+1}$ (neither present on the right), and solid curves are integration contours.  The endpoints of $\Gamma$ are the endpoints of $\Gamma_1$ and $\Gamma_2$, not shown.}
\label{Cases}
\end{figure}

In order to correctly normalize the connectivity weight $\Pi_\varsigma\in\mathscr{B}_N$ such that (\ref{duality}) is satisfied, it suffices to require that the limit (\ref{lim}) with $F=\Pi_\varsigma$ such that $(x_i,x_{i+1})$ is not a two-leg interval of $\Pi_\varsigma$ (item \ref{twolegitem}) gives another connectivity weight $\Xi_{\varsigma'}\in\mathscr{B}_{N-1}$ for some $\varsigma'\in\{1,2,\ldots,C_{N-1}\}$.  Indeed, we then have
\be\label{takefirstlim} [\mathscr{L}_\varsigma]\Pi_\varsigma=[\mathscr{M}_{\varsigma'}]\bigg(\overbrace{\lim_{x_{i+1}\rightarrow x_i}(x_{i+1}-x_i)^{6/\kappa-1}\Pi_\varsigma(\boldsymbol{x})}^{(\ref{lim})}\bigg)=[\mathscr{M}_{\varsigma'}]\Xi_{\varsigma'}=1,\ee
where $[\mathscr{M}_{\varsigma'}]\in\mathscr{B}_{N-1}^*$ is the equivalence class produced by dropping the first limit $\bar{\ell}_1$ from all elements of $[\mathscr{L}_\varsigma]$ that send $x_{i+1}\rightarrow x_i$ first.  Computing this limit involves determining the asymptotic behavior of the Coulomb gas integral (\ref{eulerintegral}) as $x_{i+1}\rightarrow x_i$ for $i\not\in\{c,c-1\}$, which is generally complicated to do.  To simplify this calculation for $\kappa\in(4,8)$, we use (\ref{Pochtostraight}) to decompose all Pochhammer contours that surround $x_i$ or $x_{i+1}$ into line segments that terminate at these points in one of the following four cases:
\begin{enumerate}
\item\label{sc1} Neither $x_i$ nor $x_{i+1}$ are endpoints of an integration contour.
\item\label{sc2} Both $x_i$ and $x_{i+1}$ are endpoints of one common contour, say $\Gamma_1$.
\item\label{sc3} $x_i$ (resp.\ $x_{i+1}$) is an endpoint of one contour, say $\Gamma_1$, and $x_{i+1}$ (resp.\ $x_i$) is not an endpoint of any contour.
\item\label{sc4} $x_i$ is an endpoint of one contour, say $\Gamma_1$, and $x_{i+1}$ is an endpoint of a different contour, say $\Gamma_2$.
\end{enumerate}
Figure \ref{Cases} illustrates cases \ref{sc1}--\ref{sc4} and shows the asymptotic behavior of the Coulomb gas integral (\ref{eulerintegral}) in each case, as was found in the proof of lemma \red{6} in \cite{florkleb3}.  (Cases \ref{sc3} and \ref{sc4} include situations in which one of the integration contours arcs over $(x_i,x_{i+1})$, not shown in figure \ref{Cases}.)  In cases \ref{sc3} and \ref{sc4}, we encounter the reciprocal of
\be\label{fugacity}n(\kappa):=-2\cos(4\pi/\kappa),\ee
called the \emph{O($n$)-model fugacity function} because loops in the loop-gas representation of the O$(n)$ model are conjectured to fluctuate to the law of SLE$_\kappa$/CLE$_\kappa$ \cite{gruz,rgbw,smir4,smir,shefwer,sheffield,doyon}, with speed $\kappa\in(8/3,8)$ related to the loop fugacity $n$ through (\ref{fugacity}).  Case \ref{sc4} yields the result of case \ref{sc3}, but with the original two contours $\Gamma_1$ and $\Gamma_2$ replaced by one terminating at the other endpoints of $\Gamma_1$ and $\Gamma_2$.

Finally, to find formulas for all of the connectivity weights in $\mathscr{B}_N$ is somewhat redundant.  Indeed, if rotating the polygon diagram for one connectivity weight $\Pi_\varsigma$ gives the diagram of another $\Pi_\vartheta$, then a similar transformation of the formula for $\Pi_\varsigma$ gives the formula for $\Pi_\vartheta$.  Therefore, we give only one  formula for each collection of connectivity weights whose polygon diagrams are identical up to rotation.

\subsection{The SLE$_\kappa$ connectivity weight ($N=1$)}
According to item \ref{item1} of the introduction \ref{intro}, the dimension of the solution space $\mathcal{S}_1$ for the system (\ref{nullstate}, \ref{wardid}) is $C_1=1$ (\ref{catalan}), so there is only one connectivity weight that spans all of $\mathcal{S}_1$.  In \cite{florkleb}, we find that every element of $\mathcal{S}_1$ is a multiple of the function
\be\label{Pi1Xing1}\Pi_1(\kappa\,|\,x_1,x_2)=(x_2-x_1)^{1-6/\kappa}.\ee 
The limit (\ref{lim}) with $i=1$ gives $\bar{\ell}_1\Pi_1=1$, so according to the duality condition (\ref{duality}), the (aptly-named) function $\Pi_1$ (\ref{Pi1Xing1}) gives the formula for the lone connectivity weight we seek.  This weight serves as the partition function for ordinary SLE$_\kappa$ in the upper half-plane and from $x_1$ to $x_2$ \cite{bbk,kl}.

\begin{figure}[b]
\centering
\includegraphics[scale=0.27]{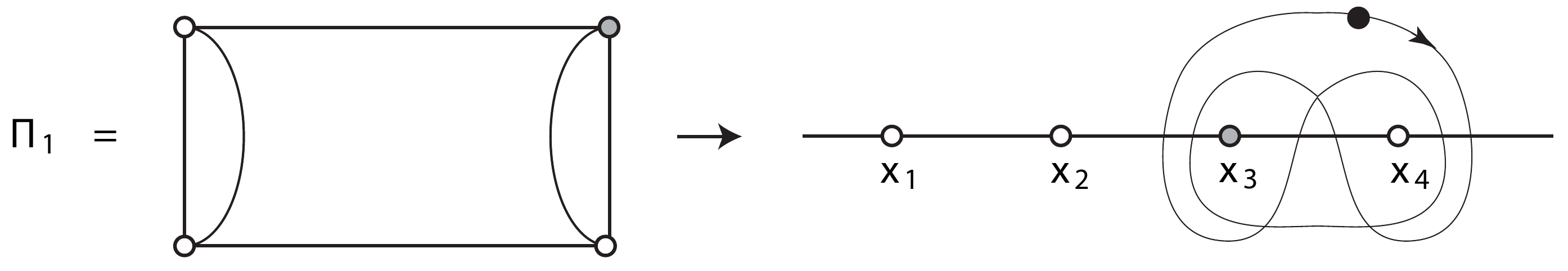}
\caption{The one topologically distinct rectangle connectivity, and the integration contour for its weight with $c=3$.  Moving counterclockwise around the rectangle corresponds to moving right along the real axis, with the bottom-left vertex sent to $x_1$.}
\label{RecXing}
\end{figure}

\subsection{Rectangle connectivity weights ($N=2$)}\label{RecXingSect}

According to item \ref{item1} of the introduction \ref{intro}, the dimension of the solution space $\mathcal{S}_2$ for the system (\ref{nullstate}, \ref{wardid}) is $C_2=2$ (\ref{catalan}), so there are two connectivity weights $\Pi_1$ and $\Pi_2$ that span $\mathcal{S}_2$.  Because their diagrams are identical up to a rotation, we find a formula for only $\Pi_1$.  An appropriate transformation then gives the formula for $\Pi_2$.

Now we find a formula for $\Pi_1$ in the form of (\ref{CGsolns}) with $N=2$.  After choosing $c=3$, we determine the integration contour of (\ref{eulerintegral}) for use in this formula.  Figure \ref{RecXing} shows the polygon diagram for $\Pi_1$, and from this diagram and item \ref{twolegitem} above, it is apparent that $(x_3,x_4)$ is a two-leg interval of $\Pi_1$.  Hence, following item \ref{ita} above, we entwine $x_3$ and $x_4$ with a Pochhammer contour $\Gamma_1=\mathscr{P}(x_3,x_4)$.  This choice of integration contour determines the formula for $\Pi_1$ up to normalization.

Assuming that $\kappa\in(4,8)$, we replace the Pochhammer contour by an integration along $[x_3,x_4]$ via (\ref{Pochtostraight}) (figure \ref{PochhammerContour}).  Thus, the Coulomb gas integral (\ref{eulerintegral}) with $N=2$ becomes (here, we implicitly order the differences in the factors of the integrand in (\ref{eulerintegral}) such that (\ref{J11}) is positive-valued)
\be\label{J11}I_3(\kappa\,|\,x_1,x_2,x_3,x_4):=\mathcal{J}\Big(\kappa\,\Big|\,[x_3,x_4]\,\Big|\,x_1,x_2,x_3,x_4\Big),\quad c=3,\quad N=2.\ee
After inserting the Coulomb gas integral (\ref{J11}) into (\ref{CGsolns}), we find the proper normalization for the resulting formula by requiring that the limit (\ref{lim2}) equals the $N=1$ connectivity weight (\ref{Pi1Xing1}) with $(x_1,x_2)\mapsto(x_2,x_3)$.  Upon using the results of sections \red{A 3} and \red{A 5} of \cite{florkleb3} to determine the asymptotic behavior of (\ref{J11}) as $x_4=-x_1\rightarrow\infty$ (figure \ref{Cases}), we find
\be\label{Pi1}\Pi_1(\kappa\,|\,x_1,x_2,x_3,x_4)=n(\kappa)\frac{\Gamma(2-8/\kappa)}{\Gamma(1-4/\kappa)^2}\Bigg(\prod_{\substack{i<j \\ i,j\neq 3}}^4(x_j-x_i)^{2/\kappa}\Bigg)\Bigg(\prod_{k\neq 3}^4|x_3-x_k|^{1-6/\kappa}\Bigg)I_3(\kappa\,|\,x_1,x_2,x_3,x_4).\ee
  If $\kappa\in(0,4]$, then the improper integral (\ref{J11}) diverges, and we regularize it via the replacement (\ref{replacePoch}) with $x_i=x_3$, $x_j=x_4$, $\beta_i=12/\kappa-2$, and $\beta_j=-4/\kappa$.

Employing the contour-integral definition of the Gauss hypergeometric function \cite{morsefesh,absteg}, we may write (\ref{Pi1}) in the alternative form 
\begin{multline}\label{Pi1hyper}\Pi_1(\kappa\,|\,x_1,x_2,x_3,x_4)=\frac{\Gamma(12/\kappa-1)\Gamma(4/\kappa)}{\Gamma(8/\kappa)\Gamma(8/\kappa-1)}(x_3-x_1)^{1-6/\kappa}(x_4-x_2)^{1-6/\kappa}\\
\times\,\lambda^{2/\kappa}(1-\lambda)^{1-6/\kappa}\,_2F_1\bigg(\frac{4}{\kappa},1-\frac{4}{\kappa};\frac{8}{\kappa}\,\bigg|\,\lambda\bigg),\quad \lambda:=\frac{(x_2-x_1)(x_4-x_3)}{(x_3-x_1)(x_4-x_2)}.\end{multline}
If $\kappa=6$, corresponding to critical percolation \cite{smir2}, then (\ref{Pi1hyper}) becomes Cardy's formula (\ref{Cardy}) \cite{c3} for the probability of a vertical percolation-cluster crossing in a rectangle with aspect ratio (bottom-side length to left-side length) $R=K(\lambda)/K(1-\lambda)$, where $K$ is the complete elliptic integral of the first kind \cite{morsefesh,absteg}.  The polygon diagram for $\Pi_1$ illustrates the corresponding boundary arc connectivity (figure \ref{RecXing}).  
Also, with the formula for $\Pi_1$ written in the form (\ref{Pi1hyper}), we may use knowledge of the behavior of the hypergeometric function as $\lambda\rightarrow0$ and $\lambda\rightarrow1$ \cite{morsefesh,absteg} to verify the duality condition (\ref{duality}).  This completes the proof that (\ref{Pi1}, \ref{Pi1hyper}) are indeed formulas for $\Pi_1$.

\begin{figure}[t]
\centering
\includegraphics[scale=0.27]{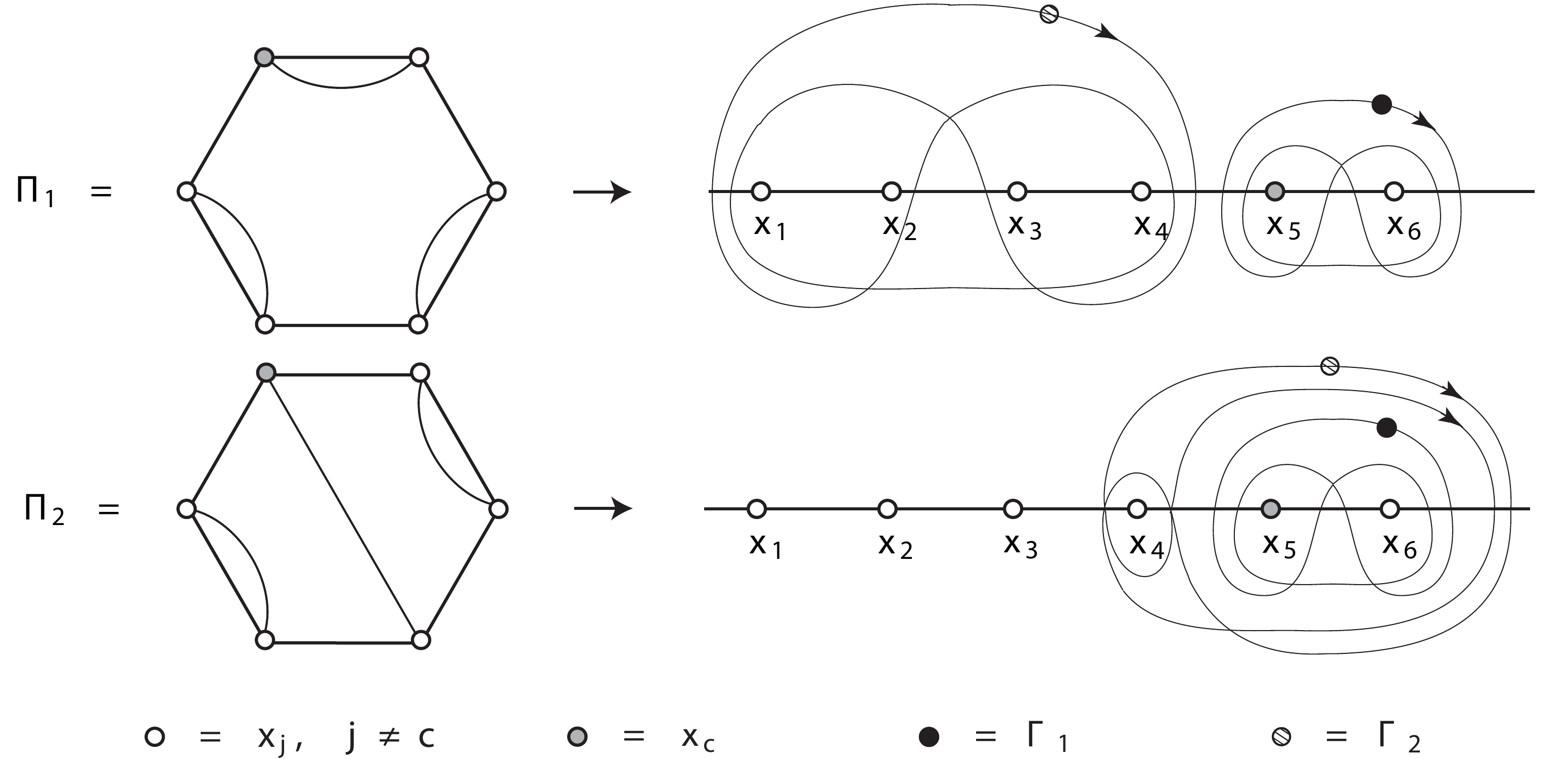}
\caption{The two topologically distinct hexagon connectivities, and the integration contours for their weights with $c=5$.  Moving counterclockwise around the hexagon corresponds to moving right along the real axis, with the bottom-left vertex sent to $x_1$.}
\label{HexXing}
\end{figure}

\subsection{Hexagon connectivity weights ($N=3$)}\label{HexXingSect}

According to item \ref{item1} of the introduction \ref{intro}, the dimension of the solution space $\mathcal{S}_3$ for the system (\ref{nullstate}, \ref{wardid}) is $C_3=5$ (\ref{catalan}), so there are five connectivity weights $\Pi_1$, $\Pi_2,\ldots,\Pi_5$ that span $\mathcal{S}_3$.  These connectivity weights sort into two disjoint sets, with the diagrams in each set identical up to a rotation.  Hence, we find formulas for only two connectivity weights $\Pi_1$ and $\Pi_2$ shown in figure \ref{HexXing}, one per set.  An appropriate transformation then gives the formulas for the other weights.

First, we find a formula for $\Pi_1$ in the form of (\ref{CGsolns}) with $N=3$.  After choosing $c=5$, we determine the two integration contours of (\ref{eulerintegral}) for use in this formula.  Figure \ref{HexXing} shows the polygon diagram for $\Pi_1$, and from this diagram and item \ref{twolegitem} above, it is apparent that $(x_1,x_2)$, $(x_3,x_4)$, and $(x_5,x_6)$ are two-leg intervals of $\Pi_1$.  Hence, following item \ref{ita} above, we entwine $x_5$ and $x_6$ with a Pochhammer contour $\Gamma_1=\mathscr{P}(x_5,x_6)$. Moreover, item \ref{itb} above requires that no integration contour crosses the other two-leg intervals.  In order to satisfy item \ref{itb}, we entwine $[x_1,x_2]$ and $[x_3,x_4]$ with a Pochhammer contour $\Gamma_2$.  These choices of integration contours determine the formula for $\Pi_1$ up to normalization.

\begin{figure}[t]
\centering
\includegraphics[scale=0.27]{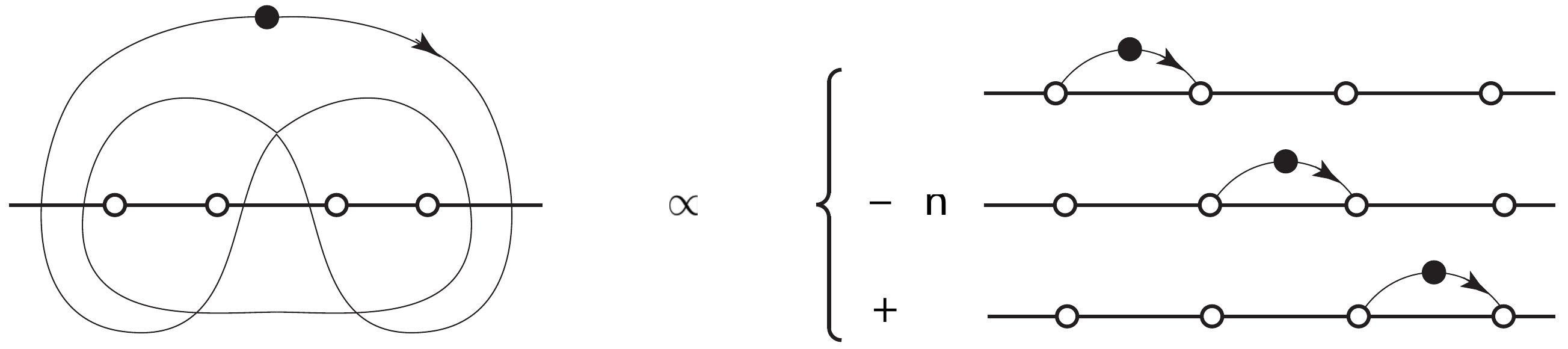}
\caption{Integration around the Pochhammer contour on the left is proportional to the combination of three integrations along line segments on the right.  We use this decomposition to find the first hexagon (\ref{Pi1N3}) and octagon (\ref{Pi1N4}) connectivity weights.}
\label{ContourId2}
\end{figure}

Assuming that $\kappa\in(4,8)$, we decompose the Coulomb gas integral (\ref{eulerintegral}) of $\Pi_1$, with $\Gamma_1$ and $\Gamma_2$ as specified, into a linear combination of the simpler definite integrals of type (here, we implicitly order the differences in the factors of the integrand in (\ref{eulerintegral}) such that (\ref{Iij}) is positive-valued, and we note that by Fubini's theorem that $I_{ij}=I_{ji}$)
\be\label{Iij}i\neq j:\quad I_{ij}(\kappa\,|\,x_1,x_2,\ldots,x_6):=\mathcal{J}\Big(\kappa\,\Big|\,[x_i,x_{i+1}], [x_j,x_{j+1}]\,\Big|\,x_1,x_2,\ldots,x_6\Big),\quad c=5,\quad N=3,\quad i,j<2N.\ee
To obtain this decomposition, we replace $\Gamma_1$ by an integration along $[x_5,x_6]$ (figure \ref{PochhammerContour}) and decompose the integration along $\Gamma_2$ into integrations along $[x_1,x_2],$ $[x_2,x_3],$ and $[x_3,x_4]$ (figure \ref{ContourId2}), finding
\begin{align}\mathcal{J}(\kappa\,|\,\Gamma_1,\Gamma_2)&\propto[e^{4\pi i/\kappa}I_{15}+e^{8\pi i/\kappa}I_{25}+e^{12\pi i/\kappa}I_{35}-e^{20\pi i/\kappa}I_{15}-e^{24\pi i/\kappa}I_{25}-e^{20\pi i\kappa}I_{35}\nonumber\\
&\hspace{1in}+e^{12\pi i/\kappa}I_{15}+e^{8\pi i/\kappa}I_{25}+e^{4\pi i\kappa}I_{35}-e^{-4\pi i/\kappa}I_{15}-e^{-8\pi i/\kappa}I_{25}-e^{-4\pi i/\kappa}I_{35}](\kappa)\nonumber\\
\label{J2decomp}&= (e^{4\pi i/\kappa}-e^{20\pi i/\kappa}+e^{12\pi i/\kappa}-e^{-4\pi i/\kappa})[I_{15}+I_{35}-nI_{25}](\kappa).\end{align}
After inserting this decomposition (\ref{J2decomp}) into (\ref{CGsolns}), we find the proper normalization for the resulting formula by requiring that the limit (\ref{lim}) with $i=2$ equals the rectangle connectivity weight  (\ref{Pi1}) with $(x_2,x_3,x_4)\mapsto(x_4,x_5,x_6)$.  Upon using the results of sections \red{A 2} and \red{A 3} in \cite{florkleb3} (figure \ref{Cases}) with the decomposition (\ref{J2decomp}) (figure \ref{ContourId2}) to determine the asymptotic behavior of $\mathcal{J}(\Gamma_1,\Gamma_2)$ as $x_3\rightarrow x_2$, we find 
\begin{multline}\label{Pi1N3}\Pi_1(\kappa\,|\,x_1,x_2,\ldots,x_6)=\frac{n(\kappa)^2}{n(\kappa)^2-2} \, \frac{\Gamma(2-8/\kappa)^2}{\Gamma(1-4/\kappa)^4}\Bigg(\prod_{\substack{i<j \\ i,j\neq 5}}^6(x_j-x_i)^{2/\kappa}\Bigg)\Bigg(\prod_{k\neq5}^6|x_5-x_k|^{1-6/\kappa}\Bigg)\\
\times[n I_{25}-I_{15}-I_{35}](\kappa\,|\,x_1,x_2,\ldots,x_6).\end{multline}
If $\kappa\in(0,4]$, then the improper integrals in (\ref{Pi1N3}) diverge, and we regularize them via the replacement (\ref{replacePoch}).  

Presently, the formula (\ref{Pi1N3}) for $\Pi_1$ is a well-motivated guess.  To prove that it is indeed correct, we must verify the duality condition (\ref{duality}) for $\vartheta=1$ and all $\varsigma\in\{1,2,3,4,5\}$.  We begin with $\varsigma=1$.  By the preceding paragraph,  $\bar{\ell}_1\Pi_1$ (\ref{lim}) with $i=2$ gives the rectangle connectivity weight (\ref{Pi1}), and thanks to (\ref{takefirstlim}), this is sufficient to confirm the duality condition (\ref{duality}) for $\varsigma=1$.  For all other $\varsigma$, we note that the polygon diagram for $[\mathscr{L}_\varsigma]$ has an arc whose two endpoints correspond with the endpoints of either $(x_1,x_2)$, $(x_3,x_4)$, or $(x_5,x_6)$.  As such, to compute $[\mathscr{L}_\varsigma]\Pi_1$, we may take the limit (\ref{lim}) respectively with $i=1$, $i=3$, or $i=5$ first.  But because all of these intervals are two-leg intervals of $\Pi_1$, these limits, and therefore $[\mathscr{L}_\varsigma]$, annihilate $\Pi_1$, confirming the duality condition (\ref{duality}) for $\varsigma\in\{2,3,4,5\}$.

Next, we find a formula for $\Pi_2$ in the form of (\ref{CGsolns}) with $N=3$.  After choosing $c=5$, we determine the two integration contours of (\ref{eulerintegral}) for use in this formula.  Figure \ref{HexXing} shows the polygon diagram for $\Pi_2$, and from this diagram and item \ref{twolegitem} above, it is apparent that $(x_1,x_2)$, $(x_2,x_3)$, $(x_4,x_5)$, and $(x_5,x_6)$ are two-leg intervals of $\Pi_2$.  Hence, following item \ref{ita} above, we entwine $x_5$ and $x_6$ with a Pochhammer contour $\Gamma_1=\mathscr{P}(x_5,x_6)$. Moreover, item \ref{itb} above requires that no integration contour crosses $(x_1,x_2)$ or $(x_2,x_3)$.  In order to satisfy item \ref{itb}, we entwine $x_4$ and $\Gamma_1$ with a Pochhammer contour $\Gamma_2$.  These choices of integration contours determine the formula for $\Pi_2$ up to normalization.

\begin{figure}[t]
\centering
\includegraphics[scale=0.27]{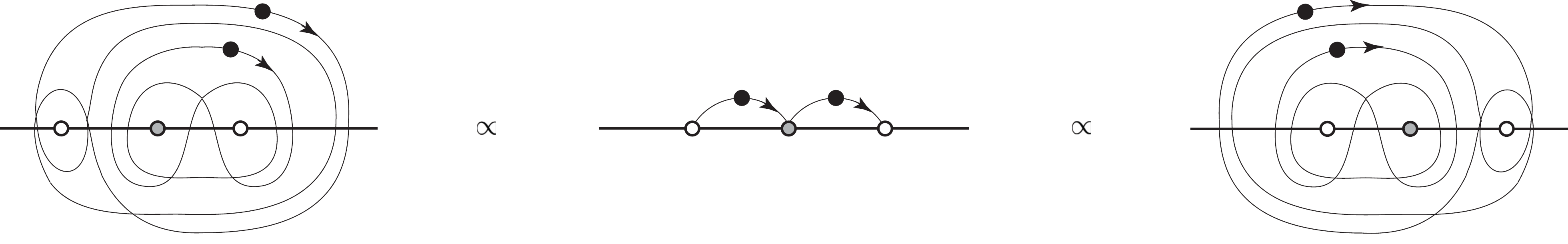}
\caption{Integration around the above nested pair of Pochhammer contours (left, right) is proportional to performing the first integration along the right interval and the second integration along the left interval (middle).  The gray circle is $x_c$ in (\ref{eulerintegral}).}
\label{ContourId1}
\end{figure}

Assuming that $\kappa\in(4,8)$, we decompose the Coulomb gas integral (\ref{eulerintegral}) of $\Pi_2$, with $\Gamma_1$ and $\Gamma_2$ as specified, into a linear combination of the simpler definite integrals of type (\ref{Iij}).  Specifically, we replace $\Gamma_1$ by an integration along $[x_5,x_6]$ (figure \ref{PochhammerContour}), freeze $u_1$ somewhere within $(x_5,x_6)$, and decompose the integration along $\Gamma_2$ into (figure \ref{Phases})
\newpage
\begin{multline}\label{J1}e^{4\pi i/\kappa}J_{x_4}^{x_5}+e^{-8\pi i/\kappa}J_{x_5}^{u_1}+e^{-16\pi i/\kappa}J_{u_1}^{x_6}-e^{-28\pi i/\kappa}J_{x_4}^{x_5}-e^{-16\pi i/\kappa}J_{x_5}^{u_1}-e^{-8\pi i/\kappa}J_{u_1}^{x_6}\\
+e^{-36\pi i/\kappa}J_{x_4}^{x_5}+e^{-24\pi i/\kappa}J_{x_5}^{u_1}+e^{-16\pi i/\kappa}J_{u_1}^{x_6}-e^{-4\pi i/\kappa}J_{x_4}^{x_5}-e^{-16\pi i/\kappa}J_{x_5}^{u_1}-e^{-24\pi i/\kappa}J_{u_1}^{x_6},\end{multline}
where $J_a^b$ is the magnitude of the integrand of $I_{ij}$ (\ref{Iij}) with $x_5<u_1<x_6$ and with $u_2$ integrated from $a$ to $b$.  We factor (\ref{J1}) into
\be\label{J2}2i\sin\left(\frac{4\pi}{\kappa}\right)(1-e^{-32\pi i/\kappa})J_{x_4}^{x_5}+(e^{-8\pi i/\kappa}-2e^{-16/\pi i/\kappa}+e^{-24\pi i/\kappa})(J_{x_5}^{u_1}-J_{u_1}^{x_6}).\ee
Now, the integrations of $J_{x_5}^{u_1}$ and $J_{u_1}^{x_6}$ with respect to $u_1$ along $[x_5,x_6]$ are equal because their integrands exchange under the switch $(u_1,u_2)\mapsto(u_2,u_1)$.  Integrating $u_1$ along $[x_5,x_6]$ in (\ref{J2}) therefore gives (figure \ref{ContourId1})
\be\label{J2decomp2}\mathcal{J}(\Gamma_1,\Gamma_2)\propto I_{45}.\ee
After inserting the decomposition (\ref{J2decomp2}) into (\ref{CGsolns}), we find the proper normalization for the resulting formula by requiring that the limit (\ref{lim}) with $i=3$ equals the rectangle connectivity weight (\ref{Pi1}) with $(x_3,x_4)\mapsto(x_5,x_6)$.  Upon using the results of section \red{A 3} in \cite{florkleb3} (figure \ref{Cases}) with the decomposition (\ref{J2decomp2}) (figure \ref{ContourId1}) to determine the asymptotic behavior of $\mathcal{J}(\Gamma_1,\Gamma_2)$ as $x_4\rightarrow x_3$, we find
\be\label{Pi2N3}\Pi_2(\kappa\,|\,x_1,x_2,\ldots,x_6)=n(\kappa)^2 \, \frac{\Gamma(2-8/\kappa)^2}{\Gamma(1-4/\kappa)^4}\Bigg(\prod_{\substack{i<j \\ i,j\neq 5}}^6(x_j-x_i)^{2/\kappa}\Bigg)\Bigg(\prod_{k\neq5}^6|x_5-x_k|^{1-6/\kappa}\Bigg)I_{45}(\kappa\,|\,x_1,x_2,\ldots,x_6).\ee
If $\kappa\in(0,4]$, then the improper integral in (\ref{Pi2N3}) diverges, and we regularize it via the replacement (\ref{replacePoch}).

Presently, the formula (\ref{Pi2N3}) for $\Pi_2$ is a well-motivated guess.  To prove that it is indeed correct, we must verify the duality condition (\ref{duality}) for $\vartheta=2$ and all $\varsigma\in\{1,2,3,4,5\}$.    We first note that $(x_4,x_5)$ is indeed a two-leg interval of $\Pi_2$ thanks to the symmetry property of the decomposition (\ref{J2decomp2}) shown in figure \ref{ContourId1}.  Then the verification proceeds similarly to that for $\Pi_1$ (\ref{Pi1N3}) above.   Section \ref{rainbow} generalizes this particular connectivity weight formula to arbitrary $N\in\mathbb{Z}^++1$.

The formulas (\ref{Pi1N3}, \ref{Pi2N3}) found for $\Pi_1$ and $\Pi_2$ respectively are singular at any $\kappa\in(0,8)\times i\mathbb{R}$ such that $8/\kappa\in\mathbb{Z}^++1$ or $12/\kappa\in\mathbb{Z}^++1$.  Also, the formula (\ref{Pi1N3}) for $\Pi_1$ is singular at any $\kappa$ such that $n(\kappa)^2=2$ (\ref{fugacity}), and both formulas appear to vanish if $n(\kappa)=0$ (that is, if $8/\kappa\in2\mathbb{Z}^++1$).  Because section \red{III} of \cite{florkleb4} shows that $\Pi_1$ and $\Pi_2$ are continuous functions of $\kappa$, these singularities must be removable.  Furthermore, because they are elements of a basis, there is no $\kappa\in(0,8)$, including those with $n(\kappa)=0$, such that $\Pi_1(\kappa)$ or $\Pi_2(\kappa)$ vanish for all $\boldsymbol{x}\in\Omega_0$.  Although we already know these facts, it is interesting to verify them directly from the formulas themselves, which we do in  Appendix \ref{appendix}.

\subsection{Connectivity weights in octagons ($N=4$)}\label{OctXingSect}

\begin{figure}[t]
\centering
\includegraphics[scale=0.27]{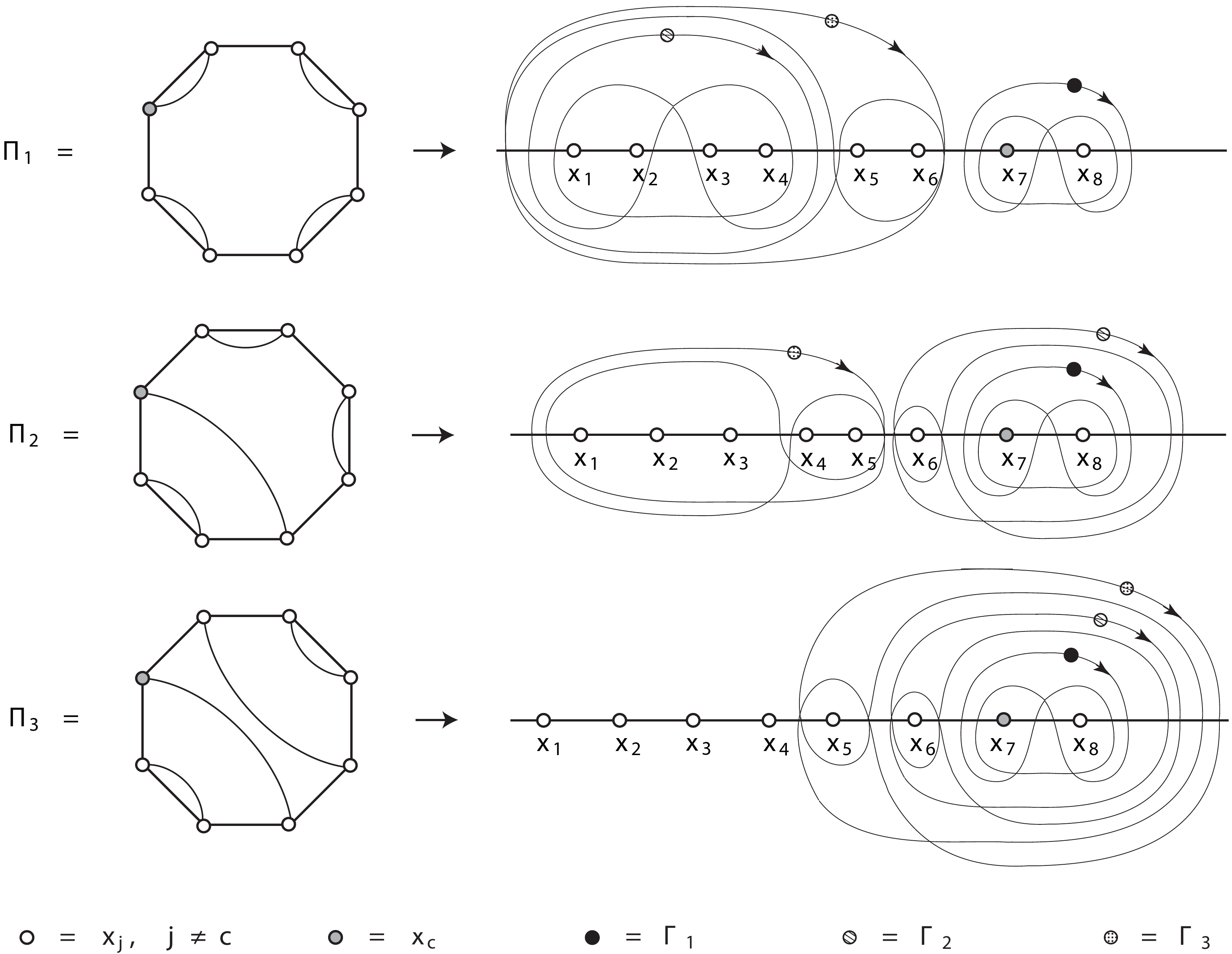}
\caption{The three topologically distinct octagon connectivities, and integration contours for their weights with $c=7$.  Moving counterclockwise around the hexagon corresponds to moving right along the real axis, with the bottom-left vertex sent to $x_1$.  In (\ref{decPi0}), we note that $\Pi_1$ (\ref{Pi1N4}) does not exactly correspond with the collection of integration contours for ``$\Pi_0$" (\ref{Pi0}) to its right.}
\label{OctXing}
\end{figure}

According to item \ref{item1} of the introduction \ref{intro}, the dimension of the solution space $\mathcal{S}_4$ for the system (\ref{nullstate}, \ref{wardid}) is $C_4=14$ (\ref{catalan}), so there are fourteen connectivity weights $\Pi_1$, $\Pi_2,\ldots,\Pi_{14}$ that span $\mathcal{S}_4$.  These connectivity weights sort into three disjoint sets, with the diagrams in each set identical up to a rotation.  Hence, we find formulas for only three connectivity weights $\Pi_1$, $\Pi_2$, and $\Pi_3$ shown in figure \ref{OctXing}, one per set.  An appropriate transformation then gives the formulas for the other weights.

Here, we first find a formula for $\Pi_2$ in the form of (\ref{CGsolns}) with $N=4$.  After choosing $c=7$, we determine the three integration contours of (\ref{eulerintegral}) for use in this formula.  Figure \ref{OctXing} shows the polygon diagram for $\Pi_2$, and from this diagram and item \ref{twolegitem} above, it is apparent that $(x_1,x_2)$, $(x_2,x_3)$, $(x_4,x_5)$, $(x_6,x_7)$, and $(x_7,x_8)$ are two-leg intervals of $\Pi_2$.  Hence, following item \ref{ita} above, we entwine $x_7$ and $x_8$ with a Pochhammer contour $\Gamma_1=\mathscr{P}(x_7,x_8)$.  Furthermore, the formula for the second hexagon connectivity weight (\ref{Pi2N3}) suggests that we entwine $\Gamma_1$ and $x_6$ with another Pochhammer contour $\Gamma_2$ in order for $(x_6,x_7)$ to be a two-leg interval of $\Pi_2$ as well.  Finally, item \ref{itb} above suggests that no integration contour crosses the other two-leg intervals.  In order to satisfy item \ref{itb}, we entwine $[x_1,x_3]$ and $[x_4,x_5]$ with a Pochhammer contour $\Gamma_3$.  These choices of integration contours determine the formula for $\Pi_2$ up to normalization.

Assuming that $\kappa\in(4,8)$, we decompose the Coulomb gas integral (\ref{eulerintegral}) of $\Pi_2$, with $\Gamma_1$, $\Gamma_2$, and $\Gamma_3$ as specified, into a linear combination of the simpler definite integrals 
\be\label{Iijk}\begin{aligned}i\neq j\neq k:&& I_{ijk}(\kappa\,|\,x_1,x_2,\ldots,x_8)&:=\mathcal{J}\Big(\kappa\,\Big|\,[x_i,x_{i+1}], [x_j,x_{j+1}], [x_k,x_{k+1}]\,\Big|\,x_1,x_2,\ldots,x_8\Big),\\
i= j\neq k:&& I_{iik}(\kappa\,|\,x_1,x_2,\ldots,x_8)&:=\mathcal{J}\Big(\kappa\,\Big|\,[x_i,u_2], [x_i,x_{i+1}], [x_k,x_{k+1}]\,\Big|\,x_1,x_2,\ldots,x_8\Big).\end{aligned}\quad \begin{gathered} c=7,\quad N=4, \\ i,j,k<2N.\end{gathered}\ee
In the definition for $I_{iik}$ on the bottom line of (\ref{Iijk}), the right endpoint of the first contour $[x_i,u_2]$ is the second integration variable $u_2$, integrated after the first ($u_1$). We implicitly order the differences in the factors of the integrand in (\ref{eulerintegral}) such that (\ref{Iijk}) is positive-valued.  Figure \ref{ContourId1} shows that integration over the nested pair $\Gamma_2\times\Gamma_1$ decomposes into integration along $[x_6,x_7]\times[x_7,x_8]$ (section \ref{HexXingSect}).  Using this fact and then decomposing the integration along $\Gamma_3$ into integrations along line segments similar to what figure \ref{ContourId2} shows, we find
\begin{align}\mathcal{J}(\kappa\,|\,\Gamma_1,\Gamma_2,\Gamma_3)&\propto\left\{\begin{aligned}&+e^{4\pi i/\kappa}I_{167}+e^{8\pi i/\kappa}I_{267}+e^{12\pi i/\kappa}I_{367}+e^{16\pi i/\kappa}I_{467}\\
&- e^{20\pi i/\kappa}I_{167}-e^{24\pi i/\kappa}I_{267}-e^{28\pi i/\kappa}I_{367}-e^{24\pi i/\kappa}I_{467}\\
&+e^{12\pi i/\kappa}I_{167}+e^{8\pi i/\kappa}I_{267}+e^{4\pi i/\kappa}I_{367}+I_{467}\\
&-e^{-4\pi i/\kappa}I_{167}-e^{-8\pi i/\kappa}I_{267}-e^{-12\pi i/\kappa}I_{367}-e^{-8\pi i/\kappa}I_{467}\end{aligned}\right\}(\kappa)\nonumber\\
\label{J3decomp}&=e^{8\pi i/\kappa}(n(\kappa)^2-4)[n(I_{167}-nI_{267})+(n^2-1)(nI_{367}-I_{467})](\kappa).\end{align}
After inserting (\ref{J3decomp}) into (\ref{CGsolns}), we find the proper normalization for the resulting formula by requiring that the limit (\ref{lim}) with $i=3$ equals the second hexagon connectivity weight (\ref{Pi2N3}) with $(x_3,x_4,x_5,x_6)\mapsto(x_5,x_6,x_7,x_8)$.  Upon using the results of sections \red{A 2} and \red{A 3} in \cite{florkleb3} (figure \ref{Cases}) with the decomposition (\ref{J3decomp}) to determine the asymptotic behavior of $\mathcal{J}(\Gamma_1,\Gamma_2,\Gamma_3)$ as $x_4\rightarrow x_3$, we find
\begin{multline}\label{Pi2N4}\Pi_2(\kappa\,|\,x_1,x_2,\ldots,x_8)=\frac{n(\kappa)^3}{n(\kappa)^4-3n(\kappa)^2+1} \, \frac{\Gamma(2-8/\kappa)^3}{\Gamma(1-4/\kappa)^6}\Bigg(\prod_{\substack{i<j \\ i,j\neq 7}}^8(x_j-x_i)^{2/\kappa}\Bigg)\Bigg(\prod_{k\neq7}^8|x_7-x_k|^{1-6/\kappa}\Bigg)\\
\times[n(I_{167}-nI_{267})+(n^2-1)(nI_{367}-I_{467})](\kappa\,|\,x_1,x_2,\ldots,x_8).\end{multline}
If $\kappa\in(0,4]$, then the improper integrals in (\ref{Pi2N4}) diverge, and we regularize them via the replacement (\ref{replacePoch}).

Presently, the formula (\ref{Pi2N4}) for $\Pi_2$ is a well-motivated guess.  To prove that it is indeed correct, we must verify the duality condition (\ref{duality}) for $\vartheta=2$ and all $\varsigma\in\{1,2,\ldots,14\}$.  We begin with $\varsigma=2$.  By the preceding paragraph, $\bar{\ell}_1\Pi_2$ (\ref{lim}) with $i=3$ gives the hexagon connectivity weight (\ref{Pi2N3}), and thanks to (\ref{takefirstlim}), this is sufficient to confirm the duality condition (\ref{duality}) for $\varsigma=2$.  For all other $\varsigma$, we note that the polygon diagram for $[\mathscr{L}_\varsigma]$ has an arc whose two endpoints correspond with the endpoints of either $(x_1,x_2)$, $(x_2,x_3)$, $(x_4,x_5)$, $(x_6,x_7)$, or $(x_7,x_8)$.  As such, to compute $[\mathscr{L}_\varsigma]\Pi_2$, we may take the limit (\ref{lim}) respectively with $i=1$, $i=2$, $i=4$, $i=6$, or $i=7$ first.  However, all of $(x_1,x_2)$, $(x_2,x_3)$, $(x_4,x_5)$, and $(x_7,x_8)$ are two-leg intervals of $\Pi_2$.  Indeed, this follows from item \ref{itb} and figure \ref{OctXing} above.  Furthermore, $(x_6,x_7)$ is a two-leg interval of $\Pi_2$, and this is manifest from item \ref{ita} above and the right illustration of figure \ref{ContourId1}.  Thus, these limits, and therefore $[\mathscr{L}_\varsigma]$, annihilate $\Pi_2$, confirming the duality condition (\ref{duality}) for $\varsigma\in\{1,3,4,\ldots,14\}$.

Next, we find a formula for $\Pi_3$ in the form of (\ref{CGsolns}) with $N=4$.  After choosing $c=7$, we determine the three integration contours of (\ref{eulerintegral}) for use in this formula.  Figure \ref{OctXing} shows the polygon diagram for $\Pi_3$, and from this diagram and item \ref{twolegitem} above, it is apparent that $(x_1,x_2)$, $(x_2,x_3)$, $(x_3,x_4)$, $(x_5,x_6)$, $(x_6,x_7)$, and $(x_7,x_8)$ are two-leg intervals of $\Pi_3$.  Hence, following item \ref{ita} above, we entwine $x_7$ and $x_8$ with a Pochhammer contour $\Gamma_1=\mathscr{P}(x_7,x_8)$.  Furthermore, the formula for the second hexagon connectivity weight (\ref{Pi2N3}) and figure \ref{ContourId1} suggest that we entwine $\Gamma_1$ and $x_6$ with another Pochhammer contour $\Gamma_2$ in order for $(x_6,x_7)$ to be a two-leg interval of $\Pi_3$ as well.  Finally, item \ref{itb} above suggests that no integration contour crosses the other two-leg intervals.  In order to satisfy item \ref{itb}, we entwine $x_5$ and $\Gamma_2$ with a Pochhammer contour $\Gamma_3$.  These choices of integration contours determine the formula for $\Pi_3$ up to normalization.

\begin{figure}[b]
\centering
\includegraphics[scale=0.27]{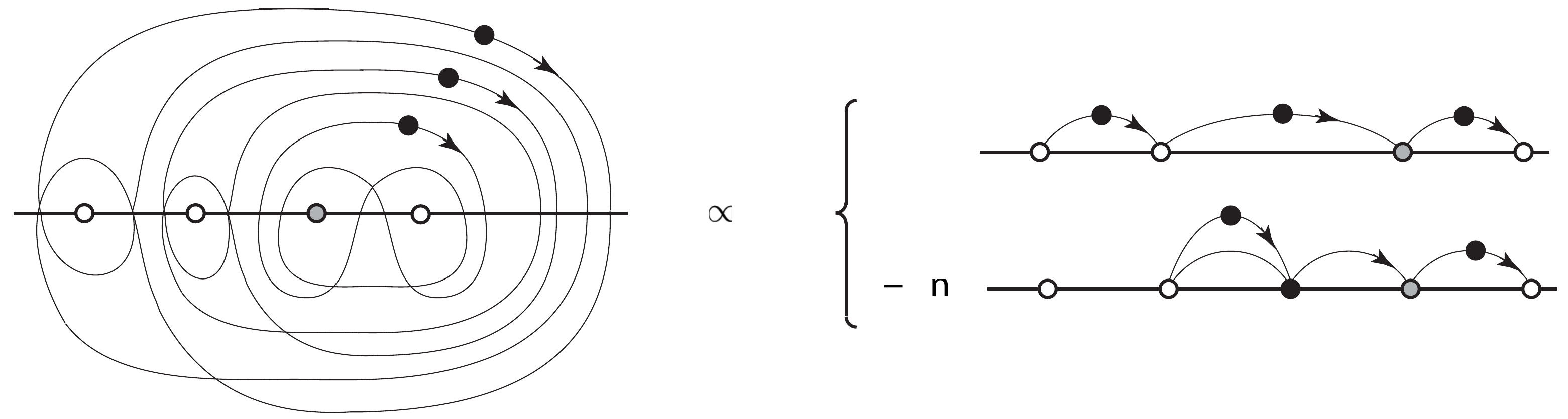}
\caption{Integration around the three nested Pochhammer contours on the left is proportional to the difference of the two integrations on the right.  The gray circle is $x_c$ in (\ref{eulerintegral}).}
\label{ContourId3}
\end{figure}

Assuming that $\kappa\in(4,8)$, we decompose the Coulomb gas integral (\ref{eulerintegral}) of $\Pi_3$, with $\Gamma_1$, $\Gamma_2$, and $\Gamma_3$ as specified, into a linear combination of the simpler definite integrals of type (\ref{Iijk}).  Figure \ref{ContourId1} shows that the integration over the nested pair $\Gamma_2\times\Gamma_1$ decomposes into integration along $[x_6,x_7]\times[x_7,x_8]$.  After inserting that decomposition into the Coulomb gas integral $\mathcal{J}(\Gamma_1,\Gamma_2,\Gamma_3)$, we freeze $u_1$ and $u_2$ at locations within $(x_7,x_8)$ and $(x_6,x_7)$ respectively, and we decompose the integration along $\Gamma_3$ into
\be\begin{aligned}\label{predecomp}&+e^{4\pi i/\kappa}J_{x_5}^{x_6}+e^{8\pi i/\kappa}J_{x_6}^{u_2}+J_{u_2}^{x_7}+e^{-12\pi i/\kappa}J_{x_7}^{u_1}+e^{-20\pi i/\kappa}J_{u_1}^{x_8}\quad\\
&-e^{-36\pi i/\kappa}J_{x_5}^{x_6}-e^{-40\pi i/\kappa}J_{x_6}^{u_2}-e^{-32\pi i/\kappa}J_{u_2}^{x_7}-e^{-20\pi i/\kappa}J_{x_7}^{u_1}-e^{-12\pi i/\kappa}J_{u_1}^{x_8}\quad\\
&+e^{-44\pi i/\kappa}J_{x_5}^{x_6}+e^{-48\pi i/\kappa}J_{x_6}^{u_2}+e^{-40\pi i/\kappa}J_{u_2}^{x_7}+e^{-28\pi i/\kappa}J_{x_7}^{u_1}+e^{-20\pi i/\kappa}J_{u_1}^{x_8}\quad\\
&-e^{-4\pi i/\kappa}J_{x_5}^{x_6}-J_{x_6}^{u_2}-e^{-8\pi i/\kappa}J_{u_2}^{x_7}-e^{-20\pi i/\kappa}J_{x_7}^{u_1}-e^{-28\pi i/\kappa}J_{u_1}^{x_8}\end{aligned},\ee
where $J_a^b$ is the magnitude of the integrand of $I_{i67}$ (\ref{Iijk}) with $x_7<u_1<x_8$ and $x_6<u_2<x_7$ and with $u_3$ integrated from $a$ to $b$.  We may factor (\ref{predecomp}) into
\begin{multline}\label{step}2i\sin\left(\frac{4\pi}{\kappa}\right)[(1-e^{-40\pi i/\kappa})J_{x_5}^{x_6}+(1-e^{-40\pi i/\kappa})(e^{4\pi i/\kappa}+e^{-4\pi i/\kappa})J_{x_6}^{u_2}\\
+(e^{-4\pi i/\kappa}-e^{-36\pi i/\kappa})(J_{u_2}^{x_7}-J_{x_6}^{u_2})+(e^{-16\pi i/\kappa}-e^{-24\pi i/\kappa})(J_{x_7}^{u_1}-J_{u_1}^{x_8})].\end{multline}
Now, the integrations of $J_{x_7}^{u_1}$ and $J_{u_1}^{x_8}$ (resp.\ $J_{u_2}^{x_7}$ and $J_{x_6}^{u_2}$) with respect to $u_1$ along $[x_7,x_8]$ (resp.\ $u_2$ along $[x_6,x_7]$) are equal because their integrands exchange under the switch $(u_1,u_3)\mapsto(u_3,u_1)$ (resp.\ $(u_2,u_3)\mapsto(u_3,u_2)$).  Integrating $u_1$ and $u_2$ along $[x_7,x_8]$ and $[x_6,x_7]$ respectively in (\ref{step}) therefore gives (figure \ref{ContourId3})
\be\label{J3decomp2}\mathcal{J}(\Gamma_1,\Gamma_2,\Gamma_3)\propto I_{567}-nI_{667}.\ee
After inserting the decomposition (\ref{J3decomp2}) into (\ref{CGsolns}), we find the proper normalization for the resulting formula by requiring that the limit (\ref{lim}) with $i=4$ equals the second hexagon connectivity weight (\ref{Pi2N3}) with $(x_4,x_5,x_6)\mapsto(x_6,x_7,x_8)$.  Upon using the results of section \red{A 3} in \cite{florkleb3} (figure \ref{Cases}) with the decomposition (\ref{J3decomp2}) (figure \ref{ContourId3}) to determine the asymptotic behavior of $\mathcal{J}(\Gamma_1,\Gamma_2,\Gamma_3)$ as $x_5\rightarrow x_4$, we find
\begin{multline}\label{Pi3N4}\Pi_3(\kappa\,|\,x_1,x_2,\ldots,x_8)=n(\kappa)^3\frac{\Gamma(2-8/\kappa)^3}{\Gamma(1-4/\kappa)^6}\Bigg(\prod_{\substack{i<j \\ i,j\neq 7}}^8(x_j-x_i)^{2/\kappa}\Bigg)\\
\times\,\Bigg(\prod_{k\neq7}^8|x_7-x_k|^{1-6/\kappa}\Bigg)[I_{567}-nI_{667}](\kappa\,|\,x_1,x_2,\ldots,x_8).\end{multline}
If $\kappa\in(0,4]$, then the improper integrals in (\ref{Pi3N4}) diverge, and we regularize them via the replacement (\ref{replacePoch}).

\begin{figure}[b]
\centering
\includegraphics[scale=0.27]{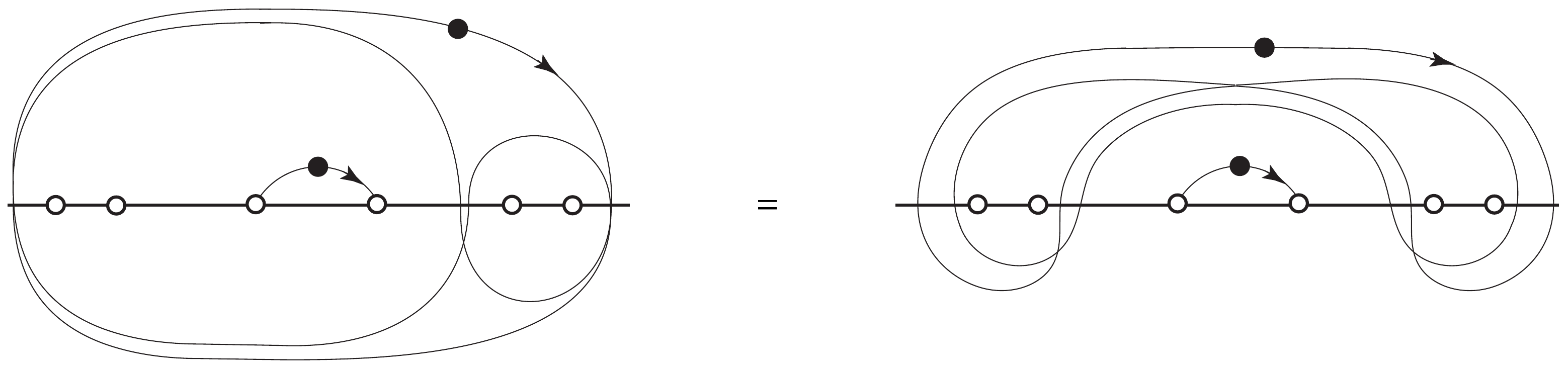}
\caption{The first octagon connectivity weight has a Pochhammer contour $\Gamma_2$ nested in another $\Gamma_3$ (figure \ref{OctXing}).
We decompose $\Gamma_2$ into three segments (figure \ref{ContourId2}), one shown on the left, and we deform $\Gamma_3$ (figure \ref{IntegrateAround}) so it does not surround those segments.}
\label{ContourId5}
\end{figure}

Presently, the formula (\ref{Pi3N4}) for $\Pi_3$ is a well-motivated guess.  To prove that it is indeed correct, we must verify the duality condition (\ref{duality}) for $\vartheta=3$ and all $\varsigma\in\{1,2,\ldots,14\}$.  This proceeds along the same lines as  that for $\Pi_2$ (\ref{Pi2N4}) above.  Section \ref{rainbow} generalizes this particular connectivity weight formula to arbitrary $N\in\mathbb{Z}^++1$.

Finally, we seek a formula for $\Pi_1$ in the form of (\ref{CGsolns}) with $N=4$.  After choosing $c=7$, we determine the three integration contours of (\ref{eulerintegral}) for use in this formula.  Figure \ref{OctXing} shows the polygon diagram for $\Pi_1$, and from this diagram and item \ref{twolegitem} above, it is apparent that $(x_1,x_2),$ $(x_3,x_4)$, $(x_5,x_6)$, and $(x_7,x_8)$ are two-leg intervals of $\Pi_1$.  Hence, following item \ref{ita} above, we entwine $x_7$ and $x_8$ with a Pochhammer contour $\Gamma_1=\mathscr{P}(x_7,x_8)$.  Moreover, item \ref{itb} above requires that no integration contour crosses the other two-leg intervals.  In order to satisfy item \ref{itb}, the formula for the first hexagon connectivity weight (\ref{Pi1N3}) (figure \ref{HexXing}) suggests that we entwine $(x_1,x_2)$ and $(x_3,x_4)$ with a Pochhammer contour $\Gamma_2$ and that we entwine $(x_5,x_6)$ and $\Gamma_2$ with another contour $\Gamma_3$.  These choices of integration contours determine the prospective formula for $\Pi_1$ up to normalization.

\begin{figure}[t]
\centering
\includegraphics[scale=0.27]{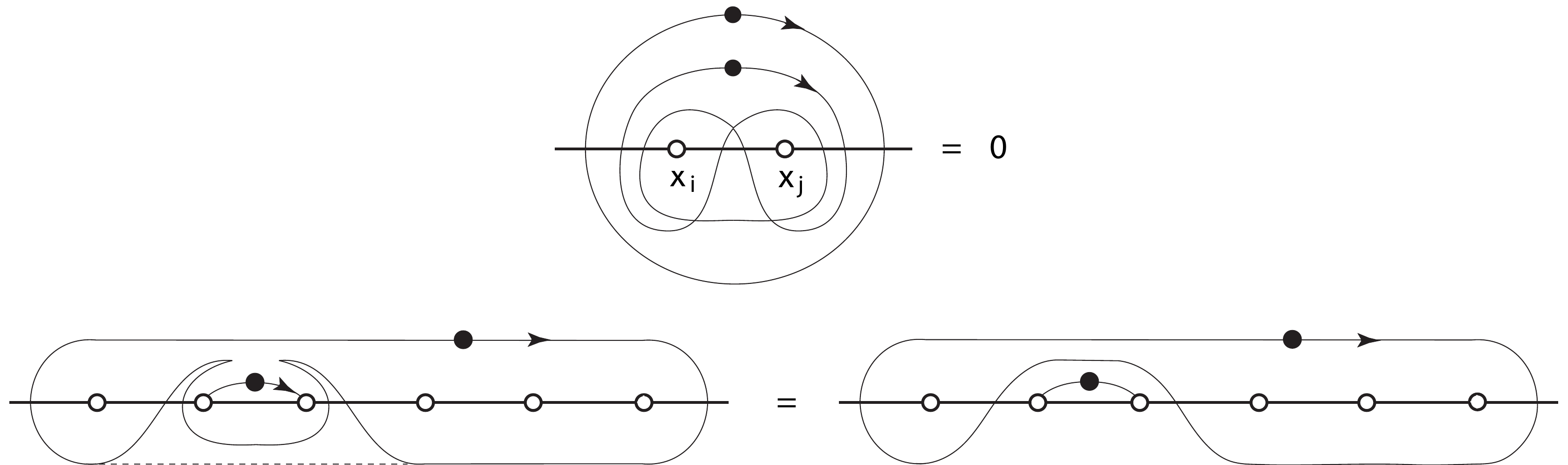}
\caption{Integrating (\ref{eulerintegral}) around an elementary Pochhammer contour $\mathscr{P}(x_i,x_j)$ (or line segment $[x_i,x_j]$) with $i,j\neq c$ and then around a loop that tightly wraps around $\mathscr{P}(x_i,x_j)$ gives zero.  Thus, we may deform an integration contour through $\mathscr{P}(x_i,x_j)$.}
\label{IntegrateAround}
\end{figure}

\begin{figure}[b]
\centering
\includegraphics[scale=0.27]{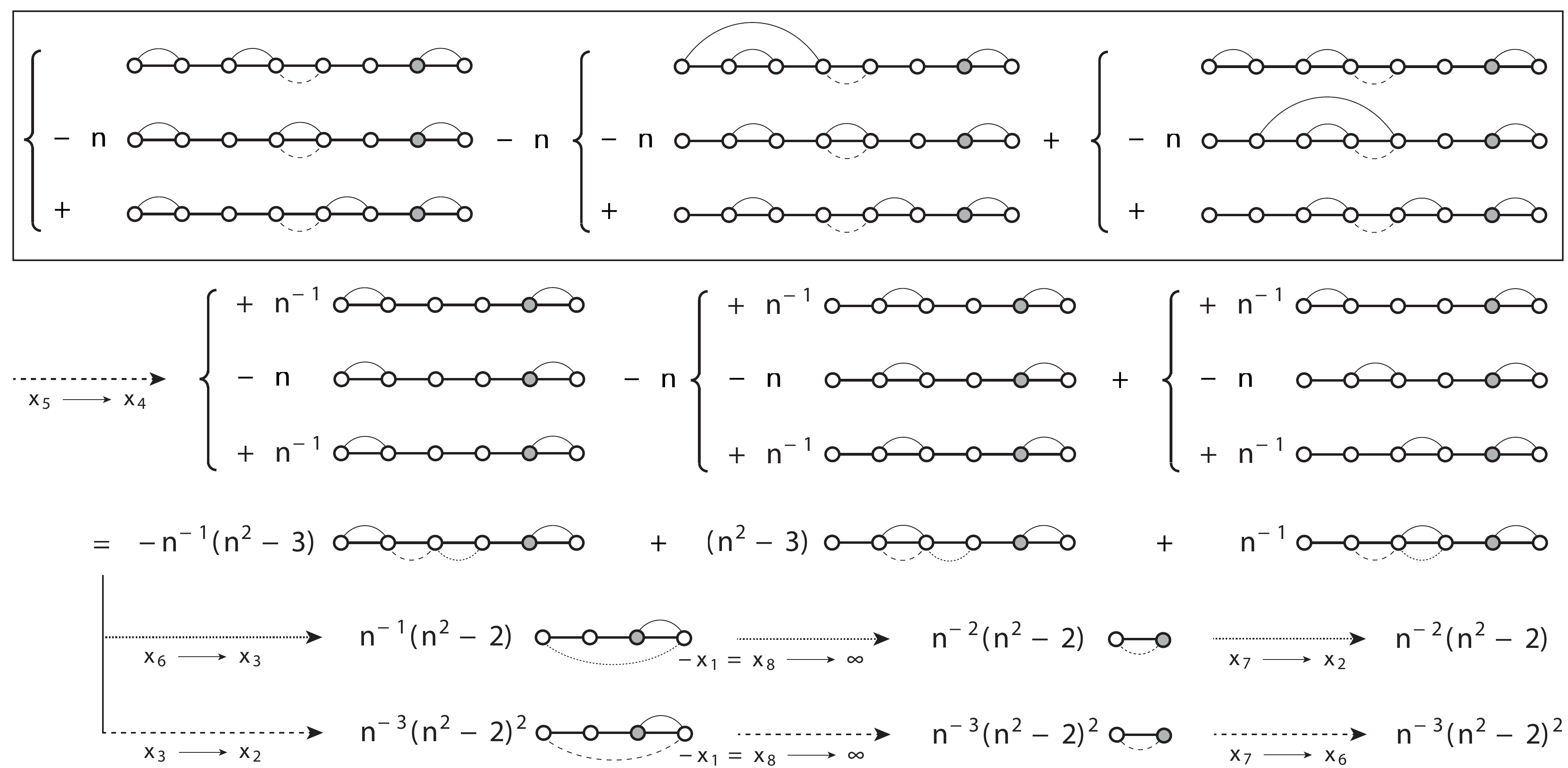}
\caption{The decomposition (\ref{J3decomp3}) of $\mathcal{J}(\Gamma_1,\Gamma_2,\Gamma_3)$ into nine terms (boxed), and the calculation of the limits (\ref{lim1}, \ref{limit2}) using figure \ref{Cases} (with Gamma functions dropped).  Course-dashed (resp.\ fine-dashed) arcs join the points brought together by $\mathscr{L}_3$ (resp.\ $\mathscr{L}_1$).}
\label{BigDecomp}
\end{figure}

\begin{figure}[b]
\centering
\includegraphics[scale=0.27]{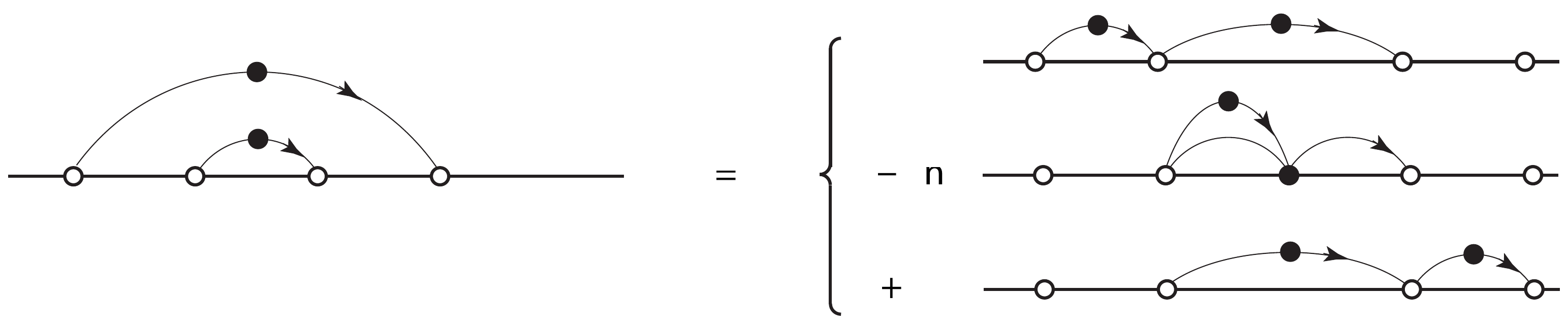}
\caption{Integration along the pair of nested contours shown on the left equals the sum of integrations shown on the right.  We use this identity to decompose the two terms in the boxed part of figure \ref{BigDecomp} that have nested contours, ultimately finding (\ref{J3decomp3}).}
\label{ContourId4}
\end{figure}

Assuming that $\kappa\in(4,8)$, we decompose the Coulomb gas integral (\ref{eulerintegral}) of $\Pi_1$, with $\Gamma_1$, $\Gamma_2$, and $\Gamma_3$ as specified, into a linear combination of the simpler definite integrals of type (\ref{Iijk}).  Specifically, we replace $\Gamma_1$ with an integration along $[x_5,x_6]$ (figure \ref{PochhammerContour}) and decompose the integration along $\Gamma_2$ into integrations along $[x_1,x_2],$ $[x_2,x_3],$ and $[x_3,x_4]$, just as we did for first the hexagon connectivity weight (\ref{Pi1N3}) (figure \ref{ContourId2}).  We call the corresponding terms in the decomposition the ``first term," the ``second term," and the ``third term" respectively.  The left illustration of figure \ref{ContourId5} shows the contours $[x_3,x_4]$ and $\Gamma_3$ for the third term.  Now, to decompose $\Gamma_3$ into simple contours, we recall identity (\red{B1}) from appendix \red{B} of \cite{florkleb3}.  This identity says that if $\Gamma$ is a line segment or an elementary Pochhammer contour with endpoints at $x_i<x_j$, if $i,j\neq c$, and if $\Gamma_0$ is a simple loop surrounding $\Gamma$, then (top illustration of figure \ref{IntegrateAround})
\be\label{oint}\int_{\Gamma}\sideset{}{_{\Gamma_0}}\oint
[\,\ldots\,\text{the integrand of $\mathcal{J}$ (\ref{eulerintegral}) \,\ldots}]\,{\rm d}u_2\,{\rm d}u_1=0.\ee
Using identity (\ref{oint}), we deform $\Gamma_3$ such that it no longer surrounds the contour originally inside it (bottom illustration of figure \ref{IntegrateAround}).  After this deformation, $\Gamma_3$ entwines $[x_3,x_4]$ (resp.\ $\{x_1,x_4\}$, resp.\ $[x_1,x_2]$) and $[x_5,x_6]$ together in the first (resp.\ second, resp.\ third) term.  For example, the right illustration of figure \ref{ContourId5} shows the contours $[x_3,x_4]$ and $\Gamma_3$ for the third term after this deformation.  Next, we decompose $\Gamma_3$ in the same way that we decomposed $\Gamma_2$ for each of the three terms (figure \ref{ContourId2}).  In so doing, we find nine terms in total, illustrated in the boxed part of figure \ref{BigDecomp}.  Some of these terms have an integration contour that arcs over another integration contour, and by decomposing that outer contour as figure \ref{ContourId4} shows, we ultimately find
\be\label{J3decomp3}\mathcal{J}(\Gamma_1,\Gamma_2,\Gamma_3)\propto I_{357}-nI_{347}+n^2I_{337}-2nI_{237}+2I_{137}-nI_{257}+n^2I_{247}+n^2I_{227}-nI_{127}+I_{157}-nI_{147}.\ee
After inserting the decomposition (\ref{J3decomp3}) into (\ref{CGsolns}), we find the proper normalization for the resulting formula by requiring that the limit (\ref{lim}) with $i=2$ equals the first hexagon connectivity weight (\ref{Pi1N3}) with $(x_2,x_3,x_4,x_5,x_6)\mapsto(x_4,x_5,x_6,x_7,x_8)$.  Upon using the results of sections \red{A 2}--\red{A 4} in \cite{florkleb3} (figure \ref{Cases}) with the boxed decomposition of figure \ref{BigDecomp} to determine the asymptotic behavior of $\mathcal{J}(\Gamma_1,\Gamma_2,\Gamma_3)$ as $x_3\rightarrow x_2$, we find the function
\begin{multline}\label{Pi0}\begin{aligned}\Pi_0(\kappa\,|\,x_1,x_2,\ldots,x_8)&=\frac{n(\kappa)^3}{[n(\kappa)^2-2]^2}\frac{\Gamma(2-8/\kappa)^3}{\Gamma(1-4/\kappa)^6}\Bigg(\prod_{\substack{i<j \\ i,j\neq 7}}^8(x_j-x_i)^{2/\kappa}\Bigg)\Bigg(\prod_{k\neq7}^8|x_7-x_k|^{1-6/\kappa}\Bigg)\\
&\times [I_{357}-nI_{347}+n^2I_{337}-2nI_{237}+2I_{137}-nI_{257}\end{aligned}\\ 
+n^2I_{247}+n^2I_{227}-nI_{127}+I_{157}-nI_{147}](\kappa\,|\,x_1,x_2,\ldots,x_8).\end{multline}

While it is a well-motivated guess, the function $\Pi_0$ (\ref{Pi0}) is unfortunately not the  connectivity weight $\Pi_1$ we seek.  Indeed, the identification of $\Pi_1$ with the collection of integration contours in the top illustration of figure \ref{OctXing} is incomplete.  More terms must be added to generate a formula for $\Pi_1$.  Perhaps this is not surprising, for to satisfy the various two-leg interval conditions for $\Pi_1$, we could have just as well entwined $[x_3,x_4]$ and $[x_5,x_6]$ together with a Pochhammer contour $\Gamma_2$ and entwined $[x_1,x_2]$ and $\Gamma_2$ together with another Pochhammer contour $\Gamma_3$.   As one might suspect,  this seemingly equally valid choice of integration contours  yields a different function.  Furthermore, neither it nor $\Pi_0$ gives a formula for $\Pi_1$.

Fortunately, however, $\Pi_0$ needs only a small correction to give a formula for $\Pi_1$.  To find this correction, we act on both sides of (\ref{Pi0}) with $[\mathscr{L}_\varsigma]$ for each $\varsigma\in\{1,2,\ldots,14\}$, beginning with $\varsigma=1$.  Above, we note that $\bar{\ell}_1\Pi_1$ (\ref{lim}) with $i=2$ gives the hexagon connectivity weight (\ref{Pi1N3}), and thanks to (\ref{takefirstlim}), this is sufficient to confirm that $[\mathscr{L}_1]\Pi_0=1$.  For all other $\varsigma\in\{2,3,4,\ldots,14\}$, we note that the polygon diagrams for all but two of the $[\mathscr{L}_\varsigma]$ with $\varsigma\geq2$ has an arc whose two endpoints correspond with the endpoints of either $(x_1,x_2)$, $(x_3,x_4)$, $(x_5,x_6)$, or $(x_7,x_8)$.  Because all of these intervals are two-leg intervals of $\Pi_0$, these $[\mathscr{L}_\varsigma]$ annihilate $\Pi_0$, confirming the duality condition (\ref{duality}) for these particular $\varsigma$.  The only two $[\mathscr{L}_\varsigma]$ with $\varsigma\geq2$ whose polygon diagram has no such arc are $[\mathscr{L}_3]$, whose polygon diagram matches that of $\Pi_3$ in figure \ref{OctXing}, and the equivalence class, call it $[\mathscr{L}_4]$, whose polygon diagram we obtain by rotating that for $[\mathscr{L}_3]$ by $\pi/2$ radians.

It is easy to show that $[\mathscr{L}_4]$ annihilates $\Pi_0$.  Indeed, we may choose an element $\mathscr{L}_4\in[\mathscr{L}_4]$ whose first limit is (\ref{lim}) with $i=2$ and whose next limit is (\ref{lim}) with $i=1$ and $i+1\mapsto4$.  Above (\ref{Pi0}), we note that the first limit sends $\Pi_0$ to the first hexagon connectivity weight (\ref{Pi1N3}) with $(x_2,x_3,x_4,x_5,x_6)\mapsto(x_4,x_5,x_6,x_7,x_8)$.  Because $(x_1,x_4)$ is a two-leg interval of this new weight, the next limit, and therefore $[\mathscr{L}_4]$, annihilates $\Pi_0$ altogether.

Next, we show that $[\mathscr{L}_3]$ does not annihilate $\Pi_0$, which is the main feature of $\Pi_0$ that distinguishes it from $\Pi_1$.  The boxed part of figure \ref{BigDecomp} shows the decomposition (\ref{J3decomp3}) of the Coulomb gas integral $\mathcal{J}(\Gamma_1,\Gamma_2,\Gamma_3)$ (\ref{eulerintegral}), and the part of this figure beneath the boxed area shows the calculation of $\mathscr{L}_3\{n^{-3}(n^2-2)^2\Pi_0\}$ and $\mathscr{L}_1\{n^{-3}(n^2-2)^2\Pi_0\}$, where $\mathscr{L}_3$ and $\mathscr{L}_1$ are allowable sequences of limits whose respective equivalence classes $[\mathscr{L}_3]$ and $[\mathscr{L}_1]$
are the only two that do not annihilate $\Pi_0$.  Their actions are respectively given by
\begin{align}\label{lim1}\begin{aligned}\mathscr{L}_3\{n^{-3}(n^2-2)^2\Pi_0\}(\kappa)&=\lim_{x_7\rightarrow x_2}\lim_{R\rightarrow\infty}\lim_{x_6\rightarrow x_3}\lim_{x_5\rightarrow x_4}[(x_7-x_2)(2R)(x_6-x_3)(x_5-x_4)]^{6/\kappa-1}\\
&\hspace{2cm}\times\,\,n(\kappa)^{-3}[n(\kappa)^2-2]^2\,\Pi_0(\kappa\,|\,x_1=-R,x_2,x_3,x_4,x_5,x_6,x_7,x_8=R),\end{aligned}\\
\label{limit2}\begin{aligned}\mathscr{L}_1\{n^{-3}(n^2-2)^2\Pi_0\}(\kappa)&=\lim_{x_7\rightarrow x_6}\lim_{R\rightarrow\infty}\lim_{x_3\rightarrow x_2}\lim_{x_5\rightarrow x_4}[(x_7-x_6)(2R)(x_3-x_2)(x_5-x_4)]^{6/\kappa-1}\\
&\hspace{2cm}\times\,\,n(\kappa)^{-3}[n(\kappa)^2-2]^2\,\Pi_0(\kappa\,|\,x_1=-R,x_2,x_3,x_4,x_5,x_6,x_7,x_8=R).\end{aligned}\end{align}
With the help of figure \ref{BigDecomp} and figure \ref{Cases}, we may find the limits (\ref{lim1}, \ref{limit2}).  Now with the image of $\Pi_0$ under all fourteen equivalence classes $[\mathscr{L}_\varsigma]\in\mathscr{B}_4^*$ known, (\ref{decompose}) gives the decomposition of $\Pi_0$ over $\mathscr{B}_4$.  The result is 
\be\label{decPi0}\mathscr{L}_\varsigma\{n^{-3}(n^2-2)^2\Pi_0\}=\left\{\begin{array}{ll}n^{-3}(n^2-2)^2, & \varsigma=1\\
n^{-2}(n^2-2),& \varsigma=3\\
0, &\varsigma\in\{2,4,5,\ldots,14\}\end{array}\right\}\quad\Longrightarrow\quad \Pi_0=\Pi_1+\left(\frac{n}{n^2-2}\right)\Pi_3.\ee
Finally, upon isolating $\Pi_1$ from (\ref{decPi0}) and substituting the formulas (\ref{Pi0}) and (\ref{Pi3N4}) for $\Pi_0$ and $\Pi_3$ respectively into the result, we find
\begin{multline}\begin{aligned}\label{Pi1N4}\Pi_1(\kappa\,|\,x_1,x_2,\ldots,x_8)&=\frac{n(\kappa)^3}{[n(\kappa)^2-2]^2} \,\frac{\Gamma(2-8/\kappa)^3}{\Gamma(1-4/\kappa)^6}\Bigg(\prod_{\substack{i<j \\ i,j\neq 7}}^8(x_j-x_i)^{2/\kappa}\Bigg)\Bigg(\prod_{k\neq7}^8|x_7-x_k|^{1-6/\kappa}\Bigg)\\
&\times [I_{357}-nI_{347}+n^2I_{337}-2nI_{237}+2I_{137}-nI_{257}+n^2I_{247}+n^2I_{227}\end{aligned}\\ 
-nI_{127}+I_{157}-nI_{147}-(n^2-2)(nI_{567}-n^2I_{667})](\kappa\,|\,x_1,x_2,\ldots,x_8).\end{multline}
If $\kappa\in(0,4]$, then the improper integrals in (\ref{Pi1N4}) diverge, and we regularize them via the replacement (\ref{replacePoch}).

The formulas (\ref{Pi2N4}, \ref{Pi3N4}, \ref{Pi1N4}) found for $\Pi_2$, $\Pi_3$, and $\Pi_1$ respectively are singular at any $\kappa\in(0,8)\times i\mathbb{R}$ such that $8/\kappa\in\mathbb{Z}^++1$ or $12/\kappa\in\mathbb{Z}^++1$.  Also, the formula (\ref{Pi2N4}) for $\Pi_2$ is singular at any $\kappa$ such that $n(\kappa)^4-3n(\kappa)^2+1=0$, the formula (\ref{Pi1N4}) for $\Pi_1$ is singular at any $\kappa$ such that $n(\kappa)^2=2$ (\ref{fugacity}), and all three formulas appear to vanish if $n(\kappa)=0$ (that is, if $8/\kappa\in2\mathbb{Z}^++1$).  Because section \red{III} of \cite{florkleb4} shows that $\Pi_1$, $\Pi_2$, and $\Pi_3$ are continuous functions of $\kappa$, these singularities must be removable.  Furthermore, because they are elements of a basis, there is no $\kappa\in(0,8)$, including those with $n(\kappa)=0$, such that $\Pi_1(\kappa)$ or $\Pi_2(\kappa)$ or $\Pi_3(\kappa)$ vanish for all $\boldsymbol{x}\in\Omega_0$.  Although we already know these facts, it is interesting to verify them directly from the formulas themselves, which we do in appendix \ref{appendix}. 

\begin{figure}[b]
\centering
\includegraphics[scale=0.2]{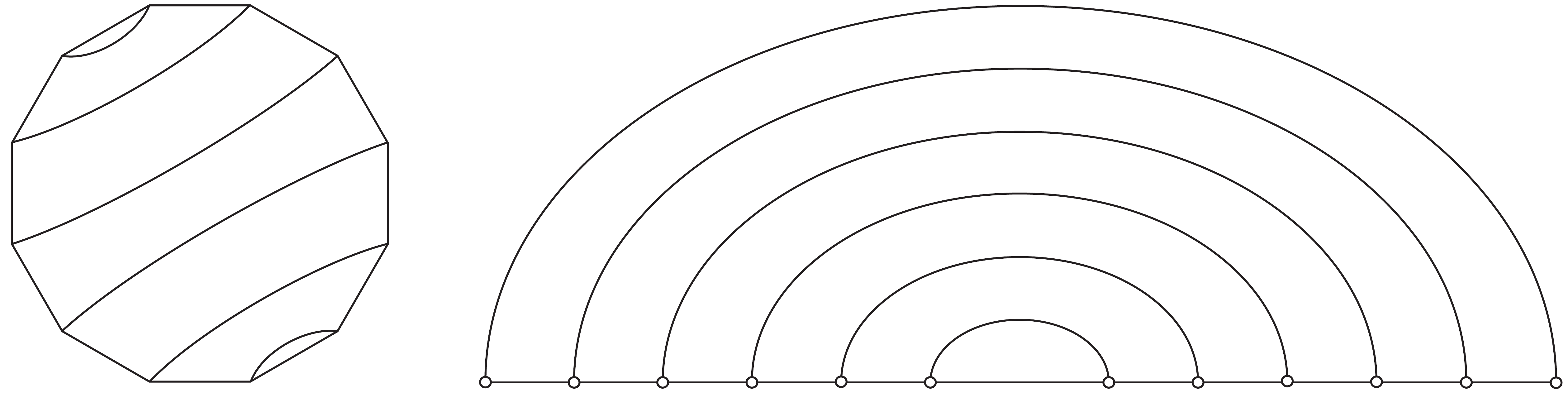}
\caption{The $N=6$ rainbow diagram (left) for the connectivity weight $\Pi_5\in\mathscr{B}_6$, and its image under a conformal map onto the upper half-plane (right).  Each but the middle interval and the outer interval containing infinity is a two-leg interval of $\Pi_5$.}
\label{12gonRainbow}
\end{figure}

\subsection{Rainbow connectivity weights for arbitrary $N$}\label{rainbow}

For $N>4$, correctly guessing a collection of integration contours for (\ref{CGsolns}) that gives a formula for some connectivity weight is generally difficult.  However, among the $C_N$ connectivity weights that span $\mathcal{S}_N$, there is a set of $N$ weights that we may always determine explicitly. (This explicit formula, however, is not always in the form of (\ref{form1}, \ref{form2}), that is, in terms of integrations along line segments or elementary Pochhammer contours.)  The polygon diagram for one of these, denoted $\Pi_{N-1}$, has an arc with its endpoints at the $j$th vertex and the $(2N-j+1)$th vertex of the polygon for all $j \in \{1,2,\ldots,N \}$, and the polygon diagrams for the other $N-1$ weights follow from rotating the diagram for $\Pi_{N-1}$.  After continuously mapping the polygon diagram for $\Pi_{N-1}$ onto the upper half-plane, we see that the image of the $j$th arc joins $x_j$ with $x_{2N-j+1}$ and that the collection of $N$ arcs sequentially nest each other in the upper half-plane to form a rainbow (figure \ref{12gonRainbow}).  For this reason, we call this polygon diagram (and the others generated by rotating it) a \emph{rainbow diagram} \cite{fgg}, and we call its associated connectivity weight a \emph{rainbow connectivity weight} \cite{florkleb3}.  As usual, we find an explicit formula only for $\Pi_{N-1}$.  An appropriate transformation then gives the formulas for the other $N-1$ rainbow connectivity weights.

Now we find a formula for $\Pi_{N-1}$ in the form (\ref{CGsolns}).  After choosing $c=2N$, we determine the integration contours of (\ref{eulerintegral}) for use in this formula.  From the rainbow diagram for $\Pi_{N-1}$, it is apparent that all intervals among $(x_1,x_2),$ $(x_2,x_3),\ldots,(x_{2N-2},x_{2N-1})$ and $(x_{2N-1},x_{2N})$, except $(x_N,x_{N+1})$, are two-leg intervals of $\Pi_{N-1}$.    Hence, following item \ref{ita} above, we entwine $x_{2N-1}$ and $x_{2N}$ with a Pochhammer contour $\Gamma_1=\mathscr{P}(x_{2N-1},x_{2N})$.  Furthermore, the formula for the second hexagon connectivity weight (\ref{Pi2N3}) suggests that we entwine $\Gamma_1$ and $x_{2N-2}$ with another Pochhammer contour $\Gamma_2$ in order for $(x_{2N-2},x_{2N-1})$ to be a two-leg interval of $\Pi_{N-1}$, and we denote this contour by $\mathscr{P}(x_{2N-2},\Gamma_1)$.  Moreover, the formula for the third octagon connectivity weight (\ref{Pi3N4}) suggests that we entwine $\Gamma_2$ and $x_{2N-3}$ with another Pochhammer contour $\Gamma_3$ in order for $(x_{2N-3},x_{2N-2})$ to be a two-leg interval of $\Pi_{N-1}$, and we denote this contour by $\mathscr{P}(x_{2N-3},\Gamma_2)$.  Repeating these selections, the $m$th contour $\Gamma_m$ is then the Pochhammer contour $\mathscr{P}(x_{2N-m},\Gamma_{m-1})$ entwining $x_{2N-m}$ with $\Gamma_{m-1}$ for all $m\in\{2,3,\ldots,N-1\}$.  No integration contour crosses the other intervals $(x_1,x_2),$ $(x_2,x_3),\ldots,(x_{N-1},x_N)$, and according to item \ref{itb} above, these are thus two-leg intervals of $\Pi_{N-1}$, as they must be.  These choices of integration contours determine the formula for $\Pi_{N-1}$ up to normalization. With $\mathcal{J}(\Gamma_1,\Gamma_2,\ldots,\Gamma_{N-1})$ defined in (\ref{eulerintegral}) with $c=2N$ and the integration contours defined above, we have
\begin{multline}\label{PiN}\Pi_{N-1}(\kappa\,|\,x_1,x_2,\ldots,x_{2N})=A_N\Bigg(\prod_{i<j}^{2N-1}(x_j-x_i)^{2/\kappa}\Bigg)\\
\times\,\Bigg(\prod_{k=1}^{2N-1}(x_{2N}-x_k)^{1-6/\kappa}\Bigg)\mathcal{J}\Big(\kappa\,\Big|\,\Gamma_1,\Gamma_2,\ldots,\Gamma_{N-1}\,\Big|\,x_1,x_2,\ldots,x_{2N}\Big),\quad c=2N,\end{multline}
where $A_N$ is a presently unspecified constant that we later choose in order to satisfy the duality condition (\ref{duality}) below.  (Equation (\ref{A}) gives the formula for $A_N$.)

Presently, the formula (\ref{CGsolns}) for $\Pi_{N-1}$ that follows from this selection of integration contours is a well-motivated guess.  To prove that it is indeed correct, we must verify the duality condition (\ref{duality}) for $\vartheta=N-1$ and all $\varsigma\in\{1,2,\ldots,C_N\}$.  For $\varsigma\neq N-1$, this verification is easy.  Indeed, every polygon diagram, except for the rainbow diagram with its $j$th arc joining the $j$th and $(2N-j+1)$th polygon vertex (i.e., the polygon diagrams of $[\mathscr{L}_{N-1}]$ and $\Pi_{N-1}$), has at least one arc with both of its endpoints among the first $N$ vertices of the polygon.  We pick one of these arcs.  There are two possibilities.  Either the arc has its endpoints at the $i$th and $(i+1)$th vertices for some $i\in\{1,2,\ldots,N-1\}$, or an arc nested within it does.  Thus, $[\mathscr{L}_\varsigma]$ with $\varsigma\neq N-1$ includes a limit (\ref{lim}) that sends $x_{i+1}\rightarrow x_i$ for some $i$ in this range.  However, because the polygon diagram of $\Pi_{N-1}$ does not have such an arc in its diagram, it follows from item \ref{twolegitem} above, and from item \ref{itb} with the previous paragraph, that $(x_i,x_{i+1})$ is a two-leg interval of $\Pi_{N-1}$.  Hence, the limit $[\mathscr{L}_\varsigma]$ annihilates $\Pi_{N-1}$ for all $\varsigma\neq N-1$.  This confirms the duality condition (\ref{duality}) for $\varsigma\neq N-1$.

\begin{figure}[b]
\centering
\includegraphics[scale=0.27]{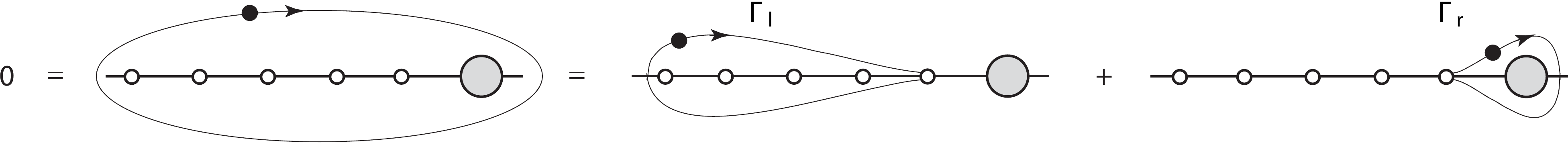}
\caption{Integrating $u_{N-1}$ of (\ref{eulerintegral}) around a loop surrounding $[x_1,x_{2N}]$ and the Pochhammer contours $\Gamma_1,$ $\Gamma_2,\ldots,\Gamma_{N-2}$, represented by the large gray circle, gives zero.  We separate the loop into a left half and right half at $x_{N+1}$.  ($N=4$ shown.)}
\label{lrloop}
\end{figure}

\begin{figure}[t]
\centering
\includegraphics[scale=0.27]{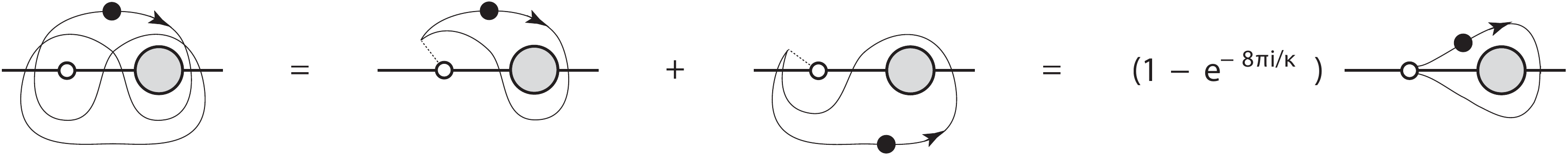}
\caption{We separate the Pochhammer contour $\Gamma_{N-1}$ that entwines $x_{N+1}$ (left point) with the Pochhammer contours $\Gamma_1,$ $\Gamma_2,\ldots,\Gamma_{N-2}$ (gray circle) into two copies of the loop $\Gamma_r$ (figure \ref{lrloop}).  The monodromy factor associated with $x_{N+1}$ is $\exp[-8\pi i/\kappa]$.}
\label{rloop}
\end{figure}

To finish the analysis, we must find a suitable normalization for $\Pi_{N-1}$ such that the duality condition (\ref{duality}) is true for $\vartheta=\varsigma=N-1$ too.  That is, we must find the normalization coefficient $A_N$ of (\ref{PiN}) such that
\begin{multline}\label{thebiglim}[\mathscr{L}_{N-1}]\Pi_{N-1}:=A_N\lim_{x_{2N}\rightarrow x_1}\lim_{x_{2N-1}\rightarrow x_2}\dotsm\lim_{x_{N+2}\rightarrow x_{N-1}}\lim_{x_{N+1}\rightarrow x_N}[(x_{2N}-x_1)(x_{2N-1}-x_2)\dotsm(x_{N+2}-x_{N-1})\,\,\times\\
(x_{N+1}-x_N)]^{6/\kappa-1}\Bigg(\prod_{i<j}^{2N-1}(x_j-x_i)^{2/\kappa}\Bigg)\Bigg(\prod_{k=1}^{2N-1}(x_{2N}-x_k)^{1-6/\kappa}\Bigg)\mathcal{J}\Big(\kappa\,\Big|\,\Gamma_1,\Gamma_2,\ldots,\Gamma_{N-1}\,\Big|\,x_1,x_2,\ldots,x_{2N}\Big)=1,
\end{multline}
where $\mathcal{J}(\Gamma_1,\Gamma_2,\ldots,\Gamma_{N-1})$ is given by (\ref{eulerintegral}) with $c=2N$ and where the integration contours are as described above (\ref{PiN}).  To compute the first limit of (\ref{thebiglim}), we find the asymptotic behavior of $\mathcal{J}(\Gamma_1,\Gamma_2,\ldots,\Gamma_{N-1}\,|\,x_1,x_2,\ldots,x_{2N})$ as $x_{N+1}\rightarrow x_N$ by replacing $\Gamma_{N-1}$ with a large, simple loop that surrounds all of $[x_1,x_{2N}]$ and $\Gamma_1,$ $\Gamma_2,\ldots,\Gamma_{N-2}$.  Because the residue at infinity of the integrand of (\ref{eulerintegral}), when viewed as a function of $u_{N-1}$, is zero, the integration around this large loop vanishes by the Cauchy integral theorem.  Assuming that $\kappa\in(4,8)$, we contract the large loop tightly around $[x_1,x_{2N}]$ and the other contours, pinch its upper and lower halves together at $x_{N+1}$, and divide it into a left loop $\Gamma_l$ and a right loop $\Gamma_r$ surrounding $\Gamma_1$, $\Gamma_2,\ldots,\Gamma_{N-1}$ (figure \ref{lrloop}).  We examine the integration around each loop.
\begin{enumerate}
\item\label{loop1} The integration around $\Gamma_r$ is proportional to what we find from integrating around the original Pochhammer contour $\Gamma_{N-1}=\mathscr{P}(x_{N+1},\Gamma_{N-2})$.  Indeed, we may decompose $\Gamma_{N-1}$ into these two simple loops (figure \ref{rloop}):  
\begin{enumerate}
\item The first loop starts at $x_{N+1}$, winds clockwise once around $\Gamma_{N-2}$, and finally ends at $x_{N+1}$.  This end point is the start point of the second loop.
\item The second loop starts at $x_{N+1}$, winds counterclockwise once around this point, then winds counterclockwise once around $\Gamma_{N-2}$, then winds clockwise once around $x_{N+1}$, and finally ends at $x_{N+1}$.
\end{enumerate}
Because it has opposite orientation and passes around $x_{N+1}$ before attaching to the first loop, the integration around the second loop gives $-\exp[2\pi i(-4/\kappa)]$ times the integration around the first loop.  We therefore have
\be\label{right}\mathcal{J}\Big(\kappa\,\Big|\,\Gamma_1,\Gamma_2,\ldots,\Gamma_{N-2},\Gamma_r\Big)=(1-e^{-8\pi i/\kappa})^{-1}\mathcal{J}\Big(\kappa\,\Big|\,\Gamma_1,\Gamma_2,\ldots,\Gamma_{N-2},\Gamma_{N-1}\Big).\ee
\item\label{loop2} The integration around $\Gamma_l$ decomposes into a sum of integrations both immediately above and below $[x_1,x_2]$, $[x_2,x_3],\ldots,[x_{N-1},x_N]$, and $[x_N,x_{N+1}]$.  Accounting for their relative phase difference, we have
\be\label{left}\mathcal{J}\Big(\kappa\,\Big|\,\Gamma_1,\Gamma_2,\ldots,\Gamma_{N-2},\Gamma_l\Big)=\sum_{j=1}^N(1-e^{2\pi ij(-4/\kappa)})\mathcal{J}\Big(\kappa\,\Big|\,\Gamma_1,\Gamma_2,\ldots,\Gamma_{N-2},[x_j,x_{j+1}]\Big).\ee
\end{enumerate}
The integration around $\Gamma_l$ combines with the integration around $\Gamma_r$ to give the integration around the large loop, which vanishes (figure \ref{lrloop}).  Therefore, upon identifying (\ref{right}) with the opposite of (\ref{left}), we find that if $\kappa\in(4,8)$, then
\begin{multline}\label{decJ}\mathcal{J}\Big(\kappa\,\Big|\,\Gamma_1,\Gamma_2,\ldots,\Gamma_{N-2},\Gamma_{N-1}\Big)=(e^{-8\pi i/\kappa}-1)\\
\times\,\sum_{j=1}^N2i\sin\left(\frac{4\pi j}{\kappa}\right)e^{-4\pi ij/\kappa}\mathcal{J}\Big(\kappa\,\Big|\,\Gamma_1,\Gamma_2,\ldots,\Gamma_{N-2},[x_j,x_{j+1}]\Big).\end{multline}
Because the left side of (\ref{decJ}) is an analytic function of all $\kappa\in(0,8)\times i\mathbb{R}$ and the equality (\ref{decJ}) holds for all $\kappa\in(4,8)$, the replacement (\ref{replacePoch}) with $\beta_i\mapsto\beta_j=-4/\kappa$ and $\beta_j\mapsto\beta_{j+1}=-4/\kappa$ gives the analytic continuation of (\ref{decJ}) to this strip, which includes $\kappa\in(0,4]$.

Now we determine the asymptotic behavior of (\ref{decJ}) as $x_{N+1}\rightarrow x_N$.  For this purpose, we write the $j$th Coulomb gas integral on the right side of (\ref{decJ}) as (assuming $\kappa\in(4,8)$, otherwise we replace $[x_j,x_{j+1}]$ with $\mathscr{P}(x_j,x_{j+1})$)
\begin{multline}\label{Jexplicit}\mathcal{J}\Big(\kappa\,\Big|\,\Gamma_1,\Gamma_2,\ldots,\Gamma_{N-2},[x_j,x_{j+1}]\,\Big|\,x_1,x_2,\ldots,x_{2N}\Big)=\oint_{\Gamma_1}{\rm d}u_1\oint_{\Gamma_2}{\rm d}u_2\dotsm\\
\begin{aligned}\dotsm&\oint_{\Gamma_{N-2}}{\rm d}u_{N-2}\,\Bigg(\prod_{l=1}^{2N-1}\prod_{m=1}^{N-2}(x_l-u_m)^{-4/\kappa}\Bigg)\Bigg(\prod_{m=1}^{N-2}(x_{2N}-u_m)^{12/\kappa-2}\Bigg)\Bigg(\prod_{p<q}^{N-2}(u_p-u_q)^{8/\kappa}\Bigg)\\
\times\,\,e^{4\pi ij/\kappa}&\int_{x_j}^{x_{j+1}}{\rm d}u_{N-1}\,\,\mathcal{N}\Bigg[\Bigg(\prod_{l=1}^{2N-1}(x_l-u_{N-1})^{-4/\kappa}\Bigg)\Big(x_{2N}-u_{N-1}\Big)^{12/\kappa-2}\Bigg(\prod_{p=1}^{N-2}(u_p-u_{N-1})^{8/\kappa}\Bigg)\Bigg],\end{aligned}\end{multline}
where the symbol $\mathcal{N}[\,\,\ldots\,\,]$ orders the enclosed terms so each difference is positive.  Now the integrand in (\ref{Jexplicit}) approaches its limit as $x_{N+1}\rightarrow x_N$ uniformly over $\Gamma_1,$ $\Gamma_2,\ldots,\Gamma_{N-2}$, and over $[x_j,x_{j+1}]$ too, unless $j \in \{N-1,N\}$.  In the former case, we find the limit of (\ref{Jexplicit}) by setting $x_{N+1}=x_N$.  If $j=N-1$ (resp.\ $j=N$), then (\red{A15}, \red{A23}) of section \red{A 3} (resp.\ (\red{A7}, \red{A9}) of section \red{A 2}) in \cite{florkleb3}, here with $i=N$ and $\beta_i=\beta_{i+1}=-4/\kappa$, give the asymptotic behavior of the integration with respect to $u_{N-1}$.  After inserting this into (\ref{Jexplicit}) and setting $x_{N+1}=x_N$ in all other factors, we find 
\begin{multline}\label{JexplicitN-1}\mathcal{J}\Bigg(\kappa\,\,\Bigg|\,\,\Gamma_1,\Gamma_2,\ldots,\Gamma_{N-2},\begin{array}{l}\,[x_{N-1},x_N] \vspace{.05cm} \\ \,[x_N,x_{N+1}]\end{array}\,\Bigg|\,x_1,x_2,\ldots,x_{2N}\Bigg)\underset{x_{N+1}\rightarrow x_N}{\sim}\left\{\begin{array}{l}e^{4\pi i(N-1)/\kappa} \, n(\kappa)^{-1} \\ e^{4\pi iN/\kappa}\end{array}\right\}e^{8(N-2)\pi i/\kappa}\\
\begin{aligned}&\times\,\Big(x_{N+1}-x_N\Big)^{1-8/\kappa}\Big(x_{2N}-x_N\Big)^{12/\kappa-2}\Bigg(\prod_{k=1}^{N-1}(x_N-x_k)^{-4/\kappa}\Bigg)\Bigg(\prod_{k=N+2}^{2N-1}(x_k-x_N)^{-4/\kappa}\Bigg)\frac{\Gamma(1-4/\kappa)^2}{\Gamma(2-8/\kappa)}\\
&\times\,\oint_{\Gamma_1}{\rm d}u_1\oint_{\Gamma_2}{\rm d}u_2\dotsm\oint_{\Gamma_{N-2}}{\rm d}u_{N-2}\,\Bigg(\prod_{\substack{l\neq N, \\ N+1}}^{2N-1}\prod_{m=1}^{N-2}(x_l-u_m)^{-4/\kappa}\Bigg)\Bigg(\prod_{m=1}^{N-2}(x_{2N}-u_m)^{12/\kappa-2}\Bigg)\Bigg(\prod_{p<q}^{N-2}(u_p-u_q)^{8/\kappa}\Bigg).\end{aligned}\end{multline}
(In particular, each of the $N-2$ identical phase factors $\exp[8\pi i/\kappa]$ that appears on the right side of (\ref{JexplicitN-1}) arises as the limit of the product $(x_N-u_p)^{-4/\kappa}(x_{N+1}-u_p)^{-4/\kappa}(u_p-u_{N-1})^{8/\kappa}$ as $x_{N+1}\rightarrow x_N$, with $u_{N-1}\in[x_N,x_{N+1}]$, for some $p\in\{1,2,\ldots,N-2\}$.)  Equation (\ref{JexplicitN-1}) gives the asymptotic behavior of the $j=N-1$ and $j=N$ terms on the right side of (\ref{decJ}) as $x_{N+1}\rightarrow x_N$. For $\kappa\in(0,8)$, the factor $(x_{N+1}-x_N)^{1-8/\kappa}$ dominates the other terms in (\ref{decJ}), which have finite limits.  After dropping these latter terms from (\ref{decJ}), replacing the $j=N$ and $j=N+1$ terms of (\ref{decJ}) with the right side of (\ref{JexplicitN-1}), inserting the right side of (\ref{decJ}) into (\ref{PiN}), and simplifying the result, we find
\begin{multline}\label{asympJ2}\Pi_{N-1}(\kappa\,|\,x_1,x_2,\ldots,x_{2N})\underset{x_{N+1}\rightarrow x_N}{\sim}(x_{N+1}-x_N)^{1-6/\kappa}4e^{4\pi i(2N-5)/\kappa}n(\kappa)^{-1}\\
\times\,\frac{A_N}{A_{N-1}}\sin\left(-\frac{4\pi}{\kappa}\right)\sin\left(\frac{4\pi(N+1)}{\kappa}\right)\frac{\Gamma(1-4/\kappa)^2}{\Gamma(2-8/\kappa)}\Xi_{N-2}(\kappa\,|\,x_1,x_2,\ldots,x_{N-1},x_{N+2},\ldots,x_{2N}),\end{multline}
where $\Xi_{N-2}$ is the rainbow connectivity weight in $\mathscr{B}_{N-1}$ whose polygon diagram follows from removing the arc joining the $N$th and $(N+1)$th vertices of the polygon diagram for $\Pi_{N-1}$.  

Returning to the task of finding the normalization $A_N$ in (\ref{thebiglim}), we use (\ref{asympJ2}) to find the first limit on the right side of (\ref{thebiglim}).  Then we repeat this process another $N-2$ times to compute all limits in (\ref{thebiglim}).  Upon doing this and applying the condition $[\mathscr{L}_{N-1}]\Pi_{N-1}=1$, we find the following normalization constant for use in (\ref{PiN}):
\be\label{A}A_N=\prod_{m=1}^{N-1}\frac{n(\kappa)e^{-4\pi i(2m-3)/\kappa} \, \Gamma(2-8/\kappa)}{4\sin(-4\pi/\kappa)\sin(4\pi(m+2)/\kappa) \, \Gamma(1-4/\kappa)^2}.\ee

This normalization constant (\ref{A}) leads to $[\mathscr{L}_{N-1}]\Pi_{N-1}=1$ only if $c=2N$, so $x_{2N}$ is the point bearing the conjugate charge.  If $c=2N-1$ instead, as it does in sections \ref{RecXingSect}--\ref{OctXingSect}, then the $m=1$ factor in (\ref{A}) must be altered.  To find $A_N$ in this situation, we perform all steps spanning (\ref{right}--\ref{asympJ2}) $N-2$ times, finding that the ratio $A_N/A_2$ equals the product of all factors in (\ref{A}), excluding the factor with  $m=1$.  Now, repeating this process one last time to determine $A_2$ would send the point $x_{2N-1}$ bearing the conjugate charge to $x_2$, producing a complicated result.  Instead, we determine $A_2$ by directly comparing (\ref{PiN}) with $N=2$ to the normalized rectangle connectivity weight formula (\ref{Pi1}) with $(x_3,x_4)\mapsto (x_{2N-1},x_{2N})$, previously found in section \ref{RecXingSect}.  We find that $A_2$ equals the $m=1$ factor in (\ref{A}) with the phase factor replaced by $\exp[-12\pi i/\kappa]$, so $A_N$ equals the product (\ref{A}) but with the phase factor in the $m=1$ factor replaced by $\exp[-12\pi i/\kappa]$.  Upon inserting this replacement into (\ref{A}, \ref{PiN}) with $N=2$, $N=3$, and $N=4$ and decomposing the integration contours as we did in sections \ref{RecXingSect}--\ref{OctXingSect}, we recover the connectivity weight formulas $\Pi_1$ (\ref{Pi1}) for the rectangle, $\Pi_2$ (\ref{Pi2N3}) for the hexagon, and $\Pi_3$ (\ref{Pi3N4}) for the octagon respectively.

For convenience, we replace the factor $(x_l-u_m)^{-4/\kappa}$ of the integrand in (\ref{PiN}) with $(u_m-x_l)^{-4/\kappa}$ for all $m\in\{1,2,\ldots,N-1\}$ and $l\in\{1,2,\ldots,2N-m\}$.  In so doing, the branch cuts anchored to points either to the left of or in the left (resp.\ points in the right) cycle of each Pochhammer contour follow the real axis leftward (resp.\ rightward) to negative (resp.\ positive) infinity, as we have in (\ref{PochDecomp}, \ref{Pochtostraight}).  For each $m$, this adjustment multiplies the $m$th factor of (\ref{A}) by $\exp[4\pi i(2N-m)/\kappa]$, or one phase factor of $\exp[4\pi i/\kappa]$ for each replacement.  Thus, we find the following alternative formula for $\Pi_{N-1}$ (\ref{PiN}),
\begin{multline}\label{RainbowPi}\Pi_{N-1}(\kappa\,|\,x_1,x_2,\ldots,x_{2N})\,=\Bigg(\prod_{m=1}^{N-1}\frac{n(\kappa)e^{4\pi i(m+3)/\kappa}\Gamma(2-8/\kappa)}{4\sin(-4\pi/\kappa)\sin(4\pi(m+2)/\kappa)\Gamma(1-4/\kappa)^2}\Bigg)\Bigg(\prod_{j<k}^{2N-1}(x_k-x_j)^{2/\kappa}\Bigg)\\
\begin{aligned}&\times\Bigg(\prod_{k=1}^{2N-1}(x_{2N}-x_k)^{1-6/\kappa}\Bigg)\oint_{\Gamma_{N-1}}{\rm d}u_{N-1}\dotsm\,\,\oint_{\Gamma_2}{\rm d}u_2\oint_{\Gamma_1}{\rm d}u_1\Bigg(\prod_{m=1}^{N-1}\prod_{l=1}^{2N-m}(u_m-x_l)^{-4/\kappa}\Bigg)\\
&\times\Bigg(\prod_{m=1}^{N-1}\prod_{l=2N-m+1}^{2N-1}(x_l-u_m)^{-4/\kappa}\Bigg)\Bigg(\prod_{m=1}^{N-1}(x_{2N}-u_m)^{12/\kappa-2}\Bigg)\Bigg(\prod_{p<q}^{N-1}(u_p-u_q)^{8/\kappa}\Bigg),\end{aligned}\end{multline}
after using the following identity to simplify the adjusted formula for $A_N$:
\be\Bigg(\prod_{m=1}^{N-1}e^{-4\pi i(2m-3)/\kappa}\Bigg)\Bigg(\prod_{m=1}^{N-1}e^{4\pi i(2N-m)/\kappa}\Bigg)=\prod_{m=1}^{N-1}e^{4\pi i(m+3)/\kappa}.\ee

Part of the prefactor in (\ref{RainbowPi}) is a natural generalization of part of the prefactor on the right side of (\ref{replacePoch}). This comes about for the following reason.  The left cycle of $\Gamma_m$ surrounds the branch point $x_{2N-m}$, so after $u_m\in\Gamma_m$ traces this cycle counterclockwise once, the integrand acquires a phase factor of $\exp[2\pi i\beta_i]$ with $\beta_i=-4/\kappa$.  And if $m>1$ (resp.\ $m=1$), then the right cycle of $\Gamma_m$ surrounds the branch points $x_j$ with $j\in\{2N-m+1,2N-m+2,\ldots,2N\}$ and $u_n$ with $n\in\{1,2,\ldots,m-1\}$ (resp.\ the one branch point $x_{2N}$).  Hence, after $u_m\in\Gamma_m$ traces this cycle counterclockwise once, the integrand acquires a phase factor of $\exp[2\pi i\beta_j]$, with
\be\label{betaj}\beta_j=\overbrace{(-4/\kappa)(m-1)}^{\{x_j\}_{j=2N-m+1}^{2N-1}}+\overbrace{(8/\kappa)(m-1)}^{\{u_n\}_{n=1}^{m-1}}+\overbrace{12/\kappa-2}^{x_{2N}}=(4/\kappa)(m+2)-2.\ee
After inserting these expressions for $\beta_i$ and $\beta_j$ (\ref{betaj}) into the prefactor on the right side of (\ref{replacePoch}), we recover part of the prefactor in (\ref{RainbowPi}):
\be\label{generalize}4e^{\pi i(\beta_i-\beta_j)}\sin\pi\beta_i\sin\pi\beta_j=4e^{-4\pi i(m+3)/\kappa}\sin(-4\pi/\kappa)\sin(4\pi(m+2)/\kappa).\ee
A natural interpretation of this result is to say that the right side of (\ref{generalize}) generalizes the prefactor on the right side of (\ref{Pochtostraight}, \ref{replacePoch}) to situations in which the right cycle of a Pochhammer contour surrounds several branch points.

\section{Summary}\label{summary}

This article presents formulas for ``connectivity weights" or ``pure SLE$_\kappa$ partition functions" $\Pi_\varsigma$ \cite{florkleb4}, special functions that arise in multiple SLE$_\kappa$ (Schramm-L\"owner Evolution) \cite{bbk,dub2,graham,kl,sakai}  and that also have a conformal field theory interpretation  \cite{florkleb4}.  The connectivity weights are the main ingredients of the conjectured formula (\ref{xing}) for the \emph{crossing probability} $P_\varsigma$  \cite{bber,florkleb4}, that is, the probability that the $2N$ growing curves of a multiple-SLE$_\kappa$ process anchored at $x_1$, $x_2,\ldots,x_{2N}$ almost surely join together pairwise in the $\varsigma$th connectivity.  (Figure \ref{Connect} shows an example.  The index  $\varsigma\in\{1,2,\ldots,C_N\}$, with $C_N$ the $N$th Catalan number, enumerates all of the possible connectivities, as explained in the introduction \ref{intro}.)

In this article, using rigorous results from \cite{florkleb,florkleb2,florkleb3,florkleb4}, we find explicit formulas for all of the connectivity weights from the solution space $\mathcal{S}_N$ of the system of PDEs (\ref{nullstate}, \ref{wardid}) with $N\in\{1,2,3,4\}$ (those with $N\in\{3,4\}$ are new), and for so-called ``rainbow connectivity weights" for all $N\in\mathbb{Z}^++1$.   These formulas are expected to give cluster-crossing probabilities for critical statistical mechanics lattice models inside rectangles ($N=2$), hexagons ($N=3$), and octagons ($N=4$) with a free/fixed side-alternating boundary condition, a topic that remains to be investigated.  In the case of critical percolation $(\kappa=6)$, $\Pi_\varsigma$ itself gives the probability that percolation clusters join the fixed sides of a polygon with $2N$ sides in the $\varsigma$th connectivity (again, with a free/fixed side-alternating boundary condition on the polygon's perimeter).  These probabilities generalize Cardy's formula (\ref{Cardy}) \cite{c3} for the probability that a percolation cluster joins the opposite sides of a rectangle $(N=2)$.
 
In section \ref{analysis}, we find formulas for the connectivity weights by choosing a suitable collection of integration contours to use in the Coulomb gas solutions (\ref{CGsolns}, \ref{eulerintegral}) for the system of equations (\ref{nullstate}, \ref{wardid}) and then verifying that the resulting formulas satisfy the duality condition (\ref{duality}).  This verification is a sufficient last step because the duality condition (\ref{duality}) uniquely defines each connectivity weight.  We give a formula for only one connectivity weight per collection of weights that are identical up to rotation of their polygon diagrams (figure \ref{PiDiagrams}).  A suitable transformation then gives formulas for the other weights in the collection.  For $N=1$, the formula for the unique connectivity weight is (\ref{Pi1Xing1}); for $N=2$, the formula is (\ref{Pi1}); for $N=3$, these formulas are (\ref{Pi1N3}, \ref{Pi2N3}); and for $N=4$, these formulas are (\ref{Pi2N4}, \ref{Pi3N4}, \ref{Pi1N4}).  In addition, we find an explicit formula (\ref{RainbowPi}) for the ``rainbow connectivity weight" (section \ref{rainbow}), generating a multiple-SLE$_\kappa$ process in which the boundary arcs join the $2N$ marked points on the real axis to form a ``rainbow" \cite{fgg, kl} in the upper half-plane (figure \ref{12gonRainbow}).  (Ref.\ \cite{kype2} obtains an alternative explicit formula for the rainbow connectivity weight using the ``spin-chain Coulomb gas correspondence," previously developed in \cite{kype}.)  In appendix \ref{appendix}, we explicitly show that all singularities in $\kappa\in(0,8)\times i\mathbb{R}$ of factors in these formulas are removable, so each connectivity weight is an analytic function of $\kappa$ in this region.  Finally, in appendix \ref{appendixB}, we study logarithmic singularities of some of the connectivity weight formulas as one or more points $x_j$ approach a common point $x_i$.  Predicted to occur by logarithmic CFT, these logarithmic singularities may arise only for certain $\kappa$ corresponding to CFT minimal models \cite{florkleb4} and for $N\in\mathbb{Z}^+$ sufficiently large.  Some of the analysis in appendix \ref{appendixB} is a natural extension of the analysis in appendix \red{A} of \cite{florkleb4}.

\section{Acknowledgements}

We thank K.\ Kyt\"ol\"a and E.\ Peltola for insightful conversations, and we thank C.\ Townley Flores for carefully proofreading the manuscript.

This work was supported by National Science Foundation Grants Nos.\ PHY-0855335 (SMF) and DMR-0536927 (PK and SMF).

\appendix{}

\section{Investigation of connectivity weights near singular $\kappa\in(0,8)\times i\mathbb{R}$}\label{appendix}

The prefactors that appear in the various connectivity weight formulas (\ref{Pi1}, \ref{Pi1N3}, \ref{Pi2N3}, \ref{Pi2N4}, \ref{Pi3N4}, \ref{Pi1N4}) have poles exclusively at certain $\kappa\in(0,8)$, yet in spite of this, the connectivity weights are analytic functions of $\kappa\in(0,8)\times i\mathbb{R}$.  Indeed, if we replace $F\in\mathcal{S}_N$ in (\ref{decompose}) with each element of the Temperley-Lieb set $\mathcal{B}_N=\{\mathcal{F}_1,\mathcal{F}_2,\ldots,\mathcal{F}_{C_N}\}$ (which is a basis for $\mathcal{S}_N$ if and only if $\kappa\in(0,8)$ is not an exceptional speed with $q\leq N+1$, see definitions \red{4} and \red{5} and lemma \red{6} of \cite{florkleb3}), then we obtain a $C_N\times C_N$ system of equations in the connectivity weights $\Pi_1$, $\Pi_2\ldots,\Pi_{C_N}$ that is invertible if and only if $\kappa\in(0,8)$ is not an exceptional speed with $q\leq N+1$.  In fact, each element of $\mathcal{B}_N$ and all coefficients of the system are analytic at any $\kappa\in(0,8)\times i\mathbb{R}$.  As such, it follows upon solving the system and applying analytic continuation that the connectivity weights are analytic at any $\kappa$ in this region where the system is invertible.  If $\kappa\in(0,8)$ is among the mentioned exceptional speeds, then the system is not invertible.  However, by using the basis from the proof of theorem \red{8} in \cite{florkleb3}, we encounter a different invertible $C_N\times C_N$ system of equations in all of the connectivity weights \cite{florkleb4}.  Then the same analysis shows that the connectivity weights are analytic at these speeds too.  

Thanks to these observations, it is not necessary to directly verify that the connectivity weight formulas derived in this article have finite, non-vanishing limits as $\varkappa\rightarrow\kappa$, where $\kappa$ is a pole of the formula prefactor.  Indeed, because we know that this limit of any particular connectivity weight exists, we infer that the corresponding connectivity weight formula with the prefactor dropped must have a zero at $\varkappa=\kappa$.   Furthermore, the order of the zero must equal the order of the pole, or else the connectivity weight would vanish at $\varkappa=\kappa$, thereby violating the duality relation (\ref{duality}) for $\varsigma=\vartheta$.  (To complete the argument supporting this last point, we must first establish that the limits of the connectivity weight formulas as $\varkappa\rightarrow\kappa$ still satisfy the duality relation (\ref{duality}).  Indeed, this follows from the fact that the limit $\varkappa\rightarrow\kappa$ commutes with all limits in every equivalence class $[\mathscr{L}_\varsigma]$ of the dual space $\mathscr{B}_N^*$.  See the paragraph beneath (\red{49}) in \cite{florkleb3}.)  

Thus, for any $N \in \mathbb{Z}^+$, all connectivity weights $\Pi_1$, $\Pi_2,\ldots,\Pi_{C_N}$ of $\mathscr{B}_N$ (\ref{BNbasis}) are analytic functions of $\kappa\in(0,8)\times i\mathbb{R}$.  The remainder of this appendix demonstrates in detail how this comes about from the formulas for the cases $N\in\{2,3,4\}$ presented in this article.  Logically, this is not necessary, and the reader may omit it.  However, it is of interest as an illustration of relevant contour integral technology and because, as we show below, the singular behavior at  the mentioned exceptional speeds sometimes occurs  for reasons related to the linear dependence of $\mathcal{B}_N$ at this speed \cite{florkleb3}.

Throughout this section, we let $\varkappa\in(0,8)\times i\mathbb{R}$ denote an arbitrary SLE$_\kappa$ speed (we consider complex $\varkappa$ only for the occasional purpose of analytic continuation), and we let $\kappa$ denote a particular SLE$_\kappa$ speed that is a singularity of a prefactor appearing  in a connectivity weight formula.

\subsection{Singularities of the rectangle connectivity weight}\label{singularrect}

Equation (\ref{Pi1}) gives a formula for the first rectangle connectivity weight.  After using (\ref{replacePoch}) to extend the integration in this formula to all speeds $\varkappa\in(0,8)\times i\mathbb{R}$, we write it as
\be\label{2Pi1}\frac{n(\varkappa) \, \Gamma(2-8/\varkappa)}{4e^{-16\pi i/\varkappa}\sin(-4\pi/\varkappa)\sin(12\pi/\varkappa) \,\Gamma(1-4/\varkappa)^2}\,\,\dotsm\,\,\oint_{\mathscr{P}(x_3,x_4)}(x_4-u)^{-4/\varkappa}(u-x_3)^{12/\varkappa-2}\dotsm\,{\rm d}u,\ee
where the ellipses stand for omitted factors analytic for $\varkappa\in(0,8)$ and over $u\in A$, where $A$ is an open region containing $\mathscr{P}(x_3,x_4)$.  Equation (\ref{fugacity}) defines the factor $n(\varkappa)$.  Among speeds $\varkappa\in(0,8)$, the prefactor in (\ref{2Pi1}) is singular at all $\kappa$ such that $8/\kappa\in\mathbb{Z}^++1$ or $12/\kappa\in\mathbb{Z}^++1$.  We divide these singularities into three cases.
\begin{enumerate}
\item\label{it1} $8/\kappa\in\mathbb{Z}^++1$ and $4/\kappa\not\in\mathbb{Z}^+$: In this case, only the factor $\Gamma(2-8/\varkappa)$  of (\ref{2Pi1}) has a pole at $\varkappa=\kappa$.  However, $\varkappa=\kappa$ is a removable singularity of the product $n(\varkappa)\Gamma(2-8/\varkappa)$ and hence of (\ref{2Pi1}).
\item\label{it2} $12/\kappa\in\mathbb{Z}^++1$ and $4/\kappa\not\in\mathbb{Z}^+$: In this case, only the reciprocal of $\sin(12\pi/\varkappa)$ has a pole at $\varkappa=\kappa$.  However, it follows  from (\ref{PochDecomp}) that $\kappa$ is a zero of the contour integral too.  As such, $\kappa$ is a removable singularity of 
\be \frac{1}{\sin(12\pi/\varkappa)}\,\oint_{\mathscr{P}(x_3,x_4)}(x_4-u)^{12/\varkappa-2}(u-x_3)^{-4/\varkappa}\dotsm\,{\rm d}u\ee
and hence of (\ref{2Pi1}).
\item\label{it3} $4/\kappa\in\mathbb{Z}^+$: In this case, almost every factor in (\ref{2Pi1}) vanishes or has a pole at $\varkappa=\kappa$.  Upon writing (\ref{2Pi1}) as
\be\label{splitPi1}\frac{n(\varkappa)}{4e^{-16\pi i/\varkappa}}\underbrace{\frac{\Gamma(2-8/\varkappa)}{\sin(-4\pi/\varkappa)\Gamma(1-4/\varkappa)^2}}_{1.}\,\,\dotsm\,\,\underbrace{\frac{1}{\sin(12\pi/\varkappa)}\oint_{\mathscr{P}(x_3,x_4)}\dotsm}_{2.}\ee
and using (\ref{PochDecomp}), we see that $\kappa$ is a removable singularity of the first and second factors in (\ref{splitPi1}) and hence of (\ref{2Pi1}).
\end{enumerate}
From these facts, we conclude that the first rectangle connectivity weight (\ref{Pi1}) is an analytic function of $\varkappa\in(0,8)\times i\mathbb{R}$.

\subsection{Singularities of the hexagon connectivity weights}\label{singularhex}

Equation (\ref{Pi1N3}) gives a formula for the first hexagon connectivity weight.  After using (\ref{replacePoch}) to extend the integration in this formula to all speeds $\varkappa\in(0,8)\times i\mathbb{R}$, we write the terms in this formula as
\begin{multline}\label{2Pi1N3}\frac{n(\varkappa)^2\Gamma(2-8/\varkappa)^2}{4^2e^{-16\pi i/\varkappa}\sin^3(-4\pi/\varkappa)\sin(12\pi/\varkappa)\Gamma(1-4/\varkappa)^4[n(\varkappa)^2-2]}\,\,\dotsm\,\,\oint_{\mathscr{P}(x_i,x_{i+1})}(x_{i+1}-u_2)^{-4/\varkappa}\dotsm\\
\dotsm\,\,(u_2-x_i)^{-4/\varkappa}\dotsm\,\,\oint_{\mathscr{P}(x_5,x_6)}(x_6-u_1)^{-4/\varkappa}(u_1-x_5)^{12/\varkappa-2}\dotsm\,{\rm d}u_1\,{\rm d}u_2,\quad i\in\{1,2,3\},
\end{multline}
where the ellipses stand for omitted factors analytic over $\varkappa\in(0,8)\times i\mathbb{R}$, over $u_1\in A$, and over $u_2\in B$, where $A$ and $B$ are open regions containing $\mathscr{P}(x_5,x_6)$ and $\mathscr{P}(x_i,x_{i+1})$ respectively.  Among speeds $\varkappa\in(0,8)\times i\mathbb{R}$, the prefactor in (\ref{2Pi1N3}) is singular at all $\kappa$ such that $8/\kappa\in\mathbb{Z}^++1$, $12/\kappa\in\mathbb{Z}^++1$, or $n(\kappa)^2=2$.  We divide these singularities into four cases.
\begin{enumerate}
\item\label{} $8/\kappa\in\mathbb{Z}^++1$ and $4/\kappa\not\in\mathbb{Z}^+$: The treatment of this case is identical to that of item \ref{it1} in section \ref{singularrect}.
\item\label{} $12/\kappa\in\mathbb{Z}^++1$ and $4/\kappa\not\in\mathbb{Z}^+$: The treatment of this case is identical to that of item \ref{it2} in section \ref{singularrect}.
\item\label{} $4/\kappa\in\mathbb{Z}^+$: In this case, almost every factor in (\ref{2Pi1N3}) vanishes or has a pole at $\varkappa=\kappa$.  Upon writing (\ref{2Pi1N3}) as
\begin{multline}\label{splitPi1N3}\frac{n(\varkappa)^2}{4^2e^{-16\pi i/\varkappa}[n(\varkappa)^2-2]}\overbrace{\frac{\Gamma(2-8/\varkappa)^2}{\sin^2(-4\pi/\varkappa)\Gamma(1-4/\varkappa)^4}}^{1.}\,\,\dotsm\\
\dotsm\,\,\underbrace{\frac{1}{\sin(-4\pi/\varkappa)}\oint_{\mathscr{P}(x_i,x_{i+1})}\dotsm}_{2.}\,\,\underbrace{\frac{1}{\sin(12\pi/\varkappa)}\oint_{\mathscr{P}(x_5,x_6)}\dotsm}_{3.}\end{multline}
and using (\ref{PochDecomp}), we see that $\kappa$ is a removable singularity of the first, second, and third factors in (\ref{splitPi1N3}) and hence of (\ref{2Pi1N3}).
\item\label{it4}$n(\kappa)^2=2$: In this case, $\kappa\in(0,8)\times i\mathbb{R}$ is any of the exceptional speeds $\kappa_{4,q'}=16/q'$ \cite{florkleb3}, where $q'>1$ is an integer coprime with four.  Here, $n(\varkappa)^2-2=O(\varkappa-\kappa)$ as $\varkappa\rightarrow\kappa$ (\ref{fugacity}).  Also, (\ref{Pi1N3}, \ref{2Pi1N3}) with the factor $[n(\varkappa)^2-2]^{-1}$ dropped is an analytic function of $\varkappa\in(0,8)\times i\mathbb{R}$.  Thus, if it vanishes at $\varkappa=\kappa$, then it is $O((\varkappa-\kappa)^m)$ as $\varkappa\rightarrow\kappa$ for some $m\in\mathbb{Z}^+$, so $\kappa$ is a removable singularity of (\ref{Pi1N3}, \ref{2Pi1N3}).  Indeed, this function does vanish at $\varkappa=\kappa$, as 
\be\label{Isvanish} [nI_{25}-I_{15}-I_{35}](\kappa)=0\quad \text{if $n(\kappa)^2=2$.}\ee

\begin{figure}[t]
\centering
\includegraphics[scale=0.27]{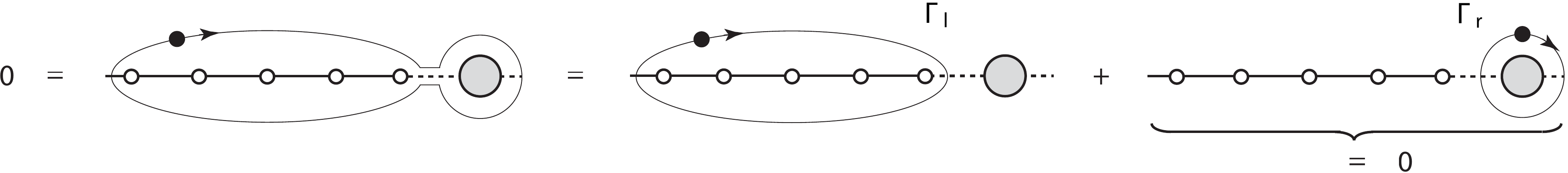}
\caption{If $\kappa=4(N+1)/q'$ for some $q'\in\mathbb{Z}^+$ coprime with $N+1$, then the integrand of (\ref{eulerintegral}) does not have a branch cut immediately to the right of $x_{N+1}$ (dashed), leading to the above decomposition.  We interpret the gray circle as in figure \ref{lrloop}.}
\label{retract}
\end{figure}

To derive (\ref{Isvanish}), we integrate $u_2$ of (\ref{2Pi1N3}) around a simple loop that surrounds $[x_1,x_6]$ and $\Gamma_1=[x_5,x_6]$, pinch the upper and lower halves of the loop together between $x_4$ and $x_5$, and divide the loop into a left half that tightly surrounds $[x_1,x_4]$ and a right half that surrounds $\Gamma_1$.  Now, as we cross the interval $(x_4,x_5)$, the integrand of (\ref{eulerintegral}) with $N=3$ and $c=5$ does not acquire a phase factor.  Indeed, it accumulates one phase factor of
\be e^{2\pi i(-4/\kappa)}=e^{2\pi i(-4q'/16)}=e^{2\pi i(-q'/4)}\ee
per each of the four rightward-pointing branch cuts respectively anchored to $x_1$, $x_2$, $x_3$, and $x_4$, and the product of all four factors is one.  Thus, the left and right halves of the simple loop separate into a left loop $\Gamma_l$ tightly wrapped around $[x_1,x_4]$ and a right loop $\Gamma_r$ surrounding $\Gamma_1$ (figure \ref{retract} with $N=3$).  Because integration around the original loop gives zero by the Cauchy integral theorem, the integration around $\Gamma_l$ gives the opposite of the integration around $\Gamma_r$.

Interestingly, the integration along $\Gamma_r\times\Gamma_1$ gives zero too.  To see why, we note that (\ref{CGsolns}) with $N=3$ gives an element of $\mathcal{S}_3$ whenever $\Gamma_1$ and $\Gamma_2$ are two nonintersecting closed contours, as is the case here with $\Gamma_2=\Gamma_r$.  (Indeed, $\Gamma_r$ closes because the integrand of (\ref{eulerintegral}) does not acquire a phase factor as this contour crosses the interval $[x_4,x_5]$.)  Then according to item \ref{itb} in section \ref{analysis}, $(x_1,x_2)$, $(x_2,x_3)$, and $(x_3,x_4)$ are two-leg intervals of this element of $\mathcal{S}_3$.  However, every equivalence class in $\mathscr{B}_3^*$ has a limit that brings together the endpoints of one of these intervals, annihilating this element.  As such, this element is in the kernel of the map $v:\mathcal{S}_N\rightarrow\mathbb{R}^{C_N}$ with $v(F)_\varsigma:=[\mathscr{L}_\varsigma]F$.  According to lemma \red{15} of \cite{florkleb}, it must therefore be zero.

It immediately follows from the previous paragraph and the last sentence of the paragraph preceding it that the integration along $\Gamma_l\times\Gamma_1$ gives zero too (figure \ref{retract} with $N=3$).  After decomposing the integration around $\Gamma_l$ into integrations along the three intervals $[x_1,x_2]$, $[x_2,x_3]$, and $[x_3,x_4]$ that $\Gamma_l$ surrounds, we find that
\be\label{match}0=\mathcal{J}(\kappa\,|\,\Gamma_1,\Gamma_l)\propto[I_{15}-nI_{25}+(n^2-1)I_{35}](\kappa),\quad n(\kappa)^2=2,\ee
with $I_{ij}$ and $n(\kappa)$ defined in (\ref{Iij}) and (\ref{fugacity}) respectively.  This result follows from factoring out phase factors in such a way that each $I_{ij}$ is real and combining terms as in the previous sections.  (If $\kappa\leq4$, then these improper integrals diverge.  We regularize them via the replacement (\ref{replacePoch}).)  Because (\ref{match}) is identical to (\ref{Isvanish}), we conclude that $\kappa$ is a removable singularity of (\ref{2Pi1N3}).

Interestingly, (\ref{match}) implies the linear dependence of the Temperley-Lieb set $\mathcal{B}_3$ for $\varkappa=\kappa=\kappa_{4,q'}$ (definitions \red{4} and \red{5} of \cite{florkleb3}), previously determined by other means in the proof of lemma \red{6} in \cite{florkleb3}.  Indeed, by integrating $u_1$ in (\ref{match}) around a loop surrounding $[x_1,x_6]$ and repeating the above steps, we may express $I_{15}$, $I_{25}$, and $I_{35}$ in (\ref{match}) as linear combinations of $I_{ij}$ with $i,j\neq5$.  After inserting these expressions into (\ref{match}) and multiplying the result by the algebraic factors preceding $\mathcal{J}(\Gamma_1,\Gamma_2)$ in (\ref{CGsolns}) with $N=3$ and $c=5$, we find a linear combination of elements in $\mathcal{B}_3$ vanishing for $\varkappa=\kappa=\kappa_{4,q'}$.  (See also corollary \red{9} of \cite{florkleb3}.)

\end{enumerate}
Thus, the first hexagon connectivity weight (\ref{Pi1N3}) is an analytic function of $\varkappa\in(0,8)\times i\mathbb{R}$.

Equation (\ref{Pi2N3}) gives a formula for the second hexagon connectivity weight.  After using (\ref{replacePoch}) to extend the integration in this formula to all speeds $\varkappa\in(0,8)\times i\mathbb{R}$, we write it as
\begin{multline}\label{2Pi2N3}\frac{n(\varkappa)^2\Gamma(2-8/\varkappa)^2}{4^2\sin^2(-4\pi/\varkappa)\sin^2(12\pi/\varkappa)\Gamma(1-4/\varkappa)^4}\dotsm\,\,\oint_{\mathscr{P}(x_4,x_5)}(x_5-u_2)^{12/\varkappa-2}\\
\times\,(u_2-x_4)^{-4/\varkappa}\dotsm\,\,\oint_{\mathscr{P}(x_5,x_6)}(x_6-u_1)^{-4/\varkappa}(u_1-x_5)^{12/\varkappa-2}\dotsm\,{\rm d}u_1\,{\rm d}u_2,
\end{multline}
where the ellipses stand for omitted factors analytic over $\varkappa\in(0,8)\times i\mathbb{R}$, over $u_1\in A$, and over $u_2\in B$, where $A$ and $B$ are open regions containing $\mathscr{P}(x_5,x_6)$ and $\mathscr{P}(x_4,x_5)$ respectively.  Among speeds $\varkappa\in(0,8)\times i\mathbb{R}$, the prefactor in (\ref{2Pi2N3}) is singular at all $\kappa$ such that $8/\kappa\in\mathbb{Z}^++1$ or $12/\kappa\in\mathbb{Z}^++1$.  We divide these singularities into three cases.
\begin{enumerate}
\setcounter{enumi}{4}
\item\label{} $8/\kappa\in\mathbb{Z}^++1$ and $4/\kappa\not\in\mathbb{Z}^+$: The treatment of this case is identical to that of item \ref{it1} in section \ref{singularrect}.
\item\label{} $12/\kappa\in\mathbb{Z}^++1$ and $4/\kappa\not\in\mathbb{Z}^+$: The treatment of this case is identical to that of item \ref{it2} in section \ref{singularrect}.
\item\label{} $4/\kappa\in\mathbb{Z}^+$: In this case, almost every factor in (\ref{2Pi2N3}) vanishes or has a pole at $\varkappa=\kappa$.  Upon writing (\ref{2Pi2N3}) as
\be\label{splitPi2N3}\frac{n(\varkappa)^2}{4^2}\underbrace{\frac{\Gamma(2-8/\varkappa)^2}{\sin^2(-4\pi/\varkappa)\Gamma(1-4/\varkappa)^4}}_{1.}\,\,\dotsm\,\,\underbrace{\frac{1}{\sin^2(12\pi/\varkappa)}\oint_{\mathscr{P}(x_4,x_5)}\dotsm\,\,\oint_{\mathscr{P}(x_5,x_6)}\dotsm}_{2.}\ee
and using (\ref{PochDecomp}), we see that $\kappa$ is a removable singularity of the first and second factors in (\ref{splitPi2N3}) and hence of (\ref{2Pi2N3}).
\end{enumerate}
From these facts, we conclude that the second hexagon connectivity weight (\ref{Pi2N3}) is an analytic function of $\varkappa\in(0,8)\times i\mathbb{R}$.

\subsection{Singularities of the octagon connectivity weights}

The content of this section is almost identical to that of sections \ref{singularrect} and \ref{singularhex}, except that the analysis of singularities of the first octagon connectivity weight $\Pi_1$ (\ref{Pi1N4}) at all $\kappa$ such that $n(\kappa)^2=2$ is considerably more involved.  There are also differences between the treatment of the other octagon weights $\Pi_2$, $\Pi_3$ and their analogues in sections \ref{singularrect} and \ref{singularhex}.  Because these differences are at times significant, we include a complete analysis of singularities of octagon connectivity weights here.

There are three distinct octagon connectivity weights.  We begin with the formula  (\ref{Pi2N4}) for the second connectivity weight $\Pi_2$.  After using (\ref{replacePoch}) to extend it to all speeds $\varkappa\in(0,8)\times i\mathbb{R}$, we find four terms, all with the form
\begin{multline}\label{2Pi2N4}\frac{n(\varkappa)^3\Gamma(2-8/\varkappa)^3}{4^3\sin^4(-4\pi/\varkappa)\sin^2(12\pi/\varkappa)\Gamma(1-4/\varkappa)^6[n(\varkappa)^4-3n(\varkappa)^2+1]}\,\,\dotsm\,\,\oint_{\mathscr{P}(x_i,x_{i+1})}(x_{i+1}-u_3)^{-4/\varkappa}(u_3-x_i)^{-4/\varkappa}\dotsm\\
\oint_{\mathscr{P}(x_6,x_7)}(x_7-u_2)^{12/\varkappa-2}(u_2-x_6)^{-4/\varkappa}\dotsm\,\,\oint_{\mathscr{P}(x_7,x_8)}(x_8-u_1)^{-4/\varkappa}(u_1-x_7)^{12/\varkappa-2}\dotsm\,{\rm d}u_1\,{\rm d}u_2\,{\rm d}u_3,\quad i\in\{1,2,3,4\},\end{multline}
where the ellipses stand for omitted factors analytic over $\varkappa\in(0,8)\times i\mathbb{R}$, over $u_1\in A$, over $u_2\in B$, and over $u_3\in C$, where $A$, $B$, and $C$ are open regions containing $\mathscr{P}(x_7,x_8)$, $\mathscr{P}(x_6,x_7)$, and $\mathscr{P}(x_i,x_{i+1})$ respectively.  Among speeds $\varkappa\in(0,8)\times i\mathbb{R}$, the prefactor in (\ref{2Pi2N4}) is singular at all $\kappa$ such that $8/\kappa\in\mathbb{Z}^++1$, $12/\kappa\in\mathbb{Z}^++1$, or $n(\varkappa)^4-3n(\varkappa)^2+1=0$.  We divide these singularities into four cases.
\begin{enumerate}
\item\label{} $8/\kappa\in\mathbb{Z}^++1$ and $4/\kappa\not\in\mathbb{Z}^+$: The treatment of this case is identical to that of item \ref{it1} in section \ref{singularrect}.
\item\label{} $12/\kappa\in\mathbb{Z}^++1$ and $4/\kappa\not\in\mathbb{Z}^+$: The treatment of this case is identical to that of item \ref{it2} in section \ref{singularrect}.
\item\label{} $4/\kappa\in\mathbb{Z}^+$: In this case, almost every factor in (\ref{2Pi2N4}) vanishes or has a pole at $\varkappa=\kappa$.  Upon writing (\ref{2Pi2N4}) as
\begin{multline}\label{splitPi2N4}\frac{n(\varkappa)^3}{4^3[n(\varkappa)^4-3n(\varkappa)^2+1]}\overbrace{\frac{\Gamma(2-8/\varkappa)^3}{\sin^3(-4\pi/\varkappa)\Gamma(1-4/\varkappa)^6}}^{1.}\,\,\dotsm\\
\dotsm\,\,\underbrace{\frac{1}{\sin(-4\pi/\varkappa)}\oint_{\mathscr{P}(x_i,x_{i+1})}\dotsm}_{2.}\,\,\underbrace{\frac{1}{\sin(12\pi/\varkappa)}\oint_{\mathscr{P}(x_6,x_7)}\dotsm}_{3.}\,\,\underbrace{\frac{1}{\sin(12\pi/\varkappa)}\oint_{\mathscr{P}(x_7,x_8)}\dotsm}_{4.}\end{multline}
and using (\ref{PochDecomp}), we see that $\kappa$ is a removable singularity of the first, second, third, and fourth factors in (\ref{splitPi2N4}) and hence of (\ref{2Pi2N4}).
\item $n(\kappa)^4-3n(\kappa)^2+1=0$: In this case, we have $n(\kappa)\in\{(1\pm\sqrt{5})/2,-(1\pm\sqrt{5})/2\}$.  That is, $\kappa\in(0,8)\times i\mathbb{R}$ is any of the exceptional speeds $\kappa_{5,q'}=20/q'$ \cite{florkleb3}, where $q'>1$ is an integer coprime with five.  Here, $n(\varkappa)^4-3n(\varkappa)+1=O(\varkappa-\kappa)$ as $\varkappa\rightarrow\kappa$ (\ref{fugacity}).  Also, (\ref{Pi2N4}, \ref{2Pi2N4}) with the factor $[n(\varkappa)^4-3n(\varkappa)+1]^{-1}$ dropped is an analytic function of $\varkappa\in(0,8)\times i\mathbb{R}$.  Thus, if it vanishes at $\varkappa=\kappa$, then it is $O((\varkappa-\kappa)^m)$ as $\varkappa\rightarrow\kappa$ for some $m\in\mathbb{Z}^+$, and $\kappa$ is a removable singularity of (\ref{Pi2N4}, \ref{2Pi2N4}).  Indeed, this function does vanish at $\varkappa=\kappa$, as
\be\label{Isvanish2} [n(I_{167}-nI_{267})+(n^2-1)(nI_{367}-I_{467})](\kappa)=0,\quad n(\kappa)^4-3n(\kappa)^2+1=0.\ee

To derive (\ref{Isvanish2}), we repeat the steps of item \ref{it4} in section \ref{singularhex}, with appropriate adjustments.  We integrate $u_3$ of (\ref{2Pi2N4}) around a simple loop that surrounds $[x_1,x_8]$, $\Gamma_1=[x_7,x_8]$, and $\Gamma_2=[x_6,x_7]$, pinch the upper and lower halves of the loop together between $x_5$ and $x_6$, and divide the loop into a left half that tightly surrounds $[x_1,x_5]$ and a right half that surrounds $\Gamma_1$ and $\Gamma_2$.  Now, as we cross the interval $(x_5,x_6)$, the integrand of (\ref{eulerintegral}) with $N=4$ and $c=7$ does not acquire a phase factor.  Indeed, the integrand accumulates one phase factor of
\be e^{2\pi i(-4/\kappa)}=e^{2\pi i(-4q'/20)}=e^{2\pi i(-q'/5)}\ee
per each of the five rightward-pointing branch cuts respectively anchored to $x_1$, $x_2,\ldots,x_5$, and the product of all five factors is one.  Thus, the left and right halves separate into a left loop $\Gamma_l$ tightly wrapped around $[x_1,x_5]$ and a right loop $\Gamma_r$ surrounding $\Gamma_1$ and $\Gamma_2$ (figure \ref{retract} with $N=4$ as shown).  Because integration around the original loop gives zero by the Cauchy integral theorem, the integration around $\Gamma_l$ gives the opposite of the integration around $\Gamma_r$.

Interestingly, the integration along $\Gamma_r\times\Gamma_2\times\Gamma_1$ gives zero as well, and the argument for why this is true is identical to the argument  made for the vanishing integration along $\Gamma_r\times\Gamma_1$ in item \ref{it4} of section \ref{singularhex}, with obvious adjustments (figure \ref{retract} with $N=4$ as shown).  Thus, we conclude that the integration along $\Gamma_l\times\Gamma_2\times\Gamma_1$ gives zero too.  After decomposing the integration around $\Gamma_l$ into integrations along the four intervals $[x_1,x_2]$, $[x_2,x_3]$, $[x_3,x_4]$, and $[x_4,x_5]$ that $\Gamma_r$ surrounds, we find that
\begin{multline}\label{match2}0=\mathcal{J}(\kappa\,|\,\Gamma_1,\Gamma_2,\Gamma_l)\\
\propto[I_{167}-nI_{267}+(n^2-1)I_{367}-n(n^2-2)I_{467})](\kappa)=0,\quad n(\kappa)^4-3n(\kappa)^2+1=0,\end{multline}
with $I_{ijk}$ and $n(\kappa)$ defined in (\ref{Iijk}) and (\ref{fugacity}) respectively.  (If $\kappa\leq4$, then these improper integrals diverge.  We regularize them via the replacement (\ref{replacePoch}).)  Because
 (\ref{match2}) is identical to (\ref{Isvanish2}), we conclude that $\kappa$ is a removable singularity of (\ref{2Pi2N4}).

Interestingly, (\ref{match2}) implies the linear dependence of the Temperley-Lieb set $\mathcal{B}_4$ for $\varkappa=\kappa=\kappa_{5,q'}$, previously determined by other means in the proof of lemma \red{6} in \cite{florkleb3}.  The argument for this claim is identical to the argument in the paragraph beneath (\ref{match}), up to obvious adjustments.

\end{enumerate}
From these facts, we conclude that the second octagon connectivity weight $\Pi_2$ (\ref{Pi2N4}) is an analytic function of $\varkappa\in(0,8)\times i\mathbb{R}$.

Equation (\ref{Pi3N4}) gives a formula for the third octagon connectivity weight $\Pi_3$.  Here, there are two terms for us to consider.  After using (\ref{replacePoch}) to extend the formula to all speeds $\varkappa\in(0,8)\times i\mathbb{R}$, we write the term of (\ref{Pi3N4}) including $I_{567}$ as (figure \ref{ContourId3})
\begin{multline}\label{21Pi3N4}\frac{n(\varkappa)^3\Gamma(2-8/\varkappa)^3}{4^3\sin^4(-4\pi/\varkappa)\sin^2(12\pi/\varkappa)\Gamma(1-4/\varkappa)^6}\,\,\dotsm\,\,\oint_{\mathscr{P}(x_5,x_6)}(x_6-u_3)^{-4/\varkappa}(u_3-x_5)^{-4/\varkappa}\dotsm\\
\dotsm\,\,\oint_{\mathscr{P}(x_6,x_7)}(x_7-u_2)^{12/\varkappa-2}\,\,(u_2-x_6)^{-4/\varkappa}\dotsm\,\,\oint_{\mathscr{P}(x_7,x_8)}(x_8-u_1)^{-4/\varkappa}(u_1-x_7)^{12/\varkappa-2}\dotsm\,{\rm d}u_1\,{\rm d}u_2\,{\rm d}u_3,
\end{multline}
and we write the term of (\ref{Pi3N4}) including $I_{667}$ as (figure \ref{ContourId3})
\begin{multline}\label{22Pi3N4}\frac{n(\varkappa)^4\Gamma(2-8/\varkappa)^3}{4^3e^{-12\pi i/\varkappa}\sin^3(-4\pi/\varkappa)\sin(8\pi/\varkappa)\sin^2(12\pi/\varkappa)\Gamma(1-4/\varkappa)^6}\,\,\dotsm\,\,\oint_{\mathscr{P}(x_6,x_7)}(x_7-u_2)^{12/\varkappa-2}(u_2-x_6)^{-4/\varkappa}\dotsm\\
\dotsm\,\,\oint_{\mathscr{P}(x_6,u_2)}(u_2-u_3)^{8/\varkappa}(u_3-x_6)^{-4/\varkappa}\dotsm\,\,\oint_{\mathscr{P}(x_7,x_8)}(x_8-u_1)^{-4/\varkappa}(u_1-x_7)^{12/\varkappa-2}\dotsm\,{\rm d}u_1\,{\rm d}u_3\,{\rm d}u_2,
\end{multline}
where the ellipses in (\ref{21Pi3N4}, \ref{22Pi3N4}) stand for omitted factors analytic over $\varkappa\in(0,8)\times i\mathbb{R}$, over $u_1\in A$, over $u_2\in B$, and over $u_3\in C$, where $A$, $B$, and $C$ are open regions containing $\mathscr{P}(x_7,x_8)$, $\mathscr{P}(x_6,x_7)$, and $\mathscr{P}(x_5,x_6)\cup\mathscr{P}(x_6,x_7)$ respectively.  Using the double-angle identity and (\ref{fugacity}) to write $\sin(8\pi/\varkappa)=-n(\varkappa)\sin(4\pi/\varkappa)$, we see that the prefactors of (\ref{21Pi3N4}) and (\ref{22Pi3N4}) are equal to within a phase factor.  Among speeds $\varkappa\in(0,8)\times i\mathbb{R}$, this common prefactor is singular at all $\kappa$ such that $8/\kappa\in\mathbb{Z}^++1$ or $12/\kappa\in\mathbb{Z}^++1$.  We divide these singularities into three cases.
\begin{enumerate}
\setcounter{enumi}{4}
\item\label{itt1} $8/\kappa\in\mathbb{Z}^++1$ and $4/\kappa\not\in\mathbb{Z}^+$: The treatment of this case is identical to that of item \ref{it1} in section \ref{singularrect}.
\item\label{itt2} $12/\kappa\in\mathbb{Z}^++1$ and $4/\kappa\not\in\mathbb{Z}^+$: The treatment of this case is identical to that of item \ref{it2} in section \ref{singularrect}.
\item\label{itt3} $4/\kappa\in\mathbb{Z}^+$: In this case, almost every factor in (\ref{21Pi3N4}) vanishes or has a pole at $\varkappa=\kappa$.  Upon writing (\ref{21Pi3N4}) as
\begin{multline}\label{1splitPi3N4}\frac{n(\varkappa)^3}{4^3}\overbrace{\frac{\Gamma(2-8/\varkappa)^3}{\sin^3(-4\pi/\varkappa)\Gamma(1-4/\varkappa)^6}}^{1.}\,\,\dotsm\\
\dotsm\,\,\underbrace{\frac{1}{\sin(-4\pi/\varkappa)}\oint_{\mathscr{P}(x_5,x_6)}\dotsm}_{2.}\,\,\underbrace{\frac{1}{\sin^2(12\pi/\varkappa)}\oint_{\mathscr{P}(x_6,x_7)}\dotsm\,\,\oint_{\mathscr{P}(x_7,x_8)}\dotsm}_{3.}\end{multline}
and using (\ref{PochDecomp}), we see that $\kappa$ is a removable singularity of the first, second, and third factors in (\ref{1splitPi3N4}) and hence of (\ref{21Pi3N4}).  Next, almost every factor in (\ref{22Pi3N4}) vanishes or has a pole at $\varkappa=\kappa$.  Upon writing (\ref{22Pi3N4}) as
\begin{multline}\label{2splitPi3N4}\frac{n(\varkappa)^4}{4^3e^{-12\pi i/\varkappa}}\overbrace{\frac{\Gamma(2-8/\varkappa)^3}{\sin^3(-4\pi/\varkappa)\Gamma(1-4/\varkappa)^6}}^{1.}\,\,\dotsm\\
\dotsm\,\,\underbrace{\frac{1}{\sin^2(12\pi/\varkappa)}\oint_{\mathscr{P}(x_6,x_7)}\dotsm\,\,\oint_{\mathscr{P}(x_7,x_8)}\dotsm}_{2.}\,\,\underbrace{\frac{1}{\sin(8\pi/\varkappa)}\oint_{\mathscr{P}(x_6,u_2)}\,\,\dotsm}_{3.}\end{multline}
and using (\ref{PochDecomp}), we see that $\kappa$ is a removable singularity of the first, second, and third factors in (\ref{2splitPi3N4}) and hence of (\ref{22Pi3N4}).  
\end{enumerate}
From these facts, we conclude that the third octagon connectivity weight $\Pi_3$ (\ref{Pi3N4}) is an analytic function of $\varkappa\in(0,8)\times i\mathbb{R}$.

Finally, (\ref{Pi1N4}) gives a formula for the first octagon connectivity weight $\Pi_1$, and as (\ref{decPi0}) shows, this formula comprises two terms.  One term is $-n(\varkappa)[n(\varkappa)^2-2]^{-1}$ multiplied by the third octagon connectivity weight $\Pi_3$ (\ref{Pi3N4}), and having found that the third weight is analytic over $\varkappa\in(0,8)\times i\mathbb{R}$, we conclude that this entire term has a pole only at the $\kappa$ values in this region with $n(\kappa)^2=2$.  Equation (\ref{Pi0}) gives the other term, denoted as $\Pi_0$.  After replacing all simple integration contours in (\ref{Pi0}) with elementary Pochhammer contours via (\ref{replacePoch}), we see that among the speeds $\varkappa\in(0,8)\times i\mathbb{R}$, the prefactor of each term in (\ref{Pi0}) is singular at all $\kappa$ such that $8/\kappa\in\mathbb{Z}^++1,$ $12/\kappa\in\mathbb{Z}^++1$, or $n(\kappa)^2=2$.  By decomposing every term in (\ref{Pi0}) as we did in items \ref{itt1}--\ref{itt3} above, we show that (\ref{Pi0}), and therefore all of (\ref{Pi1N4}) is analytic at any $\kappa$ such that $8/\kappa\in\mathbb{Z}^++1$ or $12/\kappa\in\mathbb{Z}^++1$.  Thus, we are left with determining that (\ref{Pi1N4}) is analytic at all $\kappa\in(0,8)\times i\mathbb{R}$ such that 
\be\label{ncond} n(\kappa)^2=2 \quad\Longleftrightarrow\quad\text{$\kappa=16/q'$, where $q'\in\mathbb{Z}^++1$ is coprime with four}.\ee

To prove that the first octagon connectivity weight $\Pi_1$ (\ref{Pi1N4}) is analytic at $\varkappa=\kappa$, where $\kappa$ satisfies (\ref{ncond}), is somewhat involved because showing that the bracketed linear combination of definite integrals in (\ref{Pi1N4}) vanishes there is no longer sufficient.  Indeed, because any $\varkappa=\kappa$ satisfying (\ref{ncond}) is an order-two pole of the prefactor in (\ref{Pi1N4}), we must show
\be\label{show}\overbrace{\left[\begin{array}{l}\hphantom{-}I_{357}-nI_{347}+n^2I_{337}-2nI_{237}+2I_{137}\\
-nI_{257}+n^2I_{247}+n^2I_{227}-nI_{127}+I_{157}\\
-nI_{147}-(n^2-2)(nI_{567}-n^2I_{667})\end{array}\right]}^{\text{bracketed linear combination in (\ref{Pi1N4})}}(\varkappa\,|\,x_1,x_2,\ldots,x_8)=O((\varkappa-\kappa)^2)\quad\text{as $\varkappa\rightarrow\kappa$ (\ref{ncond}),}\ee
in order to conclude that $\varkappa=\kappa$ is a removable singularity of the first octagon connectivity weight $\Pi_1$ (\ref{Pi1N4}).

\begin{figure}[b]
\centering
\includegraphics[scale=0.27]{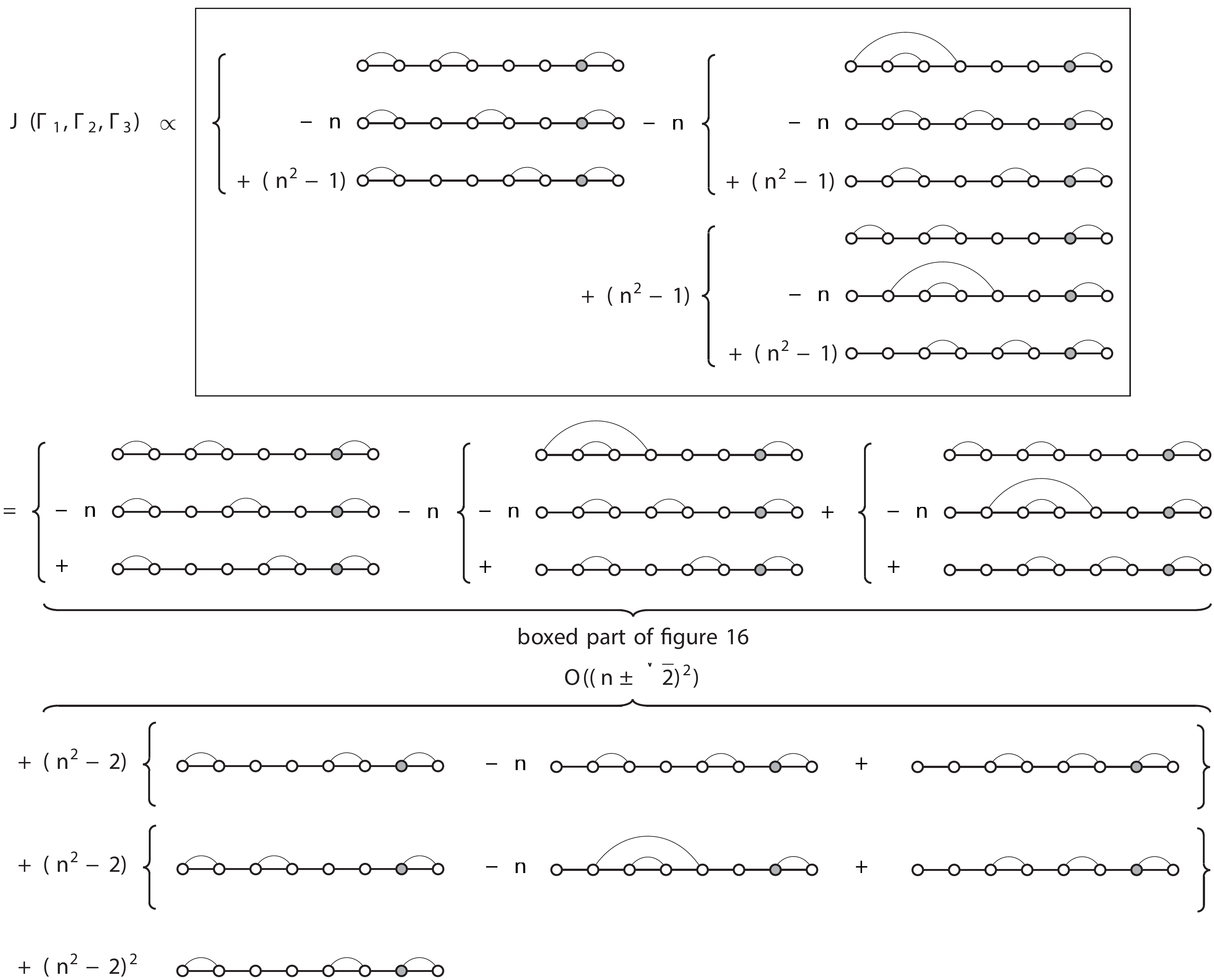}
\caption{With $\Gamma_1=\mathscr{P}(x_7,x_8)$, $\Gamma_2=\mathscr{P}(x_6,\Gamma_1)$, and $\Gamma_3=\mathscr{P}(x_5,\Gamma_2)$ (figure \ref{ContourId3}) (we explain the notation $\mathscr{P}(x,\Gamma)$ above (\ref{PiN})), $\mathcal{J}(\varkappa\,|\,\Gamma_1,\Gamma_2,\Gamma_3)$ (\ref{eulerintegral}) is proportional to the boxed part of figure \ref{BigDecomp} plus terms that are $O((\varkappa-\kappa)^2)$ as $\varkappa\rightarrow\kappa$ (\ref{ncond}).}
\label{BigDecomp2}
\end{figure}

To begin our proof of (\ref{show}), we first show that every $\kappa$ satisfying (\ref{ncond}) is a zero of the left side of (\ref{show}).  Obviously, the last two terms on the left side of (\ref{show}) vanish at $\varkappa=\kappa$ thanks to their vanishing common coefficient of $n(\kappa)^2-2$.  Moreover, the sum of the remaining terms on the left side of (\ref{show}), which is the same as the sum of bracketed terms in (\ref{Pi0}) or boxed terms in figure \ref{BigDecomp}, vanish at $\varkappa=\kappa$ too.  To see why, we produce all of these terms by inserting an elementary Pochhammer contour $\mathscr{P}(y,z)$ into the following quantity (which vanishes thanks to (\ref{Isvanish})),
\be\label{insert}\Bigg(\prod_{\substack{i<j \\ i,j\neq 5}}^6(x_j-x_i)^{2/\kappa}\Bigg)\Bigg(\prod_{i\neq5}^6|x_5-x_i|^{1-6/\kappa}\Bigg)[n I_{25}-I_{15}-I_{35}](\kappa\,|\,x_1,x_2,\ldots,x_6)=0,\quad n(\kappa)^2=2.\ee
This new contour $\mathscr{P}(y,z)$ has two new endpoints $y<z$ and is in one of three locations.  (If $\kappa>4$, then this contour is simple thanks to (\ref{Pochtostraight}).)  This insertion promotes the left side of (\ref{insert}) from an element of $\mathcal{S}_3$ to an element of $\mathcal{S}_4$.

The first location has $y<z<x_1$, and this insertion sends the left side of (\ref{insert}) to
\begin{multline}\label{inserted}\Bigg(\prod_{\substack{i<j \\ i,j\neq 5}}^6(x_j-x_i)^{2/\kappa}\Bigg)\Bigg((z-y)^{2/\kappa}\prod_{k\neq 5}^6(x_k-y)^{2/\kappa}(x_k-z)^{2/\kappa}\Bigg)\\
\Bigg((x_5-y)^{1-6/\kappa}(x_5-z)^{1-6/\kappa}\prod_{k\neq5}^6|x_5-x_k|^{1-6/\kappa}\Bigg)[n I_{147}-I_{137}-I_{157}](\kappa\,|\,y,z,x_1,x_2,\ldots,x_6),\end{multline}
where (\ref{Iij}, \ref{Iijk}) respectively define the contour integrals $I_{ij}$ and $I_{ijk}$.  The three terms in (\ref{inserted}) correspond one-to-one with the three terms to the right of the first brace, counting from the left, in the boxed part of figure \ref{BigDecomp}.  According to the discussion beneath (\red{44}) in \cite{florkleb3}, the inserted interval $(y,z)$ is an identity interval of each of these terms.

The second (resp.\ third) location of the inserted contour has $x_1<y<z<x_2$ (resp.\ $x_2<y<z<x_3$), and similar to the first, this insertion produces the three terms to the right of the second (resp.\ third) brace, counting from the left, in the boxed part of figure \ref{BigDecomp}.  In each of these three locations, $(y,z)$ is an identity interval of all generated terms.  In total, these three insertions generate nine terms (the entire boxed collection in figure \ref{BigDecomp}), and they combine to give the left side of (\ref{show}), excluding the last two terms multiplied by $n(\varkappa)^2-2$.

To finish our proof that the left side of (\ref{show}) vanishes at $\varkappa=\kappa$ satisfying (\ref{ncond}), we must show that the combination of nine terms described at the end of the previous paragraph vanishes at $\varkappa=\kappa$.  To prove this claim for the sum of the three terms with $y<z<x_1$, we assume the contrary.  Recalling that $(y,z)$ is an identity interval of each of these terms, it follows from this assumption and theorem \red{2} and corollary \red{9} of \cite{florkleb4} that $(y,z)$ is an identity interval of their sum.  In particular, $(y,z)$ is not a two-leg interval of this sum, so its limit (\ref{lim}), with $x_{i+1}\mapsto z$ and $x_i\mapsto y$, may not be zero.  But according to item \red{2} in the proof of lemma \red{6} of \cite{florkleb3}, this limit sends the sum to a multiple of (\ref{insert}), which vanishes.  From this contradiction, we conclude that the sum of three terms with $y<z<x_1$ vanishes at $\varkappa=\kappa$.  The same argument shows that the sum of three terms with either $x_1<y<z<x_2$ or $x_2<y<z<x_3$ vanishes at $\varkappa=\kappa$ too.  Because these nine terms combine to give the left side of (\ref{show}), excluding the last two terms multiplied by $n(\varkappa)^2-2$, we conclude that the entire left side of (\ref{show}) vanishes at $\varkappa=\kappa$, as desired.

Having proven that $\varkappa=\kappa$ satisfying (\ref{ncond}) is a zero of the left side of (\ref{show}), we now show that its order is no less than two.  For this purpose (and,  for the sake of brevity, omitting any motivation), we reconsider the collection of Pochhammer contours $\Gamma_1=\mathscr{P}(x_7,x_8)$, $\Gamma_2=\mathscr{P}(x_6,\Gamma_1)$, and $\Gamma_3=\mathscr{P}(x_5,\Gamma_2)$, illustrated in the left part of figure \ref{ContourId3}. (We explain the notation $\mathscr{P}(x,\Gamma)$ above (\ref{PiN}).)  Using the analysis surrounding items \ref{loop1} and \ref{loop2} of section \ref{rainbow} (figures \ref{lrloop}, \ref{rloop}), we deform $\Gamma_2$ and $\Gamma_3$ into a collection of simple contours (replaced by elementary Pochhammer contours via (\ref{replacePoch}) if $\kappa\leq4$) with endpoints among $x_1,$ $x_2,\ldots,x_6$.  We find  
\be\label{JJ}\mathcal{J}(\varkappa\,|\,\Gamma_1,\Gamma_2,\Gamma_3)=-64e^{28\pi i/\varkappa}\sin^5\left(\frac{4\pi}{\varkappa}\right)\sin\left(\frac{12\pi}{\varkappa}\right)\Big\{\text{boxed terms in figure \ref{BigDecomp2}}\Big\}(\varkappa).\ee
Next, as indicated in figure \ref{BigDecomp2}, we rewrite the boxed terms of that figure as a sum of the boxed terms in figure \ref{BigDecomp} with seven other terms.  These latter terms, appearing beneath ``$O((n\pm\sqrt{2})^2)$" in figure \ref{BigDecomp2}, sort into three groups:
\begin{enumerate} 
\item\label{row1} \emph{sum of first three terms (top row):} We may generate this sum by inserting into every term of (\ref{insert}) an elementary Pochhammer contour with two new endpoints between $x_4$ and $x_5$, as in the discussion surrounding (\ref{insert}, \ref{inserted}).  Because (\ref{insert}) vanishes at $\varkappa=\kappa$ satisfying (\ref{ncond}), it follows from the second paragraph beneath (\ref{inserted}) that this sum vanishes too.  Multiplied by $n(\varkappa)^2-2$, this sum thus has a zero at $\varkappa=\kappa$ with order at least two.
\item \emph{sum of second three terms (middle row):}   By the argument just given in the previous item \ref{row1}, but with $(x_4,x_5)\mapsto(x_2,x_3)$, this sum also has a zero at $\varkappa=\kappa$ satisfying (\ref{ncond}) with order at least two.
\item\label{row3} \emph{last term (bottom row):}  With its factor of $[n(\varkappa)^2-2]^2$, this term manifestly has a zero at $\varkappa=\kappa$ with order at least two.
\end{enumerate}
In light of items \ref{row1}--\ref{row3} immediately above, we may write (\ref{JJ}) as (figure \ref{BigDecomp2})
\be\label{JJJ}\bigg[-64e^{28\pi i/\varkappa}\sin^5\left(\frac{4\pi}{\varkappa}\right)\sin\left(\frac{12\pi}{\varkappa}\right)\bigg]^{-1}\mathcal{J}(\varkappa\,|\,\Gamma_1,\Gamma_2,\Gamma_3)=\Big\{\text{boxed terms in figure \ref{BigDecomp}}\Big\}(\varkappa)+O((\varkappa-\kappa)^2).\ee
Moreover, we may use (\ref{PiN}, \ref{A}) (with the $m=1$ phase factor of (\ref{A}) adjusted to account for the fact that $c=2N-1=7\neq 2N$ here, see the discussion beneath (\ref{A})) followed by (\ref{Pi3N4}) to write the left side of (\ref{JJJ}) as
\begin{multline}\label{bigJ}\bigg[-64e^{28\pi i/\varkappa}\sin^5\left(\frac{4\pi}{\varkappa}\right)\sin\left(\frac{12\pi}{\varkappa}\right)\bigg]^{-1}\mathcal{J}\Big(\varkappa\,\Big|\,\Gamma_1,\Gamma_2,\Gamma_3\Big|\,x_1,x_2,\ldots,x_8\Big)\\
\begin{aligned}&=\Bigg[\frac{n(\varkappa)^3e^{-12\pi i/\varkappa}e^{-4\pi i(4-3)/\varkappa}e^{-4\pi i(6-3)/\varkappa}\Gamma(2-8/\varkappa)^3}{4^3\sin^3(-4\pi/\varkappa)\sin(12\pi/\varkappa)\sin(16\pi/\varkappa)\sin(20\pi/\varkappa)\Gamma(1-4/\varkappa)^6}\Bigg]^{-1}\Pi_3(\varkappa\,|\,x_1,x_2,\ldots,x_8)\\
&\hspace{.8cm}\times\bigg[-64e^{28\pi i/\varkappa}\sin^5\left(\frac{4\pi}{\varkappa}\right)\sin\left(\frac{12\pi}{\varkappa}\right)\bigg]^{-1}\Bigg(\prod_{\substack{i<j \\ i,j\neq 7}}^8(x_j-x_i)^{-2/\varkappa}\Bigg)\Bigg(\prod_{k\neq7}^8|x_7-x_k|^{6/\varkappa-1}\Bigg)\\
&=-n(\varkappa)[n(\varkappa)^4-3n(\varkappa)^2+1][n(\varkappa)^2-2][I_{567}-nI_{667}](\varkappa\,|\,x_1,x_2,\ldots,x_8)\\
&=-n(\varkappa)[-1+O(\varkappa-\kappa)][n(\varkappa)^2-2][I_{567}-nI_{667}](\varkappa\,|\,x_1,x_2,\ldots,x_8)\\
&=\hphantom{-}n(\varkappa)[n(\varkappa)^2-2][I_{567}-nI_{667}](\varkappa\,|\,x_1,x_2,\ldots,x_8)+O((\varkappa-\kappa)^2).\end{aligned}\end{multline}
After replacing the left side of (\ref{JJJ}) with the bottom-right side of (\ref{bigJ}) and rearranging, we ultimately find
\be\label{final}\overbrace{\Big\{\text{boxed terms in figure \ref{BigDecomp}}\Big\}(\varkappa)-n(\varkappa)[n(\varkappa)^2-2][I_{567}-nI_{667}](\varkappa)}^{\text{left side of (\ref{show})}}=O((\varkappa-\kappa)^2)\quad\text{as $\varkappa\rightarrow\kappa$.}\ee
As indicated, the left side of (\ref{final}) matches the left side of (\ref{show}).  Indeed, the boxed terms in figure \ref{BigDecomp} correspond one-to-one with all of the terms on the left side of (\ref{show}), except the last two terms, which are multiplied by $n(\varkappa)^2-2$.  Also, these last two terms match the last two terms on the left side of (\ref{final}).  Thus, the left sides of (\ref{show}) and (\ref{final}) match, so this proves the claim (\ref{show}).  We conclude that all singularities  $\varkappa=\kappa$ such that $n(\kappa)^2=2$ (\ref{ncond}) of the first octagon connectivity weight formula $\Pi_1$ (\ref{Pi1N4}) are removable.  As such, $\Pi_1$ is an analytic function of $\varkappa\in(0,8)\times i\mathbb{R}$.

\section{Connectivity weights and logarithmic CFT}\label{appendixB}

In this appendix, we investigate the appearance of logarithmic terms in Frobenius series expansions of the connectivity weights in powers of the differences of neighboring coordinates from the point $\boldsymbol{x}=(x_1,x_2,\ldots,x_{2N})\in\Omega_0.$  We anticipate that for sufficiently large $N\in\mathbb{Z}^+$, such logarithmic terms may appear if and only if $\kappa$ equals an exceptional speed $\kappa_{q,q'}:=4q/q'$, where by definition, $q$ and $q'$ are coprime positive integers with $q>1$ \cite{florkleb3,florkleb4}.  (In the following discussion, we justify this expectation with CFT considerations.  If the reader is not familiar with CFT, then he/she may skip our references to it, since they are only used to motivate the results.)

The prediction that logarithmic terms appear in Frobenius series expansions of some connectivity weights if $\kappa\in(0,8)$ is an exceptional speed follows from CFT.  Indeed, such speeds correspond with CFT minimal models \cite{florkleb4}, whose operator content is restricted to the Kac table \cite{bpz,fms,henkel}.  Now, some connectivity weights do not respect this truncation in the sense that they (by hypothesis) satisfy the usual system of PDEs (\ref{nullstate}, \ref{wardid}) but not the larger ``extended system" of CFT null-state PDEs that includes the former system.   As discussed in section \red{V} of \cite{florkleb4}, the extra PDEs of the extended system arise from the presence of extra null-state conditions imposed on the affiliated CFT correlation functions, in addition to those associated with the CFT Kac operators $\psi_1=\phi_{1,2}$ or $\phi_{2,1}$ that appear in the correlation functions. By conjecture \red{17} of that section, if the extended PDEs are not satisfied, then the corresponding connectivity weights, used to calculate observables, such as crossing probabilities \cite{c3,dub,js}, in the corresponding statistical mechanics models, must arise from operator content beyond the Kac table, extending the minimal model.  Including these operators typically introduces logarithmic operators into the theory too \cite{gurarie2,gurarie,kytrid,crerid,js,rid}.  Thus, logarithmic terms should appear in the Frobenius series expansions of some connectivity weights.  

Many observables in statistical mechanics models with CFT descriptions, such as crossing probabilities,  are closely related to the connectivity weights determined in this article. From the CFT perspective, it is customary to (loosely) ``interpret" the latter functions, solving (\ref{nullstate}, \ref{wardid}) (and subject to the duality condition (\ref{duality})), as CFT correlation functions of $2N$ one-leg boundary operators $\psi_1$ (or more generally as linear combinations of such functions) \cite{florkleb}:
\be\label{2Nptfunc}F(x_1,x_2,\ldots,x_{2N})=\langle\psi_1(x_1)\psi_1(x_2)\dotsm\psi_1(x_{2N})\rangle.\ee
In section \red{IV} of \cite{florkleb4}, we observe that the exceptional speed $\kappa_{q,q'}$ with $q'>1$ corresponds to an $\mathcal{M}(p,p')$ CFT minimal model with $p=\text{max}\{q,q'\}>1$ and $p'=\text{min}\{q,q'\}>1$.  In the logarithmic extension of this model, we find a pre-logarithmic operator at the $(1,q)$ (resp.\ $(q,1)$) position of the Kac grid if $\kappa>4$ (resp.\ $\kappa\leq4$). This operator is beyond the edge of the Kac table, whose first column (resp.\ row)  terminates at $(1,q-1)$ (resp.\ $(q-1,1)$).  Loosely speaking, because we may advance by at most one position up (resp.\ left) along the leftmost column (resp.\ bottommost row) of the Kac grid through fusion with a one-leg boundary operator, we must fuse together at least $q-1$ adjacent one-leg boundary operators in order to observe the pre-logarithmic operator.  However, we may advance from the $(1,2)$ (resp.\ $(2,1)$) position in the Kac grid to at most the $(1,N+1)$ (resp.\ $(N+1,1)$) position (i.e., we may generate at most an $N$-leg boundary operator) by consecutive fusions of adjacent one-leg boundary operators within the $2N$-point function (\ref{2Nptfunc}). Hence, pre-logarithmic operators play significant roles in connectivity weights only if $q-1\leq N$. These cases correspond exactly with all of those for which the Temperley-Lieb set $\mathcal{B}_N\subset\mathcal{S}_N$ (defined in \cite{florkleb3}) of solutions for the system (\ref{nullstate}, \ref{wardid}) is not a basis of $\mathcal{S}_N$.  (See Lemma \red{6} of \cite{florkleb3}.)

Next, we find a logarithmic operator at the $(1,q+1)$ (resp.\ $(q+1,1)$) position of the Kac grid.  To arrive at this position, we must fuse together at least $q$ adjacent one-leg boundary operators, or equivalently, we must fuse a pre-logarithmic operator with an adjacent one-leg boundary operator.  By the arguments of the previous paragraph, it follows that logarithmic operators play significant roles in connectivity weights only if $q\leq N$.  In particular, logarithmic terms may appear in Frobenius series expansions of $2N$-point connectivity weights only if $q\leq N$.
 
In summary, pre-logarithmic fusions occur in all cases with $q\leq N+1$, and both pre-logarithmic and logarithmic fusions occur in all cases with $q\leq N$.  However, we should note that this last point does not imply that Frobenius series expansions of any $2N$-point connectivity weight must have logarithmic terms if $q\leq N$.  Indeed, in some cases, correlation functions (\ref{2Nptfunc}) with such pre-logarithmic and logarithmic operators may lack logarithmic terms.

Rather than considering all cases, for brevity we give examples of elements of $\mathcal{S}_N$ with Frobenius series expansions as described above for $\kappa=\kappa_{q,q'}$ with $q=2$ and for $q=3=N$ only.  (The corresponding results for Frobenius series expansions for  the case $\kappa=\kappa_{2,q'}=8/r$ with $r>1$ odd are already given in theorem \red{2} of \cite{florkleb4}.  See also appendix \red{A} of \cite{florkleb4}.) The work that we present below generalizes to the remaining cases of $q=3<N$ and $3<q\leq N$ in an evident way.  For convenience, we adopt the notation of \cite{florkleb,florkleb2,florkleb3,florkleb4}
\be\label{bdysleg}\psi_s(x)=\begin{cases}\phi_{1,s+1}(x),&\kappa>4 \\ \phi_{s+1,1}(x),&\kappa\leq 4\end{cases},\quad \theta_s:=\left\{\begin{array}{ll}h_{1,s+1}, &\kappa>4 \\ h_{s+1,1}, & \kappa\leq4 \end{array}\right\}=\frac{s(2s+4-\kappa)}{2\kappa},\ee
where $\phi_{1,s}$ (resp.\ $\phi_{s,1}$) is a CFT $(1,s)$ (resp.\ $(s,1)$) Kac operator (and we call $\psi_s$ an \emph{$s$-leg boundary operator}), and we adopt the following notation of \cite{js}:
\be\label{notation}\mathbf{1}:=\psi_0\,\,\,\text{(the identity operator)},\quad\phi:=\psi_1,\quad \upsilon:=\psi_2,\quad \psi:=\psi_3.\ee 
We reference articles in logarithmic CFT below that use the CFT central charge instead of the SLE$_\kappa$ parameter $\kappa$.  For $\kappa\neq 4$ (resp.\ $\kappa=4$), these quantities are related two-to-one (resp.\ one-to-one) via
\be\label{central}c(\kappa)=\frac{(6-\kappa)(3\kappa-8)}{2\kappa},\quad \kappa>0.\ee
Furthermore, if $c(\kappa)=c(\smash{\hat{\kappa}})$, then $\smash{\hat{\kappa}}=16/\kappa$, with one speed in the dense phase ($\kappa>4$) of SLE$_\kappa$ and the other in the dilute phase ($0<\kappa\leq 4$).  The relation between $\kappa$ and $\smash{\hat{\kappa}}$ is known as \emph{duality} in SLE$_\kappa$ \cite{rohshr,knk}.

To begin, we consider the exceptional speeds $\kappa=\kappa_{2,q'}$ (i.e., $8/\kappa=r\in2\mathbb{Z}^++1$).  One speed belonging to this case is $\kappa=8/3$, and this example is unusual because the one-leg boundary operator $\phi=\phi_{2,1}$ is pre-logarithmic, sitting just outside the Kac table of the $\mathcal{M}(3,2)$ minimal model \cite{bpz,fms,henkel}.  Thus, the first logarithmic operators for this speed appear in the operator product expansion (OPE) of $\phi(x)$ with $\phi(x+\varepsilon)$.  Using CFT arguments, V.\ Gurarie determined this OPE, finding 
\be\label{83expand}\phi(x)\phi(x+\varepsilon)\underset{\varepsilon\rightarrow0}\sim a_0\varepsilon^{-2\theta_1}\mathbf{1}+b_0\varepsilon^{-2\theta_1+\theta_2}\hat{\upsilon}(x)+\dotsm\,\,+(b_0\log\varepsilon+c_0)\varepsilon^{-2\theta_1+\theta_2}L_{-2}\mathbf{1}(x)+\dotsm,\quad \kappa=8/3.\ee
(see (\red{45}) of \cite{gurarie2}, but note that our $L_{-2}\mathbf{1}(x)$ is the stress-energy tensor $T(x)$ in \cite{gurarie2})
Here, $\smash{\hat{\upsilon}}$ is the logarithmic partner that couples with $L_{-2}\mathbf{1}$ \cite{gurarie2}, and (\ref{bdysleg}, \ref{notation}) give the other powers and operators of (\ref{83expand}).  Moreover, the ellipses following the second and third term of (\ref{83expand}) stand for descendent terms of the corresponding conformal family \cite{fms}, with factors respectively of the form $b_m\varepsilon^m$ and $\left(b_m \log\varepsilon+c_m\right)\varepsilon^m$ for all $m\in\mathbb{Z}^+$.  The regular and logarithmic operators of (\ref{83expand}) belong to the same staggered logarithmic module \cite{kytrid,crerid}.  As such, if we fix the value of $a_0,$ then the $b_m$ are fixed for all $m\in\mathbb{Z}^+\cup\{0\}$.  In particular if $a_0=0$ then $b_m=0$.

The terms of (\ref{83expand}) with the coefficients $c_m$ are algebraically independent of those with coefficients $a_0$ and $b_m$. Indeed, the invariance of the logarithmic CFT operator algebra under the shift $\smash{\hat{\upsilon}}\mapsto\smash{\hat{\upsilon}}+k L_{-2}\mathbf{1}$ induces this feature. This symmetry of the operator algebra implies that the descendants of $L_{-2}\mathbf{1}$ effectively form a second, independent fusion channel, even though they are produced from within the logarithmic fusion channel.  Here, we have glossed over a logarithmic CFT subtlety, namely that if the staggered module appears in this fusion, then $a_0 \neq0$.  However, connectivity weights, which are the subject of this article, often correspond to a difference of two staggered module blocks with different values for $c_0$, which effectively allows $a_0=b_m=0$ while $c_0\neq0$.

Another speed belonging to the case $\kappa=\kappa_{2,q'}$ is $\kappa=8$. This example is also special because $\kappa$ sits on the border of the range $(0,8)$ of SLE$_\kappa$ speeds to which the results of our previous articles \cite{florkleb,florkleb2,florkleb3,florkleb4} apply.  Assuming that theorems \red{1} and \red{2} of \cite{florkleb4} are true if $\kappa=8$, we again consider the operator product expansion (OPE) of $\phi(x)$ with $\phi(x+\varepsilon)$.  V.\ Gurarie explicitly determines this OPE in \cite{gurarie}.  Summing (\red{11}) and (\red{12}) of \cite{gurarie}, his result amounts to
\be\label{8expand}\phi(x)\phi(x+\varepsilon)\underset{\varepsilon\rightarrow0}\sim\left(a_0+b_0\log \varepsilon\right)\varepsilon^{-2\theta_1+\theta_0}\mathbf{1}+\dotsm\,\,+b_0\varepsilon^{-2\theta_1+\theta_2}\hat{\upsilon}(x)+\dotsm
,\quad \kappa=8,\ee
where the powers, operators, ellipses, and coefficients are as described beneath (\ref{83expand}), except that $\smash{\hat{\upsilon}}$ is now a logarithmic partner to $\mathbf{1}$.  (From (\ref{bdysleg}), we observe that $\theta_0=0$, and $\theta_2=0$ if $\kappa=8$.  In spite of this, we exhibit these quantities in (\ref{8expand}) to capture the structure of a two-point OPE \cite{fms}.)  Now, if we insert either of (\ref{83expand}, \ref{8expand}) with $x=x_i$ and $x+\varepsilon=x_{i+1}$ for some $i\in\{1,2,\ldots,2N-1\}$ into the correlation function (\ref{2Nptfunc}) and use the formula (\ref{bdysleg}) for the conformal weights $\theta_s$, we find an expansion of the form ($\pi_i$ projects the $i$th coordinate from $\boldsymbol{x}\in\Omega_0$, and $8/\kappa=r\in2\mathbb{Z}^+-1$)
\be\label{log}\begin{aligned}F(\boldsymbol{x})=(x_{i+1}-x_i)^{1-6/\kappa}\sum_{m=0}^{r-2} [(A_m\circ\pi_i)(\boldsymbol{x})]\,(x_{i+1}-x_i)^m\,+\,&(x_{i+1}-x_i)^{2/\kappa}\sum_{m=0}^\infty [(B_m\circ\pi_i)(\boldsymbol{x})]\,(x_{i+1}-x_i)^m\\
+\,\log(x_{i+1}-x_i)&(x_{i+1}-x_i)^{2/\kappa}\sum_{m=0}^\infty [(C_m\circ\pi_i)(\boldsymbol{x})]\,(x_{i+1}-x_i)^m.\end{aligned}\ee
(We note that, with $8/\kappa=r\in\mathbb{Z}^+$, the powers of the second and third series begin where the powers of the first series end.)  As expected, the form (\ref{log}) exactly matches the known form of the Frobenius series expansion for a generic solution $F\in\mathcal{S}_N$ with $8/\kappa=r\in2\mathbb{Z}^++1$.  (See (\red{18}) and appendix \red{A} of \cite{florkleb4}.)  As stated in theorem \red{2} of \cite{florkleb4}, if $A_0=0$ or $C_0=0$ in (\ref{log}), then we have $A_m=0$ for all $m\in\{0,1,\ldots,r-2\}$ and $C_m=0$ for all $m\in\mathbb{Z}^+\cup\{0\}$.  (Furthermore, if $A_0=B_0=0$, then $F$ is zero.)

It is interesting to observe the appearance of logarithms in the Frobenius series expansions of some of the connectivity weights in the case $\kappa=\kappa_{2,q'}$.  As the sentence ending the paragraph above (\ref{2Nptfunc}) implies, not every connectivity weight contains a logarithmic term as $x_{i+1}\rightarrow x_i$ for some fixed $i\in\{1,2,\ldots,2N-1\}$.  As an example, we let $N=2$ and consider the rectangle connectivity weight $\Pi_1(\kappa)$ (\ref{Pi1}, \ref{Pi1hyper}).  As we send $x_2\rightarrow x_1$, we observe only the second series (i.e., $A_m=C_m=0$, $B_m\neq0$) in (\ref{log}), generated solely by the regular operator in the logarithmic pair, that is, $L_{-2}\mathbf{1}$ in (\ref{83expand}) and $\mathbf{1}$ in (\ref{8expand}).  But if we consider the rectangle connectivity weight $\Pi_2(\kappa)$, obtained by changing the integration contour in (\ref{Pi1}) to $\mathscr{P}(x_2,x_3)$ (or equivalently replacing $\lambda\mapsto1-\lambda$ in (\ref{Pi1hyper})), then we observe all of the three series (i.e., $A_m,B_m,C_m\neq 0$) in (\ref{log}), including the series multiplied by a logarithm.  In particular, if $\kappa=8$, then a logarithm appears in the leading term of (\ref{log}), and if $\kappa=8/3$, then a logarithm appears in the second-to-leading term of (\ref{log}).

In appendix \red{A} of \cite{florkleb4}, we find that whether $A_m=C_m=0$ or $A_m,C_m\neq0$ in (\ref{log}) depends on how an integration contour interacts with the contracting interval $(x_i,x_{i+1})$.  Assuming that neither $x_i$ nor $x_{i+1}$ bears the conjugate charge (see below (\ref{CGsolns})), we summarize those findings as follows: with $8/\kappa\in2\mathbb{Z}^++1$, if no integration contour of $F\in\mathcal{S}_N\setminus\{0\}$ touches or surrounds either of  $x_i$ and $x_{i+1}$, or if one contour of (\ref{CGsolns}) surrounds both of $x_i$ and $x_{i+1}$, then $A_m=C_m=0$ in (\ref{log}), so no logarithm appears in the expansion (\ref{log}).  Otherwise $A_m,C_m\neq0$, so a logarithm multiplying a Frobenius series does appear in the expansion (\ref{log}).  (We expect that this is true for $\kappa=8$ too.)

Next, we consider the exceptional speeds $\kappa=\kappa_{3,q'}$ (i.e., $12/\kappa=r\in\mathbb{Z}^++1$ is coprime with three).   Just as with the previous case $\kappa=\kappa_{2,q'}$, we find logarithmic terms in Frobenius series expansions of connectivity weights.  But now (as discussed beneath (\ref{2Nptfunc})), we must bring together not just two but at least three adjacent points, $x_i<x_{i+1}<x_{i+2}$ with $i\in\{1,2,\ldots,2N-2\}$, in order to observe such a term.  To anticipate the result, we recall the OPE of three adjacent one-leg boundary operators for $\kappa=\kappa_{3,2}=6$, from appendix \red{A} of \cite{js}:
\begin{multline}\label{Jake'sOPE}
\{\phi(x)\phi(x+\lambda\varepsilon)\}_\upsilon\phi(x+\varepsilon)\underset{\varepsilon\rightarrow0}{\sim}\Pi_1(\lambda)\varepsilon^{-3\theta_1+\theta_1}\phi(x)+\varepsilon^{-3\theta_1+\theta_3}\mathcal{F}_3(\lambda)\smash{\hat{\psi}}(x)
\\
+\varepsilon^{-3\theta_1+\theta_3}\Big[\mathcal{F}_2(\lambda)+\log(c\,\varepsilon)\mathcal{F}_3(\lambda)\Big]\partial\phi(x)+\dotsm,\quad\lambda\in(0,1),\quad\kappa=6.\end{multline}
Here, the powers and operators are given in (\ref{bdysleg}, \ref{notation}), $\smash{\hat{\psi}}$ is the logarithmic partner of the first descendant $L_{-1}\phi=\partial\phi$ of $\phi$, $c$ is an unspecified constant, the notation $\{\phi(x)\phi(y)\}_\upsilon$ indicates that we allow only the two-leg channel to propagate between the enclosed one-leg boundary operators $\phi(x)$ and $\phi(y)$ (i.e., we restrict the OPE of these two operators to contain the two-leg boundary operator only), $\Pi_1(\lambda)$ is given by (\ref{Pi1hyper}) with $\kappa=6$, now strictly a function of only $\lambda$ (indeed, it is Cardy's formula \cite{c3} for critical percolation), and finally
\be \mathcal{F}_2(\lambda):=\frac{3\Gamma(5/3)}{4\Gamma(1/3)^2}\lambda^{4/3}(1-\lambda)^{1/3}\,_3F_2(2/3,1,1;2,7/3\,|\,\lambda),\quad \mathcal{F}_3(\lambda):=\frac{\sqrt{3}\Gamma(5/3)^2}{\pi\Gamma(7/3)}\lambda^{1/3}(1-\lambda)^{1/3}.\ee
(Although the conformal weight $\theta_1$ (\ref{bdysleg}) vanishes for $\kappa=6$, we include it in (\ref{Jake'sOPE}) to exhibit the formal structure of an OPE \cite{fms}.  Moreover, as the powers of (\ref{Jake'sOPE}) suggest, $\partial\phi$ is a primary operator with conformal weight $\theta_3=1$.  Its consequent close relation with $\psi_3$ is exploited to compute new percolation crossing observables in \cite{jspk}.)  

Inserting (\ref{Jake'sOPE}) with $x=x_i$, $x+\lambda\varepsilon=x_{i+1}$, and $x_{i+2}=x+\varepsilon$ for some $i\in\{1,2,\ldots,2N-2\}$, $\varepsilon>0$, and $\lambda\in(0,1)$ into (\ref{2Nptfunc}) then gives a candidate form for the expansion of an element of $\mathcal{S}_N$ if $12/\kappa=r\in\mathbb{Z}^++1$ is coprime with three.  The form is (here, $\pi_{i,i+1}$ projects the $i$th and $(i+1)$th coordinate from $\boldsymbol{x}\in\Omega_0$)
\begin{multline}\label{log2}F\Big(x_1,x_2,\ldots,x_i,\,x_{i+1}=(1-\lambda)x_i+\lambda x_{i+2},\,x_{i+2},\ldots,x_{2N}\Big)=\\
\begin{aligned}(x_{i+2}-x_i)^{1-6/\kappa}\sum_{m=0}^{r-2} \big[A_m\big(\lambda,\pi_{i,i+1}(\boldsymbol{x})\big)\big]\,(x_{i+2}-x_i)^m\,+\,&(x_{i+2}-x_i)^{6/\kappa}\sum_{m=0}^\infty \big[B_m\big(\lambda,\pi_{i,i+1}(\boldsymbol{x})\big)\big]\,(x_{i+2}-x_i)^m\\
+\,\log(x_{i+2}-x_i)&(x_{i+2}-x_i)^{6/\kappa}\sum_{m=0}^\infty \big[C_m\big(\lambda,\pi_{i,i+1}(\boldsymbol{x})\big)\big]\,(x_{i+2}-x_i)^m.\end{aligned}\end{multline}
Rather than prove that all elements of $\mathcal{S}_N$ admit a Frobenius expansion of the form (\ref{log2}) if $12/\kappa=r\in\mathbb{Z}^++1$ is coprime with three (and $N>2$, see below), for brevity we limit ourselves to showing that the hexagon connectivity weight $\Pi_1(\kappa)$ (\ref{Pi1N3}) has this feature.  The analysis closely follows the arguments of appendix \red{A} in \cite{florkleb4}.

To find a Frobenius series expansion for the hexagon connectivity weight $\Pi_1(\kappa)$ (\ref{Pi1N3}) in the form (\ref{log2}), we consider the formula for $\Pi_1(\kappa)$ in two pieces.  With $\varkappa\in(0,8)$ denoting an arbitrary SLE$_\kappa$ speed and $\kappa\in(0,8)$ denoting an exceptional speed with $12/\kappa=r\in\mathbb{Z}^++1$ coprime with three (and recalling definitions (\ref{fugacity}, \ref{Iij})), we have
\be\label{twoparts}\Pi_1(\varkappa)\propto [nI_{25}-I_{15}-I_{35}](\varkappa)\propto \underbrace{\sin(4\pi/\varkappa)I_{15}(\varkappa)+\sin(8\pi/\varkappa)I_{25}(\varkappa)}_{1.}+\underbrace{\sin(4\pi/\varkappa)I_{35}(\varkappa)}_{2.}.\ee
Next, we show that the first brace and the second brace in (\ref{twoparts}) each equal a Frobenius series or a sum of Frobenius series of the form (\ref{log2}) with $i=1$.

To begin, we study the first brace in (\ref{twoparts}).  The analysis of this term is somewhat similar to what we find in case \red{2} of appendix \red{A} in \cite{florkleb4}.  Supposing that $\varkappa=\kappa>4$ so the integration contours of $I_{ij}$ are simple, the integrand of $I_{15}(\kappa)$ (resp.\ $I_{25}(\kappa)$) (\ref{Iij}), as a function of $u_2$ (integrated along $[x_1,x_2]$ (resp.\ $[x_2,x_3]$)), has no branch cut along the segment $[x_3,x_4]$.  (Indeed, this immediately follows from the property that $12/\kappa\in\mathbb{Z}^+$.)  In light of this fact, it is easy to see that the two terms of the first brace in (\ref{twoparts}) combine to give an integration around a simple loop surrounding $[x_1,x_3]$.  (If $\kappa\leq4$, then the integration contours of $I_{ij}(\kappa)$ are not simple but are replaced by Pochhammer contours, as in (\ref{replacePoch}).  By analytically continuing the first part of the right side of (\ref{twoparts}) to the region $\varkappa\in(0,8)\times i\mathbb{R}$, we find the same result.)  Because this loop is not tangled with the points $x_1$, $x_2$, and $x_3$ inside it, the loop is not constrained to contract to a point as we send $x_3,x_2=(1-\lambda)x_1+\lambda x_3\rightarrow x_1$.  Hence, the integration around this loop, contained in the first brace of (\ref{twoparts}), is $O(1)$ in this limit.  After multiplying this integration by the power-law factors in (\ref{Pi1N3}) (which are omitted from (\ref{twoparts}), and now with $x_2=(1-\lambda)x_1+\lambda x_3$), we find that these terms contribute exclusively to the second series on the right side of (\ref{log2}) with $i=1$ if $\varkappa=\kappa$.  This outcome is similar to what we find in case \red{2} of appendix \red{A} in \cite{florkleb4}.

Next, we study the second part of the right side of (\ref{twoparts}).  The analysis of this term is somewhat similar to what we find in case \red{3} of appendix \red{A} in \cite{florkleb4}.  Again supposing that $\varkappa>4$ so the integration contours of $I_{ij}$ are simple, we use (\ref{Iij}) to write
\begin{align}\label{I35} I_{35}(\varkappa\,|\,x_1,x_2,\ldots,x_6)&=\int_{x_5}^{x_6}
\dotsm\,I_3(\varkappa\,|\,x_1,x_2,\ldots,x_5,u_1,x_6)\,{\rm d}u_1, \\
\label{Ik}I_k(\varkappa\,|\,x_1,x_2,\ldots,x_7)&:=\int_{x_k}^{x_{k+1}} \mathcal{N}\Bigg[\prod_{j=1}^7(u_2-x_j)^{\beta_j}\Bigg]\,{\rm d}u_2,\quad x_8:=x_1,\quad \int_{x_7}^{x_8}:=\int_{x_7}^\infty+\int_{-\infty}^{x_8},\end{align}
where the ellipses in (\ref{I35}) stand for omitted factors from the integrand of (\ref{Iij}),  we have relabeled $u_1\mapsto x_6$ and $x_6\mapsto x_7$ in (\ref{Ik}) (but not in (\ref{I35})),  $\mathcal{N}[\,\,\ldots\,\,]$ orders the differences of its enclosed factors so $I_k(\varkappa)$ is real-valued, and  
\be\label{powers}\beta_j=-4/\varkappa,\quad j\in\{1,2,3,4,7\},\quad \beta_5=12/\varkappa-2,\quad \beta_6=8/\varkappa.\ee
The integrand of each $I_k(\varkappa)$, as a function of $u_2$, is manifestly analytic in the upper and lower half-planes.  Furthermore, because $\beta_1+\beta_2+\dotsm+\beta_7=-2$, this integrand has no singularity at infinity either.  As such, we find that
\be\label{upperlower}\sum_{k=1}^7 e^{\pm\pi i\sum_{l=1}^k\beta_l}I_k=\sum_{k=1}^6 e^{\pm\pi i\sum_{l=1}^k\beta_l}I_k+I_7=0\ee
after integrating the integrand of $I_k$ around a large semicircle in the upper half-plane ($-$ sign in the exponent) or lower half-plane (+ sign in the exponent) with its base flush against the real axis and sending the radius of that semicircle to infinity.  Solving (\ref{upperlower}) for $I_3$ in terms of $I_1$, $I_2$, $I_4$, $I_5$, and $I_6$ gives (compare with (\red{A7}) of \cite{florkleb4})
\be\label{solveforI3}I_3=\frac{1}{\sin\pi(\beta_1+\beta_2+\beta_3)}\Bigg[-\sin(\pi\beta_1)I_1-\sin\pi(\beta_1+\beta_2)I_2-\sum_{k=4}^6\left(\sin\pi\sum_{l=1}^k\beta_l\right)I_{k}\Bigg].\ee
From (\ref{powers}), we see that the factor multiplying the bracketed terms on the right side of (\ref{solveforI3}) is $-\sin(12\pi/\varkappa)^{-1}$ and thus has a simple pole at $\varkappa=\kappa=12/r$.  But because $I_3$ is an analytic function of $\varkappa\in(0,8)\times i\mathbb{R}$ (after we extend it to this region via (\ref{replacePoch})), the bracketed factor on the right side of (\ref{solveforI3}) must vanish at $\varkappa=\kappa$.  Hence, to find $I_3(\kappa)$, we expand the terms and denominator on the right side of (\ref{solveforI3}) to first order in $\varkappa-\kappa$ and send $\varkappa\rightarrow\kappa$.  We find
\begin{align}
\label{I3dec}I_3(\kappa)&=(-1)^r\left(\frac{\kappa^2}{12\pi}\right)\sum_{k=4}^6\left[\left(\sin\pi\sum_{l=1}^k\beta_l(\kappa)\right)\partial_\varkappa I_{k}(\kappa)+\left(\cos\pi\sum_{l=1}^k\beta_l(\kappa)\right)\left(\pi\sum_{l=1}^k\partial_\varkappa\beta_l(\kappa)\right)I_k(\kappa)\right]\\
\label{I3dec2}&+(-1)^r\left(\frac{\kappa^2}{12\pi}\right)\left[\cos\left(\frac{4\pi}{\kappa}\right)\left(-\frac{4\pi}{\kappa^2}\right)I_1(\kappa)+\sin\left(\frac{4\pi}{\kappa}\right)\partial_\varkappa I_1(\kappa)\right]\\
\label{I3dec3}&+(-1)^r\left(\frac{\kappa^2}{12\pi}\right)\left[\cos\left(\frac{8\pi}{\kappa}\right)\left(-\frac{8\pi}{\kappa^2}\right)I_2(\kappa)+\sin\left(\frac{8\pi}{\kappa}\right)\partial_\varkappa I_2(\kappa)\right].
\end{align}
Because the terms in (\ref{I3dec}) do not involve an integration contour with an endpoint at $x_1$, $x_2$ or $x_3$, they are analytic at $x_1=x_2=x_3$.  Multiplied by the power-law factors in (\ref{Pi1N3}) (not included in (\ref{twoparts}), and now with $x_2=(1-\lambda)x_1+\lambda x_3$), they contribute exclusively to the second sum in (\ref{log2}) with $i=1$.  

Next, the term in (\ref{I3dec2}) (resp.\ (\ref{I3dec3})) with $I_1(\kappa)$ (resp.\ $I_2(\kappa)$) not differentiated has an integration contour with an endpoint at $x_1$, $x_2$, or $x_3$.  As a result, its asymptotic behavior is more complicated.  Substituting $u_2(t)=x_i(1-t)+x_{i+1}t$ with $i=1$ (resp.\ $i=2$) and $x_2=(1-\lambda)x_1+\lambda x_3$ into (\ref{Ik}) with $k=1$ (resp.\ $k=2$), we find that this term is $O((x_3-x_1)^{1-12/\kappa})$ as $x_3,x_2\rightarrow x_1$ with $\lambda\in(0,1)$ fixed.  Multiplied by the power-law factors in (\ref{Pi1N3}) (not included in (\ref{twoparts}), and now with $x_2=(1-\lambda)x_1+\lambda x_3$), it contributes exclusively to the first sum in (\ref{log2}) with $i=1$.

The behavior as $x_3,x_2=(1-\lambda)x_1+\lambda x_3\rightarrow x_1$ of the final terms in (\ref{I3dec2}) and (\ref{I3dec3}) with $I_1(\kappa)$ and $I_2(\kappa)$ respectively differentiated is more complicated and interesting.  Using (\ref{Ik}), after inserting $u_2(t)=(1-t)x_i+tx_{i+1}$ with $i=1$ (resp.\ $i=2$) into the former (resp.\ latter) term and inserting $x_2=(1-\lambda)x_1+\lambda x_3$ with $\lambda\in(0,1)$ into both terms, we find
\begin{align} I_1\Big(x_1,x_2=(1-\lambda)x_1+\lambda &x_3,\ldots,x_7\Big)=\lambda^{1-8/\varkappa}(x_3-x_1)^{1-12/\varkappa}\nonumber\\
\label{subform1}&\times\,\int_0^1t^{-4/\varkappa}(1-t)^{-4/\varkappa}(1-\lambda t)^{-4/\varkappa}\mathcal{N}\Bigg[\prod_{j=4}^7\big[x_j-(1-\lambda t)x_1-\lambda tx_3\big]^{\beta_j(\varkappa)}\Bigg]\,{\rm d}t,\\
I_2\Big(x_1,x_2=(1-\lambda)x_1+\lambda&x_3,\ldots,x_7\Big)=(1-\lambda)^{1-8/\varkappa}(x_3-x_1)^{1-12/\varkappa}\int_0^1[1-(1-\lambda)(1-t)]^{-4/\varkappa}t^{-4/\varkappa}(1-t)^{-4/\varkappa}\nonumber\\
\label{subform2}&\times\,\mathcal{N}\Bigg[\prod_{j=4}^7\big[x_j-(1-\lambda)(1-t)x_1-(1-(1-\lambda)(1-t))x_3\big]^{\beta_j(\varkappa)}\Bigg]\,{\rm d}t.\end{align}
With (\ref{subform1}, \ref{subform2}), the asymptotic behaviors of $\partial_\varkappa I_1(\kappa)$ and $\partial_\varkappa I_2(\kappa)$ as $x_3,x_2=(1-\lambda)x_1+\lambda x_3\rightarrow x_1$ become evident.  Indeed, differentiating (\ref{subform1}, \ref{subform2}) with respect to $\varkappa$, and setting $\varkappa=\kappa$ in the result, we find
\begin{multline}\label{lastsum}(-1)^r\left(\frac{\kappa^2}{12\pi}\right)\left[\sin\left(\frac{4\pi}{\kappa}\right)\partial_\varkappa I_1(\kappa\,|\,x_1,x_2,\ldots,x_7)+\sin\left(\frac{8\pi}{\kappa}\right)\partial_\varkappa I_2(\kappa\,|\,x_1,x_2,\ldots,x_7)\right]=\\
\begin{aligned}\frac{(-1)^r}{\pi}\log(x_3-x_1)&\overbrace{\bigg[\sin\left(\frac{4\pi}{\kappa}\right)I_1(\kappa\,|\,x_1,x_2,\ldots,x_7)+\sin\left(\frac{8\pi}{\kappa}\right)I_2(\kappa\,|\,x_1,x_2,\ldots,x_7)\bigg]}^{1.}+\,(-1)^r\left(\frac{\kappa^2}{12\pi}\right)\\
\times\,(x_3-x_1)^{1-12/\kappa}&\overbrace{\bigg[\sin\left(\frac{4\pi}{\kappa}\right)\partial_\varkappa\left\{\parbox{3.2cm}{$\lambda^{1-8/\varkappa}\times$ integral on right side of (\ref{subform1})}\right\}+\sin\left(\frac{8\pi}{\kappa}\right)\partial_\varkappa\left\{\parbox{3.4cm}{$(1-\lambda)^{1-8/\varkappa}\times$ integral on right side of (\ref{subform2})}\right\}\bigg]_{\varkappa=\kappa}}^{2.}.\end{aligned}\end{multline}
The right side of (\ref{lastsum}) gives the asymptotic behavior as $x_3,x_2=(1-\lambda)x_1+\lambda x_3\rightarrow x_1$ of the sum of terms in (\ref{I3dec2}) and (\ref{I3dec3}) that have $I_1(\kappa)$ and $I_2(\kappa)$ respectively differentiated.  As a part of $I_3(\kappa)$, these terms also form a part of $I_{35}(\kappa)$ via (\ref{I35}).  In particular, the first brace in (\ref{lastsum}) gives rise to terms of $I_{35}(\kappa)$ whose sum is proportional to the first brace in (\ref{twoparts}).  In the paragraph beneath (\ref{twoparts}), we determine that this brace contributes exclusively to the second series on the right side of (\ref{log2}).  However, $\log(x_3-x_1)$ multiplies this quantity in our present situation (\ref{lastsum}).  Hence, the first brace in (\ref{lastsum}) contributes to the third series on the right side of (\ref{log2}) with $i=1$.  In fact, these latter terms are the only terms appearing on the right side of (\ref{twoparts}) that contribute to the logarithmic series of the Frobenius series expansion (\ref{log2}) for $\Pi_1(\kappa)$ (\ref{Pi1N3}).

Finally, the terms beneath the second brace in (\ref{lastsum}) are analytic at $x_3=x_2=x_1$.  Multiplied by the preceding factor of $(x_3-x_1)^{1-12/\kappa}$ in (\ref{lastsum}) and then by the power-law factors in (\ref{Pi1N3}) (not included in (\ref{twoparts}), and now with $x_2=(1-\lambda)x_1+\lambda x_3$), these last terms evidently give a Frobenius series in powers of $x_3-x_1$ and with indicial power $1-6/\kappa$.  Because $12/\kappa=r\in\mathbb{Z}^+$, the difference $r-1$ between this power and $6/\kappa$ is a positive integer.  Hence, the terms beneath the second brace in (\ref{lastsum}) contribute exclusively to both the first and second sum in (\ref{log2}) with $i=1$.  Putting everything together, we conclude that $\Pi_1(\kappa)$ (\ref{Pi1N3}) equals a sum of Frobenius series of the form (\ref{log2}) with $i=1$.  (So far for the case where $12/\kappa=r\in\mathbb{Z}^++1$ is coprime with three, we have assumed that $\kappa>4$.  As usual, we easily extend these results to other such $\kappa\in(0,4]$ via analytic continuation (\ref{replacePoch}) to $\varkappa\in(0,8)\times i\mathbb{R}$.)

To finish, we note some interesting facts.  First, the first and third series of (\ref{log2}) (and in particular, the third series, which has a logarithm) do not arise from terms that sum to give an integration contour surrounding the interval to be collapsed, which is $[x_1,x_3]$ here.  Indeed, the terms that sum to give such a contour are those of the first brace in (\ref{twoparts}).  Rather, the first and third series of (\ref{log2}) arise from terms with an integration contour terminating at just one endpoint of the interval to be collapsed.  All of these terms arise from the second brace in (\ref{twoparts}).  (This is exactly the same circumstance that we observed in appendix \red{A} of \cite{florkleb4}, where we investigated the case with $8/\kappa\in2\mathbb{Z}^++1$.  There, we referred to these two scenarios as case \red{2} and case \red{3} respectively.)

A second observation follows from the above analysis. Either both the first sum and the third sum (containing the logarithm) are present together on the right side of (\ref{log2}) ($A_m,C_m\neq0$), or neither is present ($A_m=C_m=0$), at least if $N>2$.  (See the following paragraph for $N=2$.)  Previously, we observed that the series expansion (\ref{log}) in the case with $8/\kappa=r\in2\mathbb{Z}^+-1$ has the same property.  (See the discussion beneath (\ref{log}).)  This is natural from the following CFT point-of-view: the expansion of (\ref{Jake'sOPE}) either contains contributions for all elements of the staggered logarithmic module (i.e., $\phi$, $\smash{\hat{\psi}}$, and $\partial\phi$), or only contributions from the regular partner $\partial\phi$.  We may isolate the $\partial \phi$ contributions in the OPE (\ref{Jake'sOPE}) by altering the value of $c$ in this OPE and subtracting the result from (\ref{Jake'sOPE}).  This subtraction leaves only terms proportional to $\varepsilon^{-3\theta_1+\theta_3}\mathcal{F}_3(\lambda)\partial\phi(x)$ and its descendants, and these in turn produce only the terms of the second series in (\ref{log2}).  Thus, reasoning using logarithmic CFT explains this second observation.

For a final observation, we consider the exceptional case $N=2$ mentioned in the previous paragraph.  Indeed, if $12/\kappa=r\in\mathbb{Z}^++1$ is coprime with three, then any element of $\mathcal{S}_2$ admits the expansion (\ref{log2}) with $C_m=0$ for all $m\in\mathbb{Z}^+$, although we may have $A_m\neq0$.  This fact demonstrates the at times  wraithlike behavior of  logarithmic terms in logarithmic CFT, or more precisely, the fact that $N$ must be sufficiently large in order for such terms to appear in Frobenius series expansions of correlation functions. We can justify this claim for $\kappa=6$ and $i=1$ by considering the rectangle crossing weight $\Pi_1(\kappa=6)$ (\ref{Pi1}) and the constant solution $1$.  Taken together, these two functions span $\mathcal{S}_2$ (but only if $\kappa=6$).  Obviously, the claim is true for the latter solution, so we turn our attention to the former.  Inserting the OPE (\ref{Jake'sOPE}) into the correlation function (\ref{2Nptfunc}) with $N=2$, $x_1=x$, $x_2=x+\lambda\varepsilon$, $x_3=x+\varepsilon$, $\varepsilon>0$, and $\lambda\in(0,1)$ gives (with $\kappa=6$, we have $\theta_1=0$ and $\theta_3=1$) 
\begin{multline}\label{4ptexpand}\langle\{\phi(x)\phi(x+\lambda\varepsilon)\}_\upsilon\phi(x+\varepsilon)\phi(x_4)\rangle\underset{\varepsilon\rightarrow0}{\sim}
\Pi_1(\lambda)\langle\phi(x)\phi(x_4)\rangle+\varepsilon\mathcal{F}_3(\lambda)\langle\smash{\hat{\psi}}(x)\phi(x_4)\rangle\\
+\varepsilon\left[\mathcal{F}_2(\lambda)+\log(c\,\varepsilon)\mathcal{F}_3(\lambda)\right]\partial_x\langle\phi(x)\phi(x_4)\rangle+\dotsm.\end{multline}
From within the correlation function on the left side of (\ref{4ptexpand}), the two-leg channel exclusively propagating between $\phi(x)$ and $\phi(x+\lambda\varepsilon)$ exclusively projects out the two-leg channel between $\phi(x+\varepsilon)$ and $\phi(x_4)$.  That is,
\be\label{onechannel}
\langle\{\phi(x)\phi(x+\lambda\varepsilon)\}_\upsilon\phi(x+\varepsilon)\phi(x_4)\rangle=\langle\{\phi(x)\phi(x+\lambda\varepsilon)\}_\upsilon\{\phi(x+\varepsilon)\phi(x_4)\}_\upsilon\rangle.\ee
With only the two-leg channel propagating between the two leftmost one-leg boundary operators of (\ref{onechannel}) and between the two rightmost, the correlation functions of (\ref{onechannel}), and that of the left side of (\ref{4ptexpand}), equal Cardy's formula: $\Pi_1(\kappa=6\,|\,x,x+\lambda\varepsilon,x+\varepsilon,x_4)$ (\ref{Pi1}, \ref{Pi1hyper}) \cite{c3}.  Furthermore, with $\theta_1=0$ for $\kappa=6$, the two-point function $\langle\phi(x)\phi(y)\rangle=(y-x)^{-2\theta_1}$ is a constant. As a result, the logarithmic term in (\ref{4ptexpand}) vanishes (and so too do all of its descendants).  Although perhaps counterintuitive, it then follows that logarithms do not appear in the Frobenius series expansion (\ref{log2}) with $i=1$ of Cardy's formula, in spite of the presence of logarithmic operators in (\ref{Jake'sOPE}) \cite{rid}.  (Of course, we may also verify this claim from the explicit formula (\ref{Pi1}) for $\Pi_1(\kappa=6)$.)  This is an example of the $q-1=N$ case (here, with $q=3$ and $N=2$) attributable to pre-logarithmic operators, as discussed in the paragraphs following (\ref{2Nptfunc}).  In conclusion, because the collection $\{1,\Pi_1(\kappa=6)\}$ spans $\mathcal{S}_2$, the above arguments imply that the third series in (\ref{log2}) with $i=1$ vanishes (while the first series may or may not) for all elements of $\mathcal{S}_2$.  Invoking symmetry arguments and/or the alternative basis $\{1,\Pi_2(\kappa=6)\}$ (where $\Pi_2$ is the second rectangle connectivity weight, generated from $\Pi_1$ by replacing $\mathscr{P}(x_3,x_4)\mapsto\mathscr{P}(x_2,x_3)$ in (\ref{Pi1})), we easily extend this fact to $i=2$.

Returning to other values of $\kappa$ and $N=2$ we note that logarithmic terms do not appear in the Frobenius series expansions of solutions in $\mathcal{S}_2$ if $12/\kappa\in\mathbb{Z}^++1$ is coprime with three (which as mentioned occurs for $\kappa=\kappa_{3,q'}$), but such terms may appear for $\kappa=\kappa_{2,q'}$ if $8/\kappa\in2\mathbb{Z}^+-1$, which includes the case $\kappa=8/3$ mentioned above.  Interestingly, $\kappa=8/3$ is the SLE$_\kappa$ dual \cite{rohshr,knk} of $\kappa=6$, where logarithmic terms do not appear.

Finally, for the case $\kappa=6$, we mention that the two-point function $\langle\phi(x)\phi(y)\rangle$ is normalized to equal one in the OPE (\ref{Jake'sOPE}).  To see this, we note that the correlation function on the left side of (\ref{onechannel}), being Cardy's formula $\Pi_1(\kappa=6\,|\,x,x+\lambda\varepsilon,x+\varepsilon,x_4)$ (\ref{Pi1}, \ref{Pi1hyper}), depends exclusively on the cross-ratio
\be\label{eta}\eta=\frac{(x+\lambda\varepsilon-x)(x_4-x+\varepsilon)}{(x+\varepsilon-x)(x_4-x-\lambda\varepsilon)}\underset{\varepsilon\rightarrow0}\sim\lambda+O(\varepsilon).\ee
The asymptotic behavior given in (\ref{eta}) implies that the left side of (\ref{4ptexpand}, \ref{onechannel}), equaling $\Pi_1(\kappa=6\,|\,\eta)$, approaches $\Pi_1(\kappa=6\,|\,\lambda)$ as $\varepsilon\rightarrow0$.  Meanwhile, the right side of (\ref{4ptexpand}) approaches $\Pi_1(\kappa=6\,|\,\lambda)\langle\phi(x)\phi(x_4)\rangle$ in this limit.  Matching these two quantities gives $\langle\phi(x)\phi(x_4)\rangle=1$, as claimed.

\end{document}